\documentclass[preprint,3p]{elsarticle}
\journal{Annals of Physics}

\usepackage{ifxetex}
\ifxetex
\else
   \pdfoutput=1
\fi

\usepackage{amsmath,amsthm}
\usepackage{amsfonts,amssymb}
\usepackage{siunitx}
\usepackage{graphicx}
\graphicspath{{figures/}}         
\usepackage[labelformat=simple]{subcaption}
\usepackage{hyperref}

\newcommand{\etal}{\textit{et al}.\ }
\newcommand{\ve}{\epsilon}
\newcommand{\im}{i}

\DeclareMathOperator{\ofq}{\mathnormal{\!(\mathbf{q})\,}}
\DeclareMathOperator{\ofp}{\mathnormal{\!(\mathbf{p})\,}}
\DeclareMathOperator{\ofpq}{\mathnormal{\!(\mathbf{p} \, | \, \mathbf{q})\,}}

\newcommand{\dtwox}{{\mathrm{d}^2 x_\parallel}}

\DeclareMathOperator{\klb}{\mathnormal{\mathbf{k}_l^b}}
\DeclareMathOperator{\kma}{\mathnormal{\mathbf{k}_m^a}}

\DeclareMathOperator{\xpar}{\mathnormal{\mathbf{x}_\parallel}}

\newcommand{\Vie}[3]{\mathop{\mathbf{#1}_{#2}^{#3}}}
\newcommand{\Cie}[3]{\mathop{\mathcal{#1}_{#2}^{#3}}}

\newcommand{\BCie}[3]{\mathop{\boldsymbol{ \mathcal{#1}}_{#2}^{#3}}}
\newcommand{\ofuipi}[4]{\mathop{(\mathbf{#1}_{#2}\, |\,\mathbf{#3}_{#4})}}
\newcommand{\dtwopi}[2]{\mathop{\frac{\mathrm{d}^2{#1}_{#2}}{(2 \pi)^2}}}

\newcommand{\of}[1]{\mathop{(\mathbf{#1})}}
\newcommand{\ofuipiMulti}[6]{\mathop{(\mathbf{#1}_{#2} \, | \, \mathbf{#3}_{#4} \, | \, \mathbf{#5}_{#6})}}

\DeclareMathOperator{\st}{ | }

\begin{document}
\begin{frontmatter}

\title{Selective enhancement of Sel\'{e}nyi rings induced by the cross-correlation between the interfaces of a two-dimensional randomly rough dielectric film}
\author[1]{J.-P. Banon\corref{cor1}}
\author[1]{\O. S. Hetland}
\author[1,2]{I. Simonsen}
\cortext[cor1]{Corresponding author}
\address[1]{Department of Physics, NTNU -- Norwegian University of Science and Technology, NO-7491 Trondheim, Norway}
\address[2]{Surface du Verre et Interfaces, UMR 125 CNRS/Saint-Gobain, F-93303 Aubervilliers, France}

\begin{abstract}
  By the use of both perturbative and non-perturbative solutions of the reduced Rayleigh equation, we present a detailed study of the scattering of light from two-dimensional weakly rough dielectric films. It is shown that for several rough film configurations, Sel\'{e}nyi interference rings exist in the diffusely scattered light. For film systems supported by dielectric substrates where only one of the two interfaces of the film is weakly rough and the other planar, Sel\'{e}nyi interference rings are observed at angular positions that can be determined from simple phase arguments. For such single-rough-interface films, we find and explain by a single scattering model that the contrast in the interference patterns is better when the top interface of the film (the interface facing the incident light) is rough than when the bottom interface is rough. When both film interfaces are rough, Sel\'{e}nyi interference rings exist but a potential cross-correlation of the two rough interfaces of the film can be used to selectively enhance some of the interference rings while others are attenuated and might even disappear. This feature may in principle be used in determining the correlation properties of interfaces of films that otherwise would be difficult to access.
\end{abstract}

\end{frontmatter}

\section{Introduction}
Interference effects in the diffuse light scattered by thin and rough dielectric films can look both stunning and unexpected, and they have fascinated their observers for centuries.
First formally described in modern times as colorful rings in the diffusely scattered light originating from a dusty back-silvered mirror by Newton \cite{Newton1730}, what is today known as Qu\'{e}telet- and Sel\'{e}nyi-rings have been thoroughly analyzed theoretically \cite{DeWitte1967,Freilikher1994,Lu1998,Calvo-Perez1999,Suhr2009} and experimentally \cite{Raman1921,Kaganovskii1999}.
An example of a non-laboratory situation where one may observe this phenomenon is in light reflections from bodies of water if appropriate algae are present on the water surface. This phenomenon, modeled as a thin layer of spherical scatterers suspended on a reflecting planar surface, was investigated by Suhr and Schlichting \cite{Suhr2009}.

In a theoretical study of the scattering from one-dimensional randomly rough surfaces ruled on dielectric films on perfectly conducting substrates, Lu \etal\cite{Lu1998} concluded that the degree of surface roughness had the biggest impact on which interference phenomena could be observed. For films with a thickness on the order of several wavelengths they were able to explain the periodic fringes they observed in the mean differential reflection coefficient through simple phase arguments. The patterns in the diffusely scattered light were shown to undergo a transition, with increasing surface roughness, from an intensity pattern exhibiting fringes whose angular positions are independent of the angle of incidence (Sel\'{e}nyi rings \cite{Selenyi1911}) to one with fringes whose angular positions depend on the angle of incidence (Qu\'{e}telet rings \cite{Raman1921}) and eventually into a fringeless pattern with a backscattering peak, which is a signature of multiple scattering \cite{Simonsen2010}. Although the Sel\'{e}nyi rings are centered around the mean surface normal, with their position being independent of the angle of incidence,  their amplitude, however, is modulated by the angle of incidence.
According to the current understanding of the Sel\'{e}nyi rings, their main origin is due to the interference between light scattered back directly from the top scattering layer and light reflected by the film after being scattered within it.
In this paper we seek to complete this interpretation of the interference phenomena within a single scattering approximation, enabling a sound interpretation of the Sel\'{e}nyi rings for the previously unexplored case when the rough surface is shifted to the non-incident face of the film.

A similar system to the one studied by Lu \etal was also thoroughly studied perturbatively and experimentally by Kaganovskii \etal\cite{Kaganovskii1999}. They concluded that the long-range (smooth) component of the surface roughness, whenever present, can have a deciding effect on the interference pattern observed in the diffusely scattered light.

However, most of the relevant studies conducted on the topic so far have been restricted to investigations of scattering from a single rough interface. Allowing for more than one rough interface significantly increases the complexity of the problem both analytically and computationally, but it also opens a door to a richer set of scattering phenomena.
Such stacked, multi-layered systems will in many cases better represent the real-world scattering systems we are attempting to model~\cite{DeWitte1967}. Two or more of these randomly rough interfaces in the stack will also often be correlated, either naturally occuring, by design or by method of production~\cite{Amra1992,Pan2010}.
Since both Qu\'{e}telet- and Sel\'{e}nyi-rings may enable a practical way of remote sensing and surface characterization for certain geometries and layer thicknesses, it is important also to model the impact of such roughness cross-correlation.

\smallskip
In this paper we investigate interference effects in the light scattered diffusely from an optical system composed of two semi-infinite media separated by a single thin dielectric film where both interfaces may be rough [Fig.~\ref{fig:system}(a)]. After describing the statistical properties of the interfaces in Sec.~\ref{sec:scatt:sys}, we derive, in Sec.~\ref{sec:theory}, a set of reduced Rayleigh equations~(RREs) for the case of electromagnetic scattering from a system with two rough interfaces, inspired by the work of Soubret~\textit{et al}.~\cite{Soubret2001a}.
Although only the case of reflected light will be analyzed in detail, the RREs for both the reflection and the transmission amplitudes are given for completeness; furthermore, this also serves to show that the presented framework can easily be generalized to an arbitrary number of rough interfaces.
A perturbative method and a purely numerical method for solving the RREs are described in Sec.~\ref{sec:numerics}. Since solving the RREs for a set of two, or more,  two-dimensional randomly rough surfaces by purely numerical  means  is a highly computationally intensive task, the perturbative method will be our main investigation tool for simulating and interpreting interference effects in such geometries. In Sec.~\ref{sec:single:rough} we discuss rough film geometries where either the top interface or bottom interface of the film is allowed to be randomly rough and the other interface is planar.
For such geometries, we compare the predictions for the scattered intensities obtained on the basis of the perturbative and non-perturbative methods. After having established the apparent validity of the perturbative method for the level of roughness assumed, we continue to investigate rough film geometries where both interfaces of the film are randomly rough and have a varying cross-correlation~[Sec.~\ref{sec:two:rough}]. Section~\ref{sec:transmission} gives a brief discussion concerning additional effects one expects to observe in transmission. Finally, Sec.~\ref{Sec:Conculusions} presents the conclusions that we have drawn from this study.

\section{Scattering systems}\label{sec:scatt:sys}
\begin{figure}[t]
  \centering
  \begin{subfigure}[b]{0.5\textwidth} 
    \includegraphics[width=1\linewidth , trim= 0.cm -1.cm 0.cm 0.cm,clip]{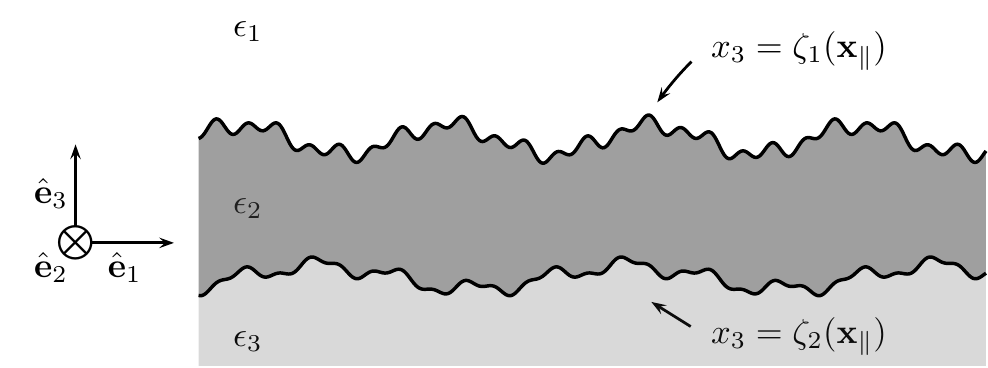}
    \caption{}
  \end{subfigure}%
  \begin{subfigure}[b]{0.5\textwidth}
    \centering
    \includegraphics[width=.85\linewidth , trim= -0.cm 0.cm 0.cm 0.cm,clip]{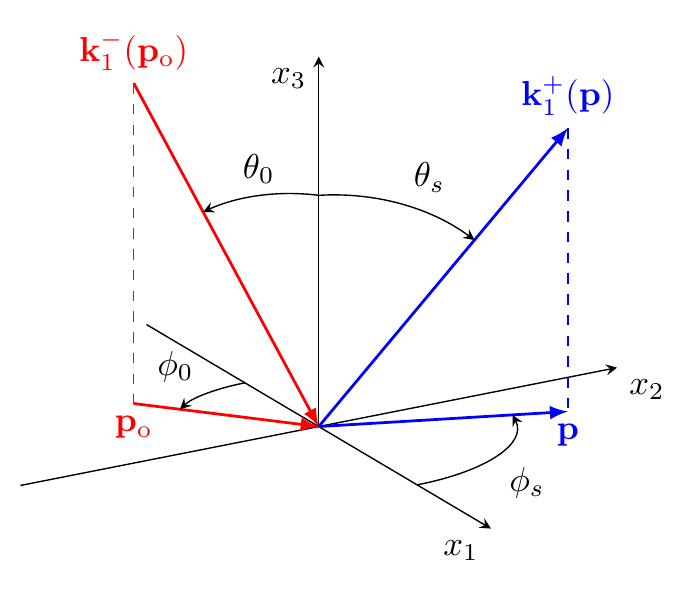}
    \caption{}
  \end{subfigure}
\caption{
  (a) Layered system with two rough interfaces. (b) Definitions of the angles of incidence and scattering and wave vectors.}
\label{fig:system}
\end{figure}

An overview of a typical system geometry is provided in Fig.~\ref{fig:system}.
We consider the case where both interfaces of the film may be randomly rough and possess non-trivial auto- and cross-correlation.
Furthermore, we will be interested in scattering systems for which the mean thickness of the film is several wavelengths so that interference fringes can be observed in the diffusely reflected or transmitted intensities.
The definition of the geometry is set in the three-dimensional space endowed with a Cartesian coordinate system $(O,\hat{\mathbf{e}}_1,\hat{\mathbf{e}}_2,\hat{\mathbf{e}}_3)$, with the vector plane $(\hat{\mathbf{e}}_1,\hat{\mathbf{e}}_2)$ parallel to the mean plane of the interfaces [Fig.~\ref{fig:system}(b)].
The origin, $O$, can be arbitrarily chosen, only affecting the complex reflection and transmission amplitudes by an overall phase factor which plays no role in the intensity of the scattered light.
The scattering system splits space into a slab of three domains, or layers, that will be denoted by the indices $j \in \{1,2,3\}$.
The mean thickness of the film will be denoted $d>0$, and the $j^\mathrm{th}$ interface separating media $j$ and $j+1$ can be described by the equation
\begin{equation}
  x_3 = \zeta_j (\mathbf{x}_\parallel) = d_j + h_j(\mathbf{x}_\parallel) \: ,
\end{equation}
for $j \in \{1,2\}$, where $\mathbf{x}_\parallel = x_1 \: \hat{\mathbf{e}}_1 + x_2 \: \hat{\mathbf{e}}_2$, $d_{j} = \langle \zeta_j \rangle$ denotes the average of the $j^\mathrm{th}$ profile (and we have $d_1 - d_2 = d$), and the term $h_j$ will be assumed to be a continuous, differentiable, single-valued, stationary, isotropic, Gaussian random process with zero mean and given auto-correlation. More specifically, the surface profile functions are assumed to satisfy the following properties
\begin{subequations}
\begin{align}
\left\langle h_j(\mathbf{x}_\parallel) \right\rangle &= 0 \\
  \left\langle h_j(\mathbf{x}_\parallel) h_j(\mathbf{x}_\parallel') \right\rangle &= \sigma_j^2 \: W(\mathbf{x}_\parallel - \mathbf{x}_\parallel').
\label{eq:auto_covariance}%
\end{align}
\end{subequations}
Here and in the following, the angle brackets denote an average over an ensemble of realizations of the stochastic process, $\sigma_j$ denotes the rms roughness of interface $j$ and $W(\mathbf{x}_\parallel)$ represents the height auto-correlation function normalized so that $W(\mathbf{0}) = 1$. For reasons of simplicity we here restrict ourselves to the situation where both interfaces are characterized by the same form of the correlation function. In particular, we will here assume a Gaussian form of the auto-correlation function that is defined by
\begin{equation}
W(\mathbf{x}_\parallel) = \exp \left( - \frac{|\mathbf{x}_\parallel|^2}{a^2} \right) \: ,
\end{equation}
where $a$ is the correlation length.
The corresponding power spectrum (defined as the Fourier transform of $W$) is then
\begin{equation}
g(\Vie{p}{}{}) = \pi a^2 \exp \left( - \frac{|\Vie{p}{}{}|^2 a^2}{4} \right) \: ,
\end{equation}
with $\Vie{p}{}{} =  p_1 \: \hat{\mathbf{e}}_1 + p_2 \: \hat{\mathbf{e}}_2$. In addition, the two interfaces will be assumed to be cross-correlated in the following way
\begin{equation}
\left\langle h_1(\mathbf{x}_\parallel) h_2(\mathbf{x}_\parallel') \right\rangle = \gamma \: \sigma_1 \sigma_2 \: W(\mathbf{x}_\parallel - \mathbf{x}_\parallel') \: ,
\label{eq:cross_covariance}
\end{equation}
where $\gamma \in [-1,1]$ is a dimensionless cross-correlation coupling variable. When $\gamma = 0$ the two interfaces are uncorrelated, and the extreme cases $\gamma = \pm 1$ and $\sigma_1 = \sigma_2$ can be viewed respectively as the second interface being a shifted copy of the first one by a vector $-d \, \hat{\mathbf{e}}_3$, or as the second interface being a symmetric copy of the first one with respect to the plane $x_3 = (d_1 + d_2) / 2$.
We can summarize the correlations expressed by Eqs.~(\ref{eq:auto_covariance}) and (\ref{eq:cross_covariance}) by the following relation
\begin{equation}
\left\langle h_i(\mathbf{x}_\parallel) h_j(\mathbf{x}_\parallel') \right\rangle = [ \delta_{ij} + \gamma (1 - \delta_{ij}) ] \: \sigma_i \sigma_j \: W(\mathbf{x}_\parallel - \mathbf{x}_\parallel')  \: ,
\label{eq:sumup:cross-covariance}
\end{equation}
where $\delta_{ij}$ denotes the Kronecker delta.


\section{Formulation of the problem}\label{sec:theory}

The theoretical approach used in this work to study the scattering of light from the systems of interest is based on the so-called reduced Rayleigh equations. A reduced Rayleigh equation is an integral equation in which the integral kernel encodes the materials and geometry of the scattering system and the unknowns are the reflection or transmission amplitudes for each polarization.
In the following, in order to establish the notation and highlight the main assumptions of the method, we will briefly recall the key ideas of the derivation of the reduced Rayleigh equations for a system composed of three media separated by two disjoint rough interfaces.
We will use, to our knowledge, the most general form of the reduced Rayleigh equations for a single interface derived by Soubret \etal in Ref.~\cite{Soubret2001a} and used by these authors in Refs.~\cite{Soubret2001a,Soubret2001} in the case of a single interface system and a film geometry. Once the general framework is established, we will apply it to the specific geometries of interest.

\subsection{The reduced Rayleigh equations}
All physical quantities introduced hereafter will be indexed with respect to the medium (domain) they belong to. The electromagnetic response of the media is modeled by non-magnetic, homogeneous, isotropic, linear constitutive relations in the frequency domain, i.e. that \emph{a priori} each medium is characterized by frequency dependent scalar complex dielectric functions, $\epsilon_j(\omega)$, where $\omega$ denotes the frequency of the electromagnetic wave excitation.
We consider the presence of an electromagnetic field $(\mathbf{E},\mathbf{H})$ in the whole space.
The fields will be denoted by a subscript $j$ depending on their containing medium. As an example, the electric field evaluated at a point $\mathbf{x}$ in medium 1 at time $t$ is denoted $\mathbf{E}_1 (\mathbf{x},t) = \mathbf{E}_1 (\mathbf{x},\omega)\exp(-i\omega t)$.
The source free Maxwell equations, together with homogeneous, linear and isotropic constitutive relations in the frequency domain, result in the electric and magnetic fields satisfying the Helmholtz equation in each region. Namely, for all $j \in \{1,2,3\}$,
\begin{equation}
	\nabla^2 \mathbf{E}_j (\Vie{x}{}{},\omega) + \epsilon_j (\omega) \: \left(\frac{\omega}{c} \right)^2 \, \mathbf{E}_j (\Vie{x}{}{},\omega) = \mathbf{0} \: \mathrm{,}
	\label{helmoholtz}
\end{equation}
and a similar equation satisfied for $\mathbf{H}$.
Here, $\nabla^2$ denotes the Laplace operator and $c$ represents the speed of light in vacuum. In the following, we will drop the time, or frequency, dependence, since we assume a stationary regime where time contributes only by an overall phase factor $\exp(- i \omega t)$.
It is known that a solution to the Helmholtz equation can be written as a linear combination of plane waves, thus the representation of the electric field in each region can be written as
%
%
\begin{equation}
	\mathbf{E}_j (\mathbf{x}) = \sum_{a=\pm} \: \int_{\mathbb{R}^2} \: \left[ \Cie{E}{j,p}{a} \ofq \,
  \hat{\mathbf{e}}_{p,j}^{a} \ofq  + \Cie{E}{j,s}{a} \ofq \, \hat{\mathbf{e}}_s \ofq  \right] \,
  \exp \left( i \, \Vie{k}{j}{a} \ofq \cdot \mathbf{x} \right) \dtwopi{q}{} \: \mathrm{,}
	\label{fieldexpansion}
\end{equation}
%
where 
\begin{subequations}
\begin{align}
	\alpha_j \ofq &= \sqrt{\epsilon_j \left(\frac{\omega}{c} \right)^2 - \mathbf{q}^2}, \qquad \mathrm{Re} \, (\alpha_j), \mathrm{Im} \, (\alpha_j) \geq 0 \: ,\\
	\Vie{k}{j}{\pm} \ofq &= \mathbf{q} \pm \alpha_j \ofq  \hat{\mathbf{e}}_3	\: , \\
\hat{\mathbf{e}}_s \ofq &=  \hat{\mathbf{e}}_3 \times \hat{\mathbf{q}} \: ,\\
\hat{\mathbf{e}}_{p,j}^\pm \ofq &= \frac{c}{\sqrt{\epsilon_j} \omega} \left(\pm \alpha_j\ofq \, \hat{\mathbf{q}} - |\mathbf{q}| \, \hat{\mathbf{e}}_3 \right) .
\end{align}
\label{basis}%
\end{subequations}
Here a caret over a vector indicates that the vector is a unit vector.
Note that the wave vector $\Vie{k}{j}{\pm} \ofq$ of an elementary plane wave is decomposed into its projection $\mathbf{q}$ in the lateral vector plane $(\hat{\mathbf{e}}_1, \hat{\mathbf{e}}_2)$ and the component $\pm \alpha_j \ofq$ along $\hat{\mathbf{e}}_3$.
The sum for $a=\pm$ takes into account both upwards and downwards propagating and evanescent (and possibly growing) waves. The field amplitude is decomposed in the \emph{local polarization basis} $(\hat{\mathbf{e}}_{p,j}^a \ofq,\hat{\mathbf{e}}_s \ofq)$, so that $\Cie{E}{j,\alpha}{a} \ofq$ denotes the component of the field amplitude in the polarization state $\alpha$ of the mode characterized by $a$ and $\Vie{q}{}{}$.
In this basis, the directions given by $\hat{\mathbf{e}}_{p,j}^\pm \ofq$, and $\hat{\mathbf{e}}_s \ofq$ are respectively the directions of the $p$- and $s$-polarization of the electric field amplitude. Furthermore, the electromagnetic fields have to satisfy the boundary conditions ($j \in \{1,2\}$)
\begin{subequations}
\begin{align}
	\mathbf{n}_j(\mathbf{x}_\parallel) \times \Big[ \mathbf{E}_{j+1}(\mathbf{s}_j (\mathbf{x}_\parallel) ) - \mathbf{E}_j(\mathbf{s}_j (\mathbf{x}_\parallel)) \Big] &= \mathbf{0}\\
	\mathbf{n}_j(\mathbf{x}_\parallel) \times \Big[ \mathbf{H}_{j+1}(\mathbf{s}_j (\mathbf{x}_\parallel)) - \mathbf{H}_j(\mathbf{s}_j (\mathbf{x}_\parallel)) \Big] &= \mathbf{0} \: ,
\end{align}\label{BC}%
\end{subequations}
where $\mathbf{n}_j(\mathbf{x}_\parallel)$ is a vector that is normal to surface $j$ at the surface point $\mathbf{s}_j (\mathbf{x}_\parallel) = \mathbf{x}_\parallel + \zeta_j(\mathbf{x}_\parallel) \hat{\mathbf{e}}_3$, and given by
\begin{equation}
	\mathbf{n}_j(\mathbf{x}_\parallel) = \hat{\mathbf{e}}_3 - \frac{ \partial \zeta_j}{\partial x_1} (\mathbf{x}_\parallel)\: \hat{\mathbf{e}}_1 - \frac{\partial \zeta_j}{\partial x_2} (\mathbf{x}_\parallel) \: \hat{\mathbf{e}}_2	\: \mathrm{.}
\end{equation}
Here, $\partial/ \partial x_k$ denotes the partial derivative along the direction $\hat{\mathbf{e}}_k$.
Following Soubret~\emph{et al.}~\cite{Soubret2001a}, for a given surface indexed by $j$, by substituting the field expansion Eq.~(\ref{fieldexpansion}) into Eq.~(\ref{BC}) and by a clever linear integral combination of the boundary conditions, one can show that the upward or downward field amplitudes in medium $j+1$ can be linked to the upward and downward field amplitudes in medium $j$ via the following integral equation defined for $a_{j+1} = \pm$, $j \in \{1,2\}$, and $\Vie{p}{}{}$ in the vector plane $(\hat{\mathbf{e}}_1,\hat{\mathbf{e}}_2)$:
\begin{equation}
\sum_{a_j=\pm} \int \: \Cie{J}{j+1,j}{a_{j+1},a_j}\ofpq \: \Vie{M}{j+1,j}{a_{j+1},a_{j}}\ofpq \:  \BCie{E}{j}{a_j} \ofq \dtwopi{q}{} = \frac{2 \, a_{j+1} \, \sqrt{\epsilon_{j} \epsilon_{j+1}} \,\alpha_{j+1}\ofp}{\epsilon_{j+1} - \epsilon_{j}} \, \BCie{E}{j+1}{a_{j+1}} \ofp		\: \mathrm{.}
\label{rrefinal}
\end{equation}
Here $\BCie{E}{j}{a} \ofq = (\Cie{E}{j,p}{a} \ofq,\Cie{E}{j,s}{a} \ofq)^\mathrm{T}$ denotes a column vector of the polarization components of the field amplitude in medium $j$.
Moreover, $\Vie{M}{l,m}{b,a} \ofpq$ is a 2$\times$2 matrix which originates from a change of coordinate system between the local polarization basis  $(\hat{\mathbf{e}}_{p,l}^b \ofp,\hat{\mathbf{e}}_s \ofp)$ and $(\hat{\mathbf{e}}_{p,m}^a \ofq,\hat{\mathbf{e}}_s \ofq)$, defined for $a = \pm$, $b = \pm$, and $l, m \in \{j, j+1\}$ such that $l \neq m$ as
%
\begin{equation}
	\Vie{M}{l,m}{b,a} \ofpq = \begin{pmatrix}
	|\Vie{p}{}{}| |\Vie{q}{}{}| + a b \: \alpha_l\ofp \alpha_m\ofq \hat{\mathbf{p}} \cdot \hat{\mathbf{q}}&
	-b \: \sqrt{\epsilon_m} \: \frac{\omega}{c} \: \alpha_l\ofp [\hat{\mathbf{p}} \times \hat{\mathbf{q}}] \cdot \hat{\mathbf{e}}_3\\
	a \: \sqrt{\epsilon_l} \: \frac{\omega}{c} \: \alpha_m\ofq [\hat{\mathbf{p}} \times \hat{\mathbf{q}}] \cdot \hat{\mathbf{e}}_3 &
		\sqrt{\epsilon_l \epsilon_m} \: \frac{\omega^2}{c^2} \: \hat{\mathbf{p}} \cdot \hat{\mathbf{q}}
  \end{pmatrix}.
  \label{Mdef}
\end{equation}
The kernel scalar factor $\Cie{J}{l,m}{b,a}\ofpq$ encodes the surface geometry and is defined as
\begin{equation}
\Cie{J}{l,m}{b,a} \ofpq = \left( b \alpha_l\ofp - a \alpha_m\ofq \right)^{-1} \int  \: \exp \left[-i(\klb \ofp - \kma \ofq) \cdot (\xpar + \zeta_{j}(\xpar) \: \hat{\mathbf{e}}_3) \right] \: \dtwox .
\label{Iintdef}
\end{equation}
Notice that, as already pointed out in Ref.~\cite{Soubret2001a}, due to the symmetry of the boundary conditions, one may also show in the same way that
%
\begin{equation}
\sum_{a_{j+1}=\pm} \int \: \Cie{J}{j,j+1}{a_j,a_{j+1}}\ofpq \: \Vie{M}{j,j+1}{a_j,a_{j+1}}\ofpq \:  \BCie{E}{j+1}{a_{j+1}}\ofq \dtwopi{q}{} = \frac{2 \, a_{j} \, \sqrt{\epsilon_{j} \epsilon_{j+1}} \,\alpha_{j}\ofp}{\epsilon_{j} - \epsilon_{j+1}} \, \BCie{E}{j}{a_{j}}\ofp \: \mathrm{,}
\label{rrefinalswitch}
\end{equation}
%
which can be obtained from Eq.~(\ref{rrefinal}) by interchanging $j$ and $j+1$. Typically, Eq.~(\ref{rrefinal}) is appropriate to solve the problem of reflection whereas Eq.~(\ref{rrefinalswitch}) is appropriate to solve the problem of transmission, as we will see later.
In the following, it will be convenient to define
\begin{equation}
\Vie{\Theta}{j+1,j}{a_{j+1},a_j} \ofpq = \alpha_{j+1}^{-1} (\Vie{p}{}{}) \Cie{J}{j+1,j}{a_{j+1},a_j}\ofpq \: \Vie{M}{j+1,j}{a_{j+1},a_j} \ofpq
\end{equation}
and
\begin{equation}
\Vie{\Theta}{j,j+1}{a_j,a_{j+1}} \ofpq = \alpha_{j}^{-1} (\Vie{p}{}{}) \Cie{J}{j,j+1}{a_{j},a_{j+1}}\ofpq \: \Vie{M}{j,j+1}{a_{j},a_{j+1}} \ofpq
\end{equation}
which we will refer to as the forward and backward \emph{single interface transfer kernels} between media $j$ and $j+1$, respectively.
Our aim is to study reflection from and transmission through the whole system, i.e. we need to relate the field amplitudes in regions~$1$ and $3$ without having to explicitly consider the field amplitudes in region~$2$. To this end, we have to combine Eq.~(\ref{rrefinal}) for $j=1$ and $j=2$ in order to eliminate $\BCie{E}{2}{\pm}$.
A systematic way of doing this, and which can be generalized to an arbitrary number of layers, is presented below. The key observation lies in the fact that one can choose the sign $a_{j+1}$ in Eq.~(\ref{rrefinal}) and therefore Eq.~(\ref{rrefinal}) contains two vector equations for a given $j$.
For reasons that will soon become clear, the variable $\Vie{p}{}{}$ that appears in Eq.~(\ref{rrefinal}) is renamed $\Vie{p}{2}{}$. By left-multiplying both sides of Eq.~(\ref{rrefinal}) taken at $j=1$ by $a_{2} \: \Vie{\Theta}{3,2}{a_3,a_2} \ofuipi{p}{}{p}{2}$, where $a_3 = \pm$ can be arbitrarily chosen, we obtain
\begin{equation}
\sum_{a_1=\pm} a_2 \: \int  \:  \Vie{\Theta}{3,2}{a_3,a_2}
\ofuipi{p}{}{p}{2}
\: \Vie{\Theta}{2,1}{a_2,a_1} \ofuipi{p}{2}{q}{}
\:  \BCie{E}{1}{a_1} \ofq \dtwopi{q}{} = \frac{2 \,  \sqrt{\epsilon_{1} \epsilon_2}}{\epsilon_2 - \epsilon_{1}}
\: \Vie{\Theta}{3,2}{a_3,a_2}\ofuipi{p}{}{p}{2} \: \BCie{E}{2}{a_2} (\Vie{p}{2}{})	\: \mathrm{.} \nonumber
\end{equation}
By integrating this equation over $\Vie{p}{2}{}$ divided by $(2\pi)^2$ and summing over $a_2 = \pm$, one obtains that the right-hand-side of the resulting equation is, up to a constant factor, equal to the left-hand-side of Eq.~(\ref{rrefinal}) evaluated for $j=2$. In this way we obtain
\begin{equation}
\sum_{a_1=\pm}  \int  \: \Vie{\Theta}{3,1}{a_3,a_1} \ofuipi{p}{}{q}{} \:  \BCie{E}{1}{a_1} \ofq \dtwopi{q}{} = a_3 \: \frac{4 \,  \sqrt{\epsilon_{1} \epsilon_2^2 \epsilon_3} }{(\epsilon_3 - \epsilon_2)(\epsilon_2 - \epsilon_{1})} \, \BCie{E}{3}{a_3} (\Vie{p}{}{})	\: \mathrm{,}
\label{rrefinal2}
\end{equation}
where the \emph{forward two-interface transfer kernel} $\Vie{\Theta}{3,1}{a_3,a_1}(\Vie{p}{}{}|\Vie{q}{}{})$ is defined by the composition rule
\begin{equation}
\Vie{\Theta}{3,1}{a_3,a_1} \ofuipi{p}{}{q}{} =  \sum_{a_2=\pm} a_2 \: \int  \: \Vie{\Theta}{3,2}{a_3,a_2} \ofuipi{p}{}{p}{2} \: \Vie{\Theta}{2,1}{a_2,a_1} \ofuipi{p}{2}{q}{} \: \dtwopi{p}{2} .
\label{rre2kernel}
\end{equation}
By a similar method and by the use of Eq.~(\ref{rrefinalswitch}), we obtain the backward relation
\begin{equation}
\sum_{a_3=\pm}  \int \:  \Vie{\Theta}{1,3}{a_1,a_3} \ofuipi{p}{}{q}{} \:  \BCie{E}{3}{a_3} \ofq \dtwopi{q}{} = a_1 \: \frac{4 \,  \sqrt{\epsilon_{1} \epsilon_2^2 \epsilon_3} }{(\epsilon_1 - \epsilon_2)(\epsilon_2 - \epsilon_3)} \, \BCie{E}{1}{a_1} (\Vie{p}{}{})	\: \mathrm{,}
\label{rrefinal2b}
\end{equation}
where the \emph{backward two-interface transfer kernel} $\Vie{\Theta}{1,3}{a_1,a_3}(\Vie{p}{}{}|\Vie{q}{}{})$ is defined as
\begin{equation}
\Vie{\Theta}{1,3}{a_1,a_3} \ofuipi{p}{}{q}{} =  \sum_{a_2=\pm} a_2 \: \int  \: \Vie{\Theta}{1,2}{a_1,a_2} \ofuipi{p}{}{p}{2} \: \Vie{\Theta}{2,3}{a_2,a_3}\ofuipi{p}{2}{q}{} \: \dtwopi{p}{2} .
\label{rre2kernelb}
\end{equation}
Let us now make a few remarks on Eqs.~(\ref{rrefinal2}) and (\ref{rre2kernel}). Equation~(\ref{rrefinal2}) is an integral equation of the same form as Eq.~(\ref{rrefinal}) but it only relates the field amplitudes in medium~$1$ and $3$.
Our aim of eliminating the field amplitudes in the intermediary medium is therefore achieved. However, this comes at a cost since the new transfer kernel $\Vie{\Theta}{3,1}{a_3,a_1}(\Vie{p}{}{}\,|\,\Vie{q}{}{})$ is defined as an integral of the product of two single interface kernels as can be seen in Eq.~(\ref{rre2kernel}).
We will see that this pays off in the case where one of the interfaces is flat, but that the cost can be significant in terms of computational load when both interfaces are rough.

\medskip
So far, we have stayed general and simply assumed the presence of an electromagnetic field decomposed in propagating and non-propagating waves in each region.
Therefore, there is no uniqueness in the solutions to the transfer equations, Eqs.~(\ref{rrefinal2}) and (\ref{rrefinal2b}). To ensure a unique solution, one needs to impose some constraints on the field.
First, we need to introduce an incident field to our model. This will split the field expansion into a sum of an incident field, which is given by our model of the problem, and a scattered field.
Note that within this framework, the incident field may be chosen to be in either medium, or to be a combination of excitations incident from different media. Second, we need to impose the Sommerfeld radiation condition at infinity. This implies that the non-propagating waves are indeed only evanescent waves in the media unbounded in the $\hat{\mathbf{e}}_3$-direction and that the propagating ones are directed outwards.

In our case, the incident field will be taken as a plane wave incident from medium 1 and defined as
\begin{equation}
  \Vie{E}{0}{} (\mathbf{x})
  =
  \left[ \Cie{E}{0,p}{} \, \hat{\mathbf{e}}_{p,1}^{\mathnormal{-}} (\Vie{p}{0}{})  + \Cie{E}{0,s}{} \, \hat{\mathbf{e}}_s  (\Vie{p}{0}{}) \right]
  \exp \left( i \Vie{k}{1}{-} (\Vie{p}{0}{}) \cdot \mathbf{x} \right),
  \label{incField}
\end{equation}
where $\Vie{p}{0}{}$ is the projection of the incident wave's wave vector in the $(\hat{\mathbf{e}}_1,\hat{\mathbf{e}}_2)$ plane, with the property $| \Vie{p}{0}{} | \leq \sqrt{\epsilon_1} \: \omega/c$, i.e. we consider an incident propagating wave.
The fact that this is the only incident wave considered, together with the Sommerfeld radiation condition at infinity, gives, apart from the incident field, that the only elementary waves allowed in the scattered field are those with wave vectors of the form $\Vie{k}{1}{+}(\Vie{p}{}{})$ and $\Vie{k}{3}{-}(\Vie{p}{}{})
$ in medium~$1$ and $3$, respectively. This property can be expressed by defining the field amplitudes
\begin{subequations}
\begin{align}
  \BCie{E}{1}{-} (\Vie{q}{}{}) &= (2 \pi)^2 \: \delta (\Vie{q}{}{} - \Vie{p}{0}{}) \: \BCie{E}{0}{}\: ,\\
  \BCie{E}{3}{+} (\Vie{q}{}{}) &= \mathbf{0} \: ,
\end{align}
\label{sommerfeld}%
\end{subequations}
where $\BCie{E}{0}{} = (\Cie{E}{0,p}{},\Cie{E}{0,s}{})^\mathrm{T}$.
Next, we assume that the scattered field amplitudes are linearly related to the incident field amplitude $\BCie{E}{0}{}$ via the reflection and transmission amplitudes, $\Vie{R}{}{} ( \Vie{q}{}{} \,|\, \Vie{p}{0}{})$ and $\Vie{T}{}{} ( \Vie{q}{}{} \,|\, \Vie{p}{0}{})$, defined as
\begin{subequations}
\begin{align}
  \BCie{E}{1}{+} (\Vie{q}{}{}) &= \Vie{R}{}{} ( \Vie{q}{}{}| \Vie{p}{0}{})  \BCie{E}{0}{},\\
  \BCie{E}{3}{-} (\Vie{q}{}{}) &= \Vie{T}{}{} ( \Vie{q}{}{}| \Vie{p}{0}{}) \BCie{E}{0}{}.
\end{align}
\label{RTdef}%
\end{subequations}
The reflection and transmission amplitudes are therefore described by 2$\times$2 matrices, i.e. for $\Vie{X}{}{} = \Vie{R}{}{}$ or $\Vie{T}{}{}$
\begin{equation}
\Vie{X}{}{} = \begin{pmatrix}
  X_{pp} & X_{ps}  \\
  X_{sp} & X_{ss}
 \end{pmatrix} .
\end{equation}
From a physical point of view, the coefficient $R_{\alpha \beta}( \Vie{q}{}{}| \Vie{p}{0}{})$ (resp. $T_{\alpha \beta}( \Vie{q}{}{}| \Vie{p}{0}{})$) for $\alpha, \beta \in \{p,s\}$ is the field amplitude for the reflected (resp. transmitted) light with lateral wave vector $\Vie{q}{}{}$ in the polarization state $\alpha$ from a unit incident field with lateral wave vector $\Vie{p}{0}{}$  in the polarization state $\beta$.
The reflection and transmission amplitudes are then the unknowns in our scattering problem. The equations we need to solve are deduced from the general equations Eqs.~(\ref{rrefinal2}) and (\ref{rrefinal2b}) by applying them respectively at $a_3= +$ and $a_1=-$ and by using Eqs.~(\ref{sommerfeld}) and (\ref{RTdef}) for the model of the field expansion. This yields the following two decoupled integral equations for the reflection or transmission amplitudes, the so-called reduced Rayleigh equations, that can be written in the following general form, for $\Vie{X}{}{} = \Vie{R}{}{}$ or $\Vie{T}{}{}$~\cite{Brown1984}
\begin{equation}
\int  \mathbf{M}_\mathbf{X} (\Vie{p}{}{}|\Vie{q}{}{}) \: \mathbf{X} (\Vie{q}{}{}|\Vie{p}{0}{}) \dtwopi{q}{}= - \mathbf{N}_\mathbf{X} (\Vie{p}{}{}|\Vie{p}{0}{}) \: ,
\label{RREint}
\end{equation}
where the matrices $\mathbf{M}_\mathbf{X}$ and $\mathbf{N}_\mathbf{X}$ are given by
\begin{subequations}
\begin{align}
\mathbf{M}_\mathbf{R} (\Vie{p}{}{}|\Vie{q}{}{}) &= \Vie{\Theta}{3,1}{+,+}(\Vie{p}{}{}|\Vie{q}{}{})\\
\mathbf{M}_\mathbf{T} (\Vie{p}{}{}|\Vie{q}{}{}) &= \Vie{\Theta}{1,3}{-,-}(\Vie{p}{}{}|\Vie{q}{}{})\\
\mathbf{N}_\mathbf{R} (\Vie{p}{}{}|\Vie{q}{}{}) &= \Vie{\Theta}{3,1}{+,-}(\Vie{p}{}{}|\Vie{q}{}{})\\
\mathbf{N}_\mathbf{T} (\Vie{p}{}{}|\Vie{q}{}{}) &= \frac{4 \: \sqrt{\epsilon_1 \epsilon_2^2 \epsilon_3} }{(\epsilon_1 -\epsilon_2)(\epsilon_2 -\epsilon_3)} \: (2 \pi)^2 \: \delta(\Vie{p}{}{} - \Vie{q}{}{}) \: \Vie{I}{2}{} ,
\end{align}
\label{matrices}%
\end{subequations}
with $\Vie{I}{2}{}$ denoting the 2$\times$2 identity matrix. In the cases where only one interface is rough and the other interface is planar, the complexity associated with the transfer kernels is equivalent to that of a single rough interface separating two media.
For instance, if the second interface is planar and the first interface is rough, we can choose the origin of the coordinate system such that $\zeta_2 (\xpar) = d_2 = 0$, and Eq.~(\ref{Iintdef}) yields, for $l, m \in \{2,3\}$ and $l \neq m$,
\begin{equation}
\Cie{J}{l,m}{b,a} \ofpq = \frac{(2\pi)^2 \: \delta(\Vie{p}{}{} - \Vie{q}{}{})}{b \alpha_l\ofp - a \alpha_m\ofq} .
\label{Jreduction}
\end{equation}
The Dirac distribution then simplifies the wave vector integration present in the two-interface transfer kernels and one gets
\begin{subequations}
\begin{equation}
\Vie{\Theta}{3,1}{a_3,a_1} \ofuipi{p}{}{q}{} =  \sum_{a_2=\pm} a_2 \:  \frac{\Vie{M}{3,2}{a_3,a_2} \ofuipi{p}{}{p}{} \: \Vie{\Theta}{2,1}{a_2,a_1}\ofuipi{p}{}{q}{}}{\alpha_3(\Vie{p}{}{}) \: \left[ a_3 \alpha_3\ofp - a_2 \alpha_2\ofp \right]}  ,
\end{equation}
and
\begin{equation}
\Vie{\Theta}{1,3}{a_1,a_3} \ofuipi{p}{}{q}{} =  \sum_{a_2=\pm} a_2 \:  \frac{\Vie{\Theta}{1,2}{a_1,a_2} \ofuipi{p}{}{q}{} \: \Vie{M}{2,3}{a_2,a_3} \ofuipi{q}{}{q}{}}{\alpha_2(\Vie{q}{}{}) \: \left[ a_2 \alpha_2\ofq - a_3 \alpha_3\ofq \right]} .
\end{equation}
\label{transferKernelsSecondFlat}%
\end{subequations}
If the first interface is planar and the second interface rough, we can choose the origin of the coordinate system such that $\zeta_1 (\xpar) = d_1 = 0$, and Eq.~(\ref{Jreduction}) holds for $l, m \in \{1,2\}$ and $l \neq m$, and the two-interface transfer kernels read
\begin{subequations}
\begin{equation}
\Vie{\Theta}{3,1}{a_3,a_1} \ofuipi{p}{}{q}{} =  \sum_{a_2=\pm} a_2 \:  \frac{ \Vie{\Theta}{3,2}{a_3,a_2} \ofuipi{p}{}{q}{} \: \Vie{M}{2,1}{a_2,a_1} \ofuipi{q}{}{q}{}}{\alpha_2(\Vie{q}{}{}) \: \left[ a_2 \alpha_2\ofq - a_1 \alpha_1\ofq \right]} \: ,
\end{equation}
and
\begin{equation}
\Vie{\Theta}{1,3}{a_1,a_3}\ofuipi{p}{}{q}{} =  \sum_{a_2=\pm} a_2 \:  \frac{\Vie{M}{1,2}{a_1,a_2}\ofuipi{p}{}{p}{} \: \Vie{\Theta}{2,3}{a_2,a_3}\ofuipi{p}{}{q}{}}{\alpha_1(\Vie{p}{}{}) \: \left[ a_1 \alpha_1\ofp - a_2 \alpha_2\ofp \right]}  .
\end{equation}
\label{transferKernelsFirstFlat}
\end{subequations}

\subsection{Observables}

The observable of interest in this study is the so-called incoherent (or diffuse) component of the \emph{mean differential reflection coefficient}~(DRC) that we denote $\left\langle \partial R_{\alpha \beta} (\Vie{p}{}{} | \Vie{p}{0}{}) / \partial \Omega_s \right\rangle_{\mathrm{incoh}}$.
It is defined as the ensemble average over realizations of the surface profile function of the incoherent component of the radiated reflected flux of an $\alpha$-polarized wave around direction $\Vie{\hat{k}}{1}{+} (\Vie{p}{}{})$,  per unit incident flux of a $\beta$-polarized plane wave of wave vector $\Vie{k}{1}{-} (\Vie{p}{0}{})$, and per unit solid angle.
The precise mathematical definition and the derivation of the expression for the mean DRC as a function of the reflection amplitudes is given in \ref{AppendixB}.
%

\section{Numerical methods}\label{sec:numerics}
Solutions of the reduced Rayleigh equation, Eq.~\eqref{RREint}, are obtained via both a perturbative and a non-perturbative numerical approach. In this work we investigate systems with two interfaces; For the case when one of these interfaces is planar we are able to employ both approaches, but when both interfaces are rough we will exclusively use the perturbative approach due to the high computational cost of the non-perturbative approach.

\subsection{Perturbative method}
The approximated solution of Eq.~\eqref{RREint} for the reflection amplitudes, and to first order in product of surface profiles, obtained by small amplitude perturbation theory (SAPT) is derived in \ref{AppendixA} and given by
\begin{subequations}
\begin{align}
  \mathbf{R} \ofuipi{p}{}{p}{0} &\approx  \Vie{R}{}{(0)} \ofuipi{p}{}{p}{0} - i \Vie{R}{}{(1)} \ofuipi{p}{}{p}{0} \: , \\
  \Vie{R}{}{(1)} \ofuipi{p}{}{p}{0} &= \hat{h}_1 (\Vie{p}{}{} -\Vie{p}{0}{}) \Vie{\boldsymbol \rho}{1}{} \ofuipi{p}{}{p}{0} + \, \hat{h}_2 (\Vie{p}{}{} -\Vie{p}{0}{}) \Vie{\boldsymbol \rho}{2}{} \ofuipi{p}{}{p}{0} .
\end{align}
\label{1storder:amplitude}%
\end{subequations}
Here $\Vie{R}{}{(0)} \ofuipi{p}{}{p}{0}$ is the response from the corresponding system with planar interfaces (i.e.~that of a Fabry-Perot interferometer), $\hat{h}_j$ are the Fourier transforms of the stochastic component of the surface profiles and $\Vie{\boldsymbol \rho}{j}{} \ofuipi{p}{}{p}{0}$ are matrix-valued amplitudes depending \emph{only} on the mean film thickness, the dielectric constants of all media and the wave vectors of incidence and scattering.
The explicit expressions for these matrices are given in~\ref{AppendixA} (see Eq.~(\ref{def:rho})). The corresponding expression for the incoherent component of the mean differential reflection coefficient reads (\ref{AppendixA} and \ref{AppendixB})
\begin{align}
  \left\langle \frac{\partial R_{\alpha \beta} (\Vie{p}{}{} | \Vie{p}{0}{})}{\partial \Omega_s} \right\rangle_{\mathrm{incoh}}
  &=
    \: \epsilon_1 \left(\frac{\omega}{2 \pi c} \right)^2 \: \frac{\cos^2 \theta_s}{\cos \theta_0} \: g (\Vie{p}{}{} -\Vie{p}{0}{})
  %
    \Big[
      \sigma_1^2 \: |\rho_{1,\alpha \beta} \ofuipi{p}{}{p}{0} |^2
    + \sigma_2^2 \: |\rho_{2,\alpha \beta} \ofuipi{p}{}{p}{0} |^2
    \nonumber \\
  & \qquad 
    + 2 \gamma \sigma_1 \sigma_2\:
    \mathrm{Re} \left\{ \rho_{1,\alpha \beta} \ofuipi{p}{}{p}{0}
    \rho_{2,\alpha \beta}^* \ofuipi{p}{}{p}{0}  \right\}
    \Big] \: ,
\label{eq:incoMDRC:final}
\end{align}
where the wave vectors
\begin{subequations}
\begin{align}
  \Vie{p}{}{} &=
      \sqrt{\epsilon_1} \frac{\omega}{c} \sin \theta_s
      (\cos \phi_s \, \hat{\mathbf{e}}_1 + \sin \phi_s\,\hat{\mathbf{e}}_2)
\end{align}
and
\begin{align}
  \Vie{p}{0}{}
    = \sqrt{\epsilon_1} \frac{\omega}{c} \sin \theta_0
      (\cos \phi_0 \, \hat{\mathbf{e}}_1 + \sin \phi_0 \, \hat{\mathbf{e}}_2 )
\end{align}
\end{subequations}

\noindent are defined in terms of the angles of scattering $(\theta_s,\phi_s)$ and incidence $(\theta_0,\phi_0)$, respectively [see Fig.~\ref{fig:system}].
The three terms present in the angular brackets of Eq.~\eqref{eq:incoMDRC:final} can be interpreted as follows. The term containing $\sigma_1^2 \: |\rho_{1,\alpha \beta} \ofuipi{p}{}{p}{0} |^2 $ (resp. $\sigma_2^2 \: |\rho_{2,\alpha \beta} \ofuipi{p}{}{p}{0} |^2 $) corresponds to the contribution to the diffuse intensity of the associated system for which the first (resp. second) interface would be rough and the other planar.
Indeed, this would be the only remaining term if we were to set $\sigma_2 = 0$ (resp. $\sigma_1 = 0$) in Eq.~(\ref{eq:incoMDRC:final}). The sum of the two first terms would correspond to the sum of intensity of the aforementioned associated systems, which would be the expected overall response if the two interface \emph{were not} correlated, i.e. if $\gamma = 0$.
The last term in Eq.~\eqref{eq:incoMDRC:final}, which does not vanish for $\gamma \neq 0$, can be interpreted physically as taking into account the interference between paths resulting from single scattering events on the top interface and those resulting from single scattering events on the bottom interface.
Note that this last term, in contrast to the two first, may take positive \emph{and} negative values as the incident and scattering wave vectors are varied, and hence may result in cross-correlation induced constructive and destructive interference.
It is clear from the derivation, however, that the overall incoherent component of the mean differential coefficient remains non-negative, as is required for any intensity.

\subsection{Nonperturbative method}
\label{sec:nonperturbative_method}
Solutions of Eq.~\eqref{RREint} were also obtained in a rigorous, purely numerical, nonperturbative manner according to the method described in detail in Ref.~\cite{Nordam2013a}; only a brief summary of the method is presented here.
This method has previously been used for the investigations of the two-dimensional rough surface scattering of light from metallic or perfectly conducting surfaces~\cite{Nordam2013a,Leskova2011,Simonsen2012-05}; from and through single dielectric interfaces~\cite{Leskova2011,Hetland2016a,Hetland2017} and film geometries~\cite{Nordam2012,Gonzalez-Alcalde2016,Simonsen2016-11}.
In this method, an ensemble of realizations of the surface profile function $\zeta_j (\mathbf{x}_\parallel)$ is generated by the use of the Fourier filtering method \cite{Maradudin1990} on a square grid of $N_x\times N_x$ surface points, covering an area of $S=L^2$ in the $(\hat{\mathbf{e}}_1,\hat{\mathbf{e}}_2)$-plane.
The integral equation, Eq.~\eqref{RREint}, is solved numerically with finite limits $\pm Q$ and discretization $\Delta q=2\pi/L$ with $N_q\times N_q$ points in wave vector space according to the Nyquist sampling theorem given the spatial discretization of the surface. On evaluating the kernel scalar factors $\Cie{J}{l,m}{b,a}\ofpq$, defined in Eq.~\eqref{Iintdef}, we first expand the integrand in powers of $\zeta_j (\mathbf{x}_\parallel)$, truncate this expansion after \num{20} terms, and integrate the resulting sum term-by-term. The Fourier integral of $\zeta_j^n(\mathbf{x}_\parallel)$ that remains now only depends on the surface profile function and the difference in lateral wave vectors
$\mathbf{p} - \mathbf{q}$, and not on $\alpha_l(\mathbf{p})$ and $\alpha_m(\mathbf{q})$. These Fourier integrals are therefore calculated only once, on a $\mathbf{p} - \mathbf{q}$ grid, for every surface realization by the use of the fast Fourier transform.
The resulting matrix equations are then solved by LU factorization and back substitution, using the ScaLAPACK library~\cite{scalapack}. This process is repeated for a large number $N_p$ of realizations of the surface profile function, enabling the calculation of the ensemble averaged observables of interest; like the mean DRC.


It remains to mention that Eqs.~(\ref{transferKernelsSecondFlat}) and (\ref{transferKernelsFirstFlat}), giving the transfer kernels in the case where only one of the interfaces is rough and the other planar, have been written in a rather compact form.
Numerically, these expressions tend to lead to instabilities due to factors of the form $\exp(- i \alpha_2(\Vie{q}{}{}) d)$ or $\exp(-i \alpha_2(\Vie{p}{}{}) d)$ which grow for evanescent waves inside the film. This technical issue is resolved by using the following two ideas: (i) expanding the two terms in the kernels (i.e. for $a_2 = \pm$) and factorizing out the troublesome exponential factor and canceling it on both sides of the reduced Rayleigh equation (if the exponential factor is a function of the variable $\Vie{p}{}{}$) or (ii) making a change of variable such that the troublesome exponential factor is absorbed into the reflection or transmission amplitudes (if the exponential factor is a function of the variable $\Vie{q}{}{}$). One may also shift the $x_3$-axis in order to facilitate the aforementioned steps. We chose here not to give more details on the explicit implementation, as these modifications are to be done in a case by case basis depending on which surface is planar and whether the reflected or transmitted light is considered.

\begin{figure*}[t]
  \captionsetup[subfigure]{justification=centering}
  \centering
  \begin{subfigure}{0.5\textwidth} 
    \includegraphics[width=1.0\linewidth,trim= 0cm 0.cm 0.0cm 0.0cm,clip]{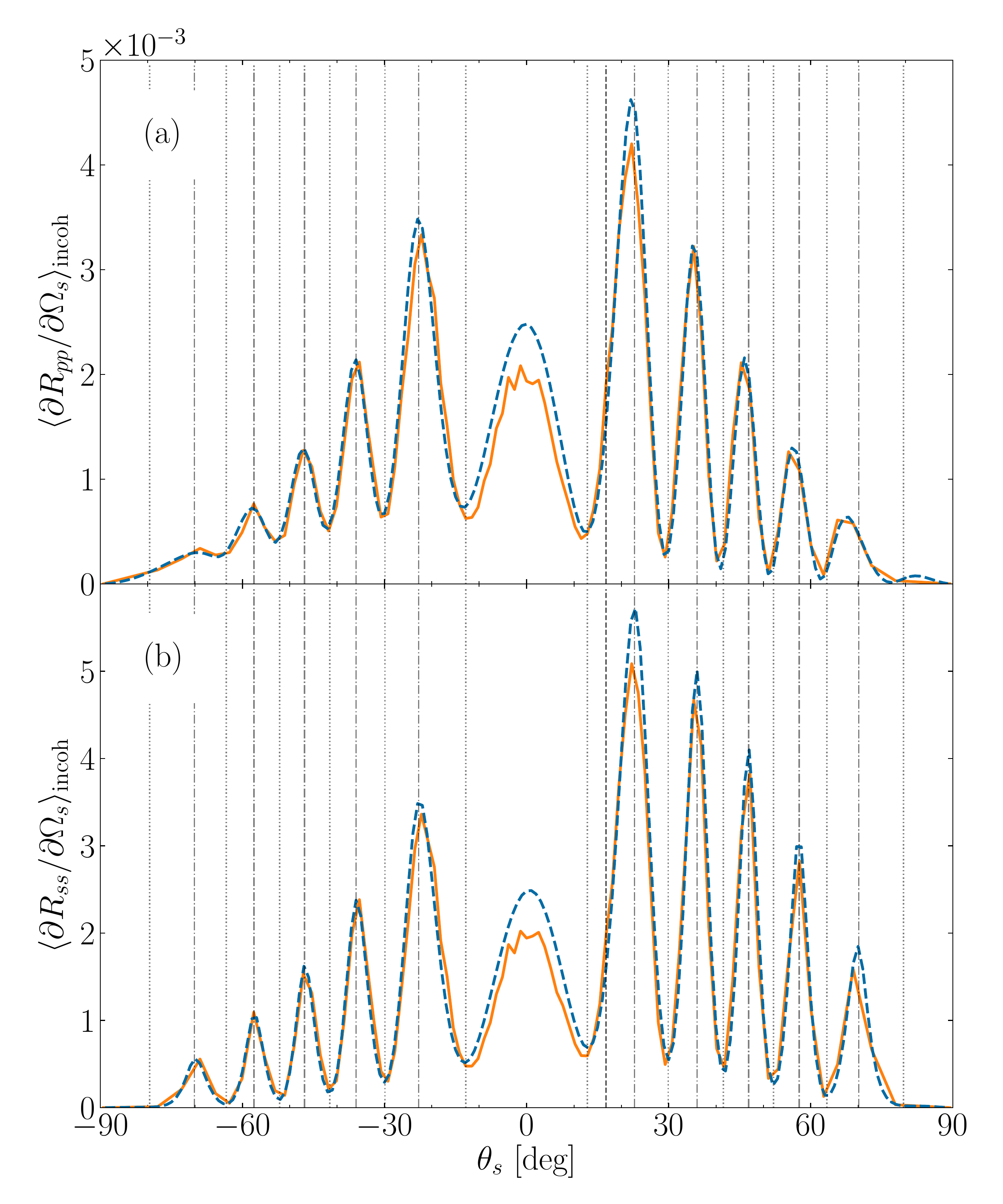}
    \phantomcaption
    \label{fig:mdrc_cut_rough-on-top}
  \end{subfigure}%
  \begin{subfigure}{0.5\textwidth}
    \includegraphics[width=1.0\linewidth,trim= 0cm 0.cm 0cm 0.0cm,clip]{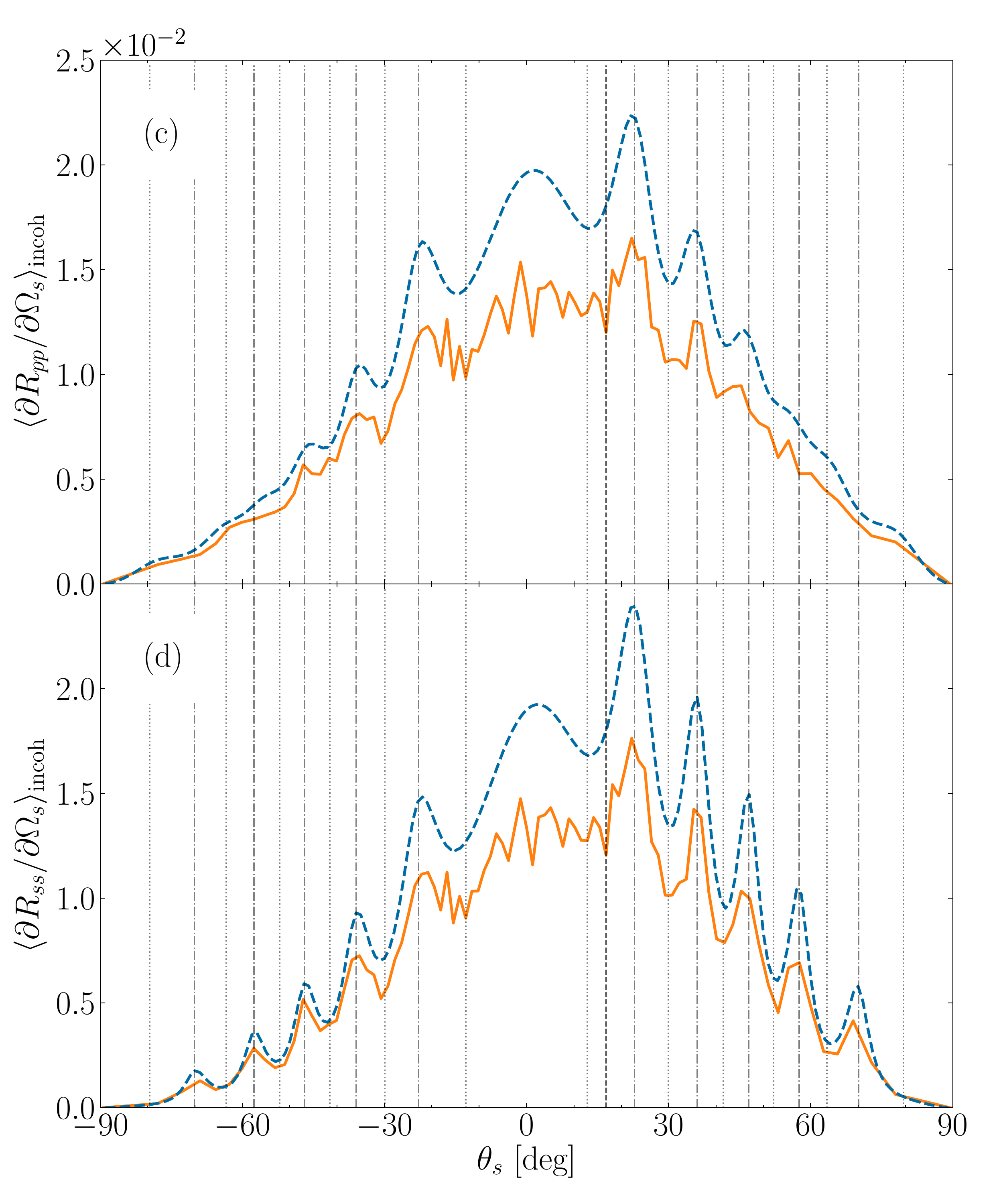}
    \phantomcaption
    \label{fig:mdrc_cut_rough-on-bottom}
\end{subfigure}
\caption{
  Incoherent components of the mean DRCs for in-plane co-polarized scattering as  functions of the polar angle of scattering, $\theta_s$ (note the convention $\theta_s < 0$ for $\phi_s = \phi_0 + \ang{180}$).
  The light of wavelength $\lambda = 632.8~\si{nm}$ was incident from vacuum on the rough photoresist film supported by a silicon substrate~[$\ve_1=1.0$, $\ve_2=2.69$, $\ve_3=15.08 + 0.15\im$]. 
  The surface-height correlation length of the rough Gaussian correlated surface was $a=\lambda/3$, the mean film thickness was $d=8\lambda$, and the angles of incidence were $(\theta_0,\phi_0)=(\ang{16.8},\ang{0})$ in all cases. Panels~(a) and (b) correspond to cases where only the top interface was rough, while panels~(c) and (d) presents the results for a film where only the bottom interface of the film is rough.
  In both cases, the rms-roughness of the rough interface was set to $\sigma = \lambda / 30$. The results obtained on the basis of the non-perturbative method are shown as solid lines while those obtained with the perturbative method, Eq.~(\ref{eq:incoMDRC:final}), are shown as dashed lines.
  The position of the specular direction in reflection is indicated by the vertical dashed lines.
  The vertical dash-dotted and dotted lines indicate the angular positions of the maxima and minima predicted by Eq.~(\ref{extrema}), respectively.
}
\label{fig:mdrc_cut_vps_bothfigs}
\end{figure*}

\section{Results and discussion}
\subsection{Single rough interface}\label{sec:single:rough}

As a direct comparison between results obtained by the perturbative and nonperturbative solutions of Eq.~\eqref{RREint}, Fig.~\ref{fig:mdrc_cut_vps_bothfigs} shows the angular distributions of the co-polarized ($\alpha=\beta$) incoherent contribution to the mean DRC for light incident from vacuum ($\epsilon_1=1$) that is reflected diffusively into the plane of incidence (i.e. $|\Vie{\hat{p}}{}{} \cdot \Vie{\hat{p}}{0}{}| = 1$) from a randomly rough dielectric film (photoresist, $\ve_2 = 2.69$) deposited on a silicon substrate ($\ve_3 = 15.08 + 0.15\im$) for the cases where only one of the interfaces is rough and the other planar.
Results for the case where only the top interface (the interface facing the medium of incidence) is rough ($\sigma_2=0$) and where only the bottom interface is rough ($\sigma_1=0$) are shown in Figs.~\ref{fig:mdrc_cut_vps_bothfigs}(a)--(b) and \ref{fig:mdrc_cut_vps_bothfigs}(c)--(d), respectively.
Light was incident on the dielectric film from the vacuum side in the form of a plane wave of wavelength $\lambda=632.8~\si{\nm}$ with angles of incidence $(\theta_0,\phi_0)=(\ang{16.8}, \ang{0})$.
The two interfaces were characterized by rms-roughness $\sigma_1=\lambda/30, \sigma_2=0$ [Figs.~\ref{fig:mdrc_cut_vps_bothfigs}(a)--(b)] or $\sigma_1=0, \sigma_2=\lambda/30$ [Figs.~\ref{fig:mdrc_cut_vps_bothfigs}(c)--(d)], correlation length $a=\lambda/3$, and the film thickness was assumed to be $d=8\lambda \approx 5~\si{\um}$.
The scattering system was chosen in order to highlight the interference phenomena and to purposely deviate from the more historically typical scattering system of a dielectric film on a perfect electric conductor.
The dashed curves in Fig.~\ref{fig:mdrc_cut_vps_bothfigs} display the results of computations of the perturbative solution of the RRE, Eq.~(\ref{eq:incoMDRC:final}), to leading order, while the solid curves in Fig.~\ref{fig:mdrc_cut_vps_bothfigs} show the non-perturbative solutions of the RRE, Eq.~(\ref{RREint}).
In obtaining these latter results the following parameters, defined in Sec.~\ref{sec:nonperturbative_method}, were used: $N_x=449$, $L=45\lambda$, $N_q=225$ and $N_p=325$, implying integration limits in wavevector space $Q=\pm 2.5 \omega/c$. Since these non-perturbative results for the mean DRC are obtained through an ensemble average over a finite number of surface realizations, they are less smooth than their perturbative counterparts, for which the averaging is performed analytically.
Using a larger number of surface realizations in obtaining the ensemble average would have produced smoother results, but we have chosen not to do so here due to the high associated computational cost.

\begin{figure*}[t]
\vspace{-1cm}
\centering
\includegraphics[width=0.85\linewidth,trim= 2.cm 0.cm 5.cm 1.5cm,clip]{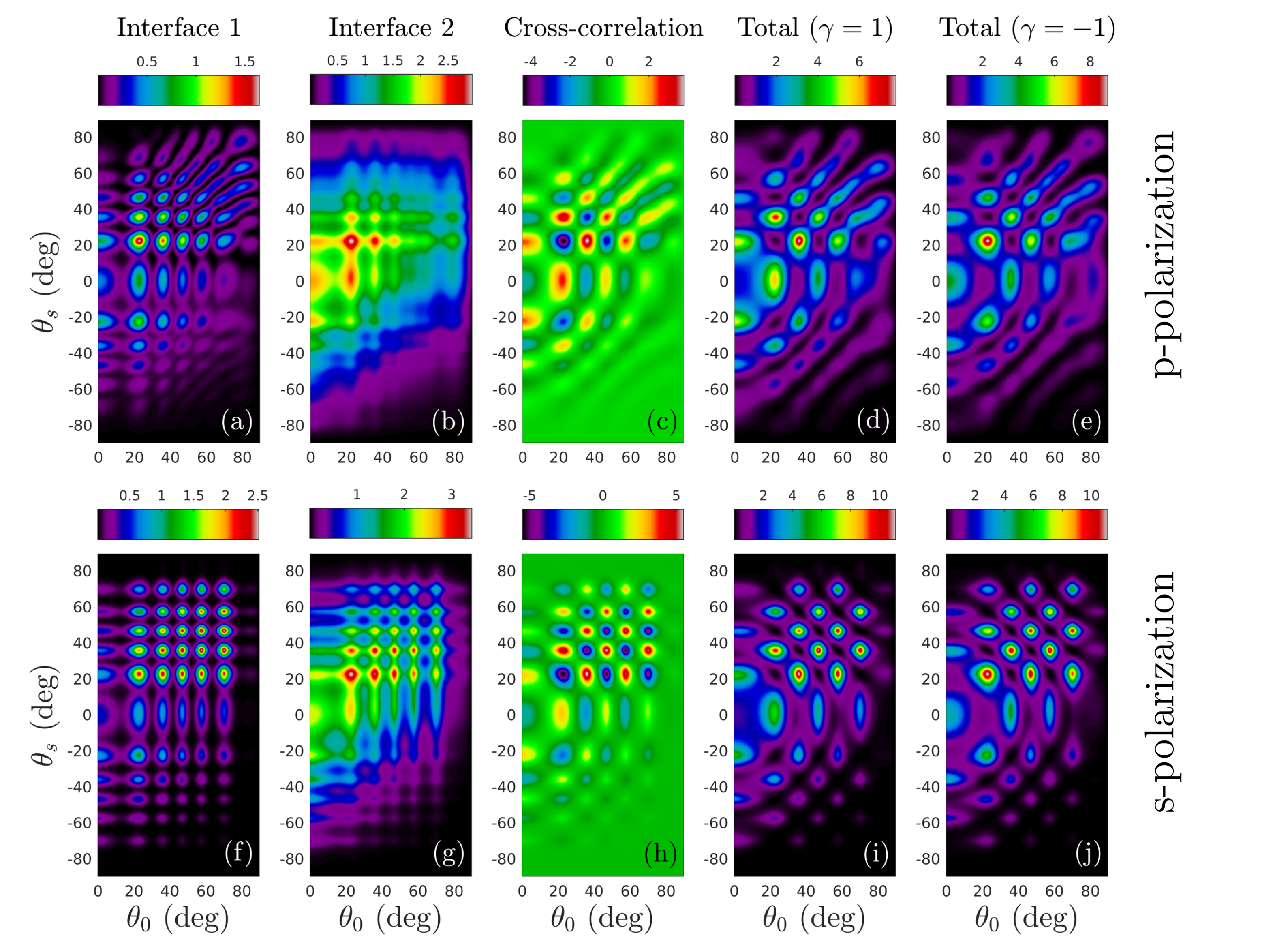}
\caption{
  Scaled incoherent component of the mean DRCs for in-plane co-polarized scattering, $100 \times \left\langle \partial R_{\alpha \alpha} / \partial \Omega_s \right\rangle_\mathrm{incoh}$, as functions of the polar angle of incidence $\theta_0$ and the polar angle of scattering $\theta_s$ obtained on the basis of Eq.~\eqref{eq:incoMDRC:final}. The first row of sub-figures [Figs.~\ref{fig:decomposition}(a)--(e)] corresponds to $p$-polarized light (as marked in the figure), while the second row [Figs.~\ref{fig:decomposition}(f)--(j)]  corresponds to $s$-polarized light. These results were obtained under the assumption that the wavelength in vacuum was $\lambda = 632.8$~nm, the mean film thickness was $d = 8 \lambda$, and the dielectric constants of the media were $\ve_1=1.0$, $\ve_2=2.69$, $\ve_3=15.08 + 0.15\im$.
  The rms-roughness of the rough interfaces of the film were assumed to be $\sigma_1 = \sigma_2 = \lambda/30$, and the Gaussian correlation functions were characterized by the correlation length  $a=\lambda/3$. The first column of sub-figures presents contour plots of the mean DRCs for a film geometry where only the top interface of the film is rough and the bottom interface planar. The second column shows similar results when the top film interface is planar and the bottom film interface is rough. In the third column,  contour plots of \textit{only} the cross-correlation term in Eq.~\eqref{eq:incoMDRC:final} --- that is, the contribution to the mean DRC produced by the last term in the square brackets of this equation ---  are depicted assuming a perfect correlation [$\gamma=1$] between the rough top and rough bottom interface of the film. Finally, in the forth and fifth column, contour plots of the total mean DRCs obtained on the basis of Eq.~\eqref{eq:incoMDRC:final} are presented for two-rough-interface film geometries characterized by $\gamma=1$ and $\gamma=-1$, respectively.}
\label{fig:decomposition}
\end{figure*}

\begin{figure*}[t]
\centering
  \begin{subfigure}{0.42\textwidth} 
    \includegraphics[width=1.00\linewidth,trim= 0.cm 1.5cm 0.cm 2.cm,clip]{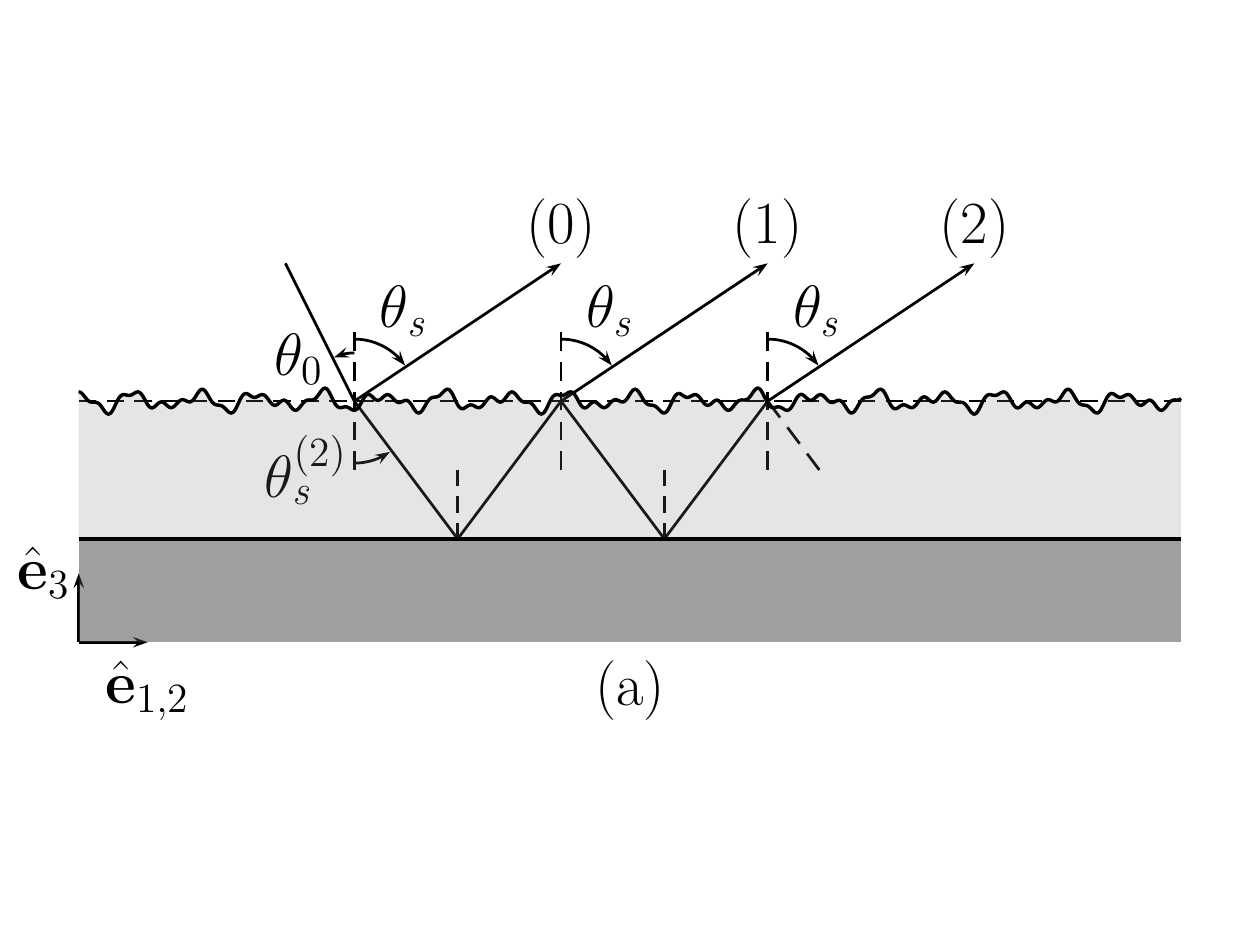}
    \includegraphics[width=1.00\linewidth,trim= 0.cm 1.5cm 0.cm 2.cm,clip]{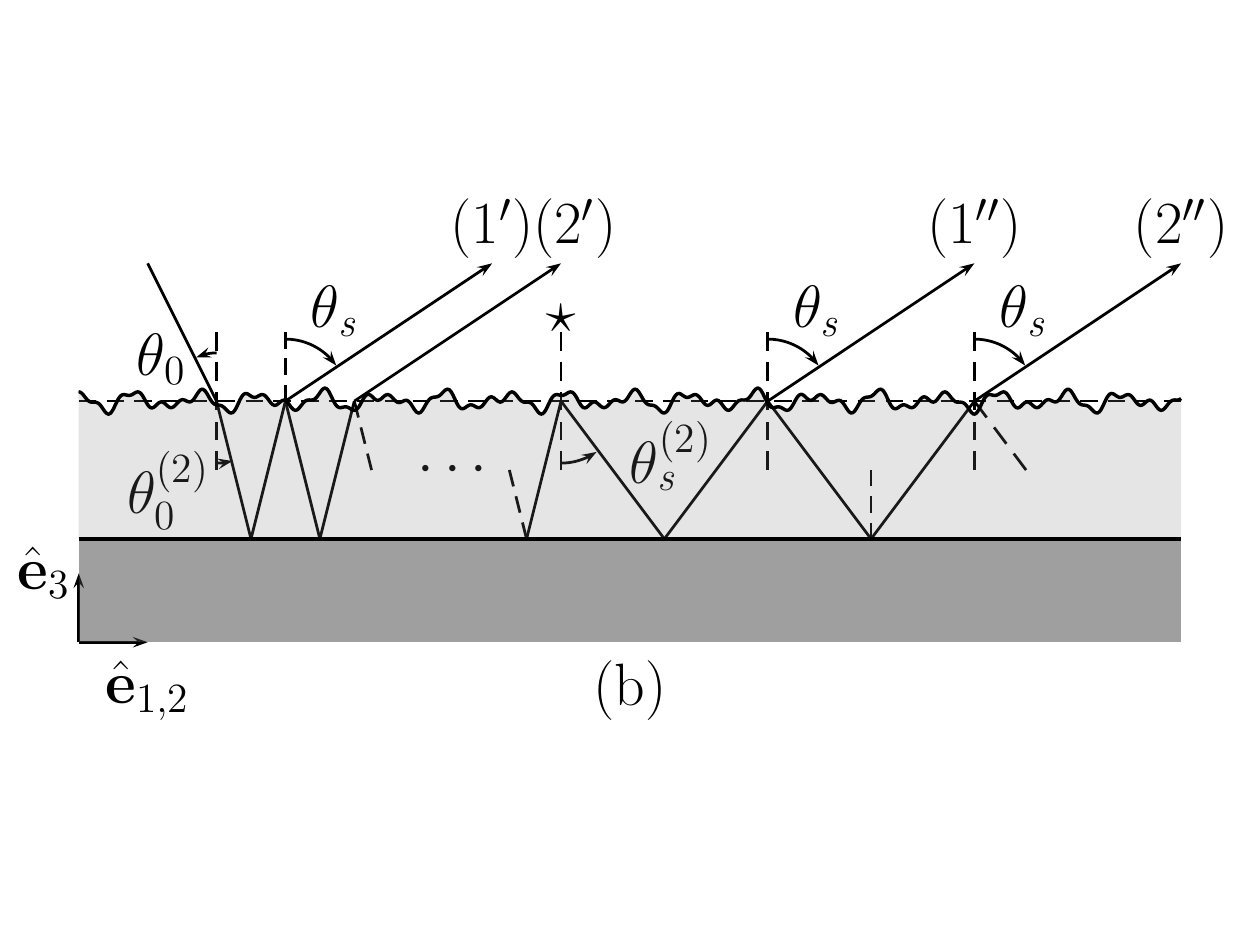}
    \includegraphics[width=1.00\linewidth,trim= 0.cm 2.cm 0.cm 2.cm,clip]{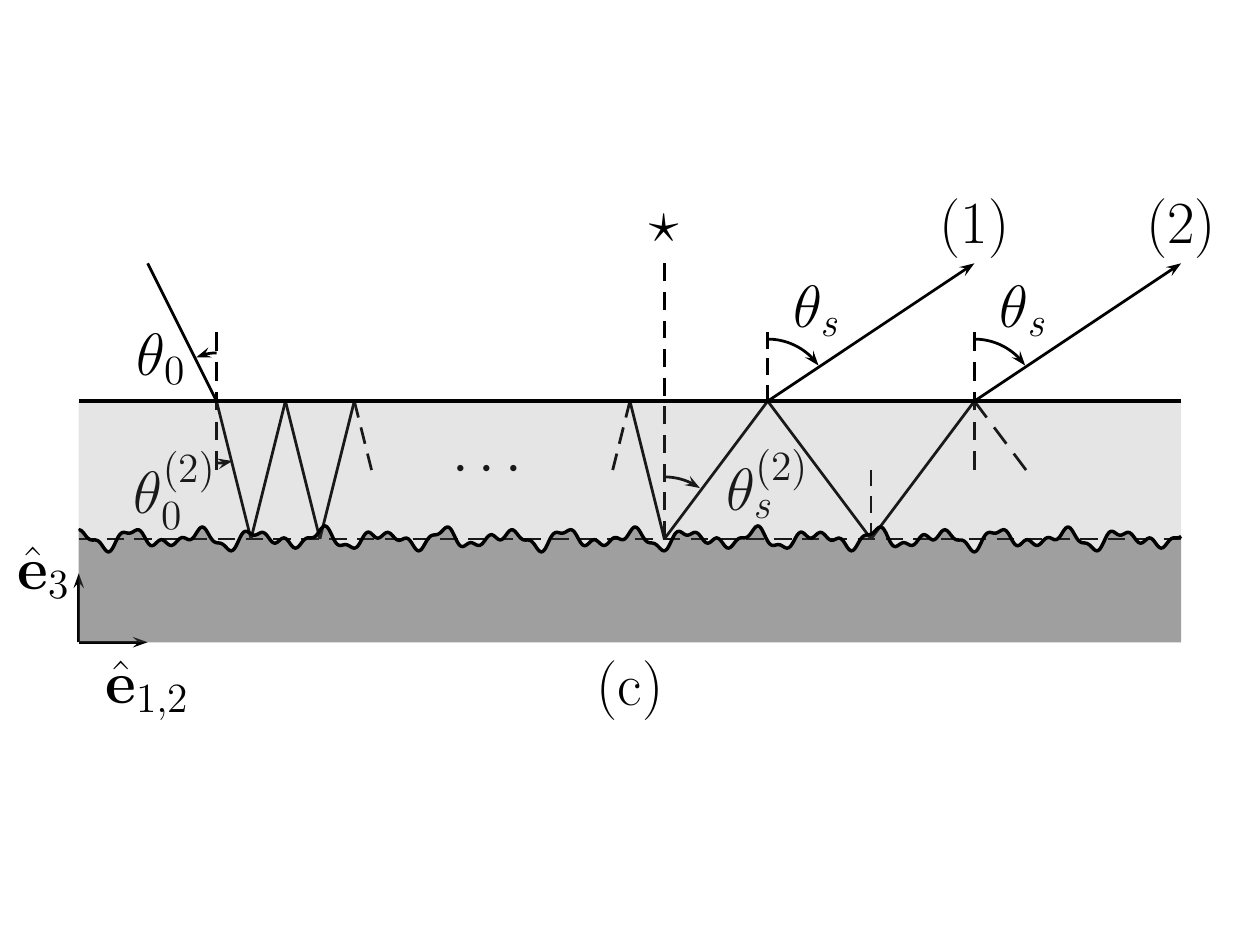}
  \end{subfigure}%
  \begin{subfigure}{0.58\textwidth} 
    \includegraphics[width=1.0\linewidth,trim= -1.cm 0.cm 0.cm 0.cm,clip]{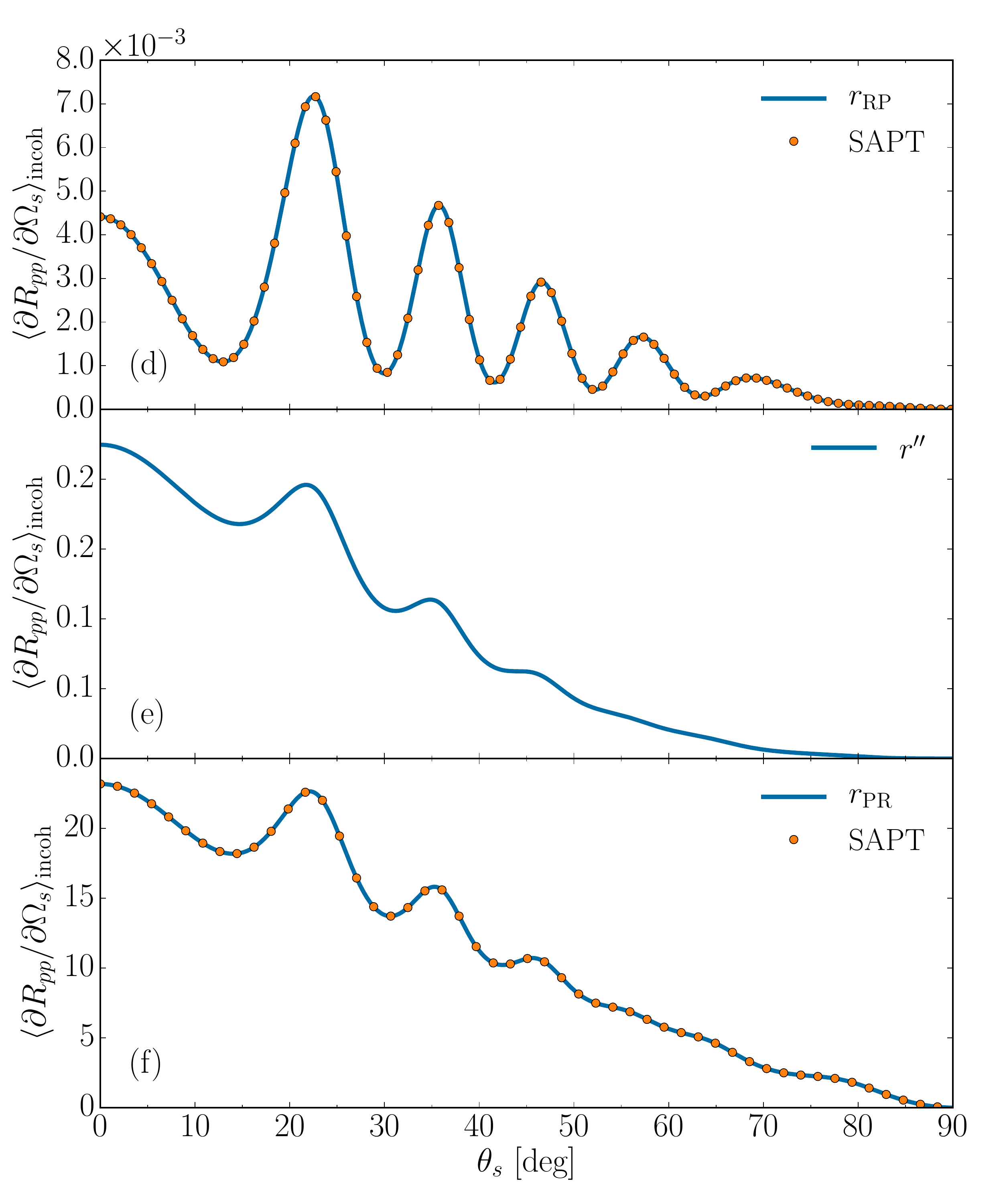}
  \end{subfigure}%
  \caption{Sketch of the optical paths involved in the single scattering model in the case of scattering from the top surface (a) and (b), or from the bottom interface (c). Incoherent component of the mean differential reflection coefficient for in-plane co-polarized scattering as a function of the polar angle of scattering for normal incidence for $p$-polarization (d) to (f).
  Apart from the angle of incidence the remaining parameters are the same as those from Fig.~\ref{fig:mdrc_cut_vps_bothfigs}. In panels (d) and (f), the results were obtained from SAPT (circles), and from the single scattering model Eqs.~(\ref{eq:toy:rf})(d) and (\ref{eq:toy:fr})(e) (solid line) respectively for the cases illustrated in (a-b) and (c).
  In panel (e), only the contribution of $r''$ (Eq.~(\ref{eq:toy:rf:2})) to the incoherent component of the mean DRC is shown.}
\label{fig:toy}
\end{figure*}

Figures~\ref{fig:mdrc_cut_vps_bothfigs}(a)--(b) shows excellent agreement between the results for the mean DRC obtained by the analytical perturbative method and the corresponding results obtained by a full solution of the RRE for the chosen parameters for the case where only the upper interface is rough. In particular, the fringes observed in these figures are consistently predicted by both calculation methods for the set of parameters assumed and their angular positions agree well with the expected angular positions (dashed-dotted  vertical lines in Figs.~\ref{fig:mdrc_cut_vps_bothfigs}(a)--(b)).
When the lower surface is rough, the results presented in Figs.~\ref{fig:mdrc_cut_vps_bothfigs}(c)--(d) show that the agreement between the two calculation methods is still satisfactory, but a larger discrepancy between them is now observed relative to what was found when the upper surface was rough. This larger discrepancy might be due to the fact that the error between the perturbative solution and the exact solution grows with the ratio of the dielectric constants of the media that are separated by the rough interface.
Since the dielectric contrast between the silicon substrate and the photoresist film is larger than that between the photoresist film and vacuum, the corresponding error is also larger. Since the perturbative method is employed only to leading order, these agreements overall indicate that the physical phenomena that give rise to the scattered intensity distributions are well approximated as single scattering phenomena, at least for weakly rough surfaces.

We identify the interference fringes in Fig.~\ref{fig:mdrc_cut_vps_bothfigs} as in-plane scattering distributions of Sel\'{e}nyi rings \cite{Selenyi1911}.
These rings are known to be centered around the mean surface normal, with their \textit{angular position} being independent of the angle of incidence. Their \textit{amplitude}, however, is modulated by the angle of incidence. This can indeed be observed if we vary the angle of incidence and record the resulting in-plane co-polarized angular scattering distributions, presented as contour plots in the first two columns of Fig.~\ref{fig:decomposition}.
Figures~\ref{fig:decomposition}(a)--(b) present, for $p$-polarized  light, contour plots of the  $(\theta_0,\theta_s)$ dependence of the in-plane co-polarized incoherent component of the mean DRC when the top or bottom interface of the film is rough, respectively. Similar results but for $s$-polarized light are presented in Figs.~\ref{fig:decomposition}(f)--(g).
For both configurations, the co-polarized incoherent component of the mean DRC exhibits maxima that occur on a regular grid of $(\theta_0,\theta_s)$-points for $s$-polarized light [Figs.~\ref{fig:decomposition}(f)--(g)]. A similar pattern is observed for $p$-polarized light in Figs.~\ref{fig:decomposition}(a)--(b), although the grid of maxima appears to lose some of its regularity for the larger polar angles of incidence and scattering [Figs.~\ref{fig:decomposition}(a)--(b)].
We speculate that this is due to a Brewster effect, both in its traditional sense and through the Brewster scattering angles~\cite{Hetland2016a,Hetland2017,Kawanishi1997}, but we will not delve further on this behaviour here.
In addition, by comparing the results presented in Figs.~\ref{fig:mdrc_cut_vps_bothfigs}, \ref{fig:decomposition}(a)--(b), and \ref{fig:decomposition}(f)--(g), we note that the contrast in the interference pattern is better for the configurations where the top interface is rough than for those where the bottom interface is rough.
In the following we will explain these observations in terms  of a single scattering model which is an extension of the model previously proposed by Lu and co-workers~\cite{Lu1998}.

Lu~\textit{et al}.\ suggested that, for sufficiently small roughness, the main effect of the rough interface is to produce scattered waves that cover a wide range of scattering angles both inside and outside the film, and the film may then be considered to approximate a planar waveguide for subsequent reflections and refractions within the film.
This claim is supported by the observed agreement between the mean DRC distributions obtained through the perturbative solution to leading order, whose physical interpretation is to take only single scattering events into account, and the full solutions of the RRE in Fig.~\ref{fig:mdrc_cut_vps_bothfigs}, since the latter method allows for the full range of multiple scattering events.
As the incident light interacts with the rough interface, whether it is located at the top or bottom interface, multiple wave components are generated in the film. These waves then undergo multiple specular reflections within the film while also being partially refracted back into the medium of incidence.
Since Lu \etal only investigated the case where the rough interface is on top, their results were adequately explained under the assumption that the incident light was scattered by the rough interface during its first encounter with the interface.
However, a more detailed analysis of the possible optical paths in the system is necessary in order to fully understand the case where the rough interface is at the bottom of the film, as illustrated by the more complete depiction of optical paths in Figs.~\ref{fig:toy}(a)--(c).
We will now analyze the different optical paths involving a single scattering event in the two configurations in more detail, and also construct a model for the resulting reflection amplitudes.
Let $r_{ji} (\Vie{p}{}{} \st \Vie{p}{0}{})$ and $t_{ji}(\Vie{p}{}{} \st \Vie{p}{0}{})$ denote the reflection and transmission amplitudes obtained by small amplitude perturbation theory to first order in the surface profile separating \emph{two} media with dielectric constants $\epsilon_i$ and $\epsilon_j$ (with the incident wave in medium $i$).
Note that these amplitudes are different from those obtained for the full system considered in this paper. The expressions for these reflection amplitudes can be found e.g. in Refs.~\cite{Hetland2016a,Hetland2017}.
Moreover, let $r_{ji}^\mathrm{(F)}(\Vie{p}{}{})$ and $t_{ji}^\mathrm{(F)}(\Vie{p}{}{})$ represent the corresponding Fresnel amplitudes. All the amplitudes considered here may represent either $p$-polarization or $s$-polarization as we treat in-plane co-polarized scattering for simplicity.

In the case where only the top interface is rough the scattering event may occur on the first intersection between the path and the top interface, yielding a reflected scattered path denoted (0) in Fig.~\ref{fig:toy}(a). Alternatively, on the first intersection the scattering event may yield a refracted (and scattered) wave in the film. Since the single scattering event allowed in our analysis has then occurred, subsequent reflections within the film and refractions through the top interface are treated according to Snell's law of reflection and refraction, resulting in the paths denoted (1) and (2) (and so on) in Fig.~\ref{fig:toy}(a).
With each such non-scattering interaction with an interface, the reflection/transmission amplitude associated with the path is given by the Fresnel amplitude. Following the different paths depicted in Fig.~\ref{fig:toy}(a) and summing the corresponding (partial) reflection amplitudes we obtain the following reflection amplitude:
\begin{align}
r (\Vie{p}{}{} | \Vie{p}{0}{}) &= r_{21} (\Vie{p}{}{}|\Vie{p}{0}{}) + t_{12}^\mathrm{(F)} (\Vie{p}{}{}) \, r_{32}^\mathrm{(F)}(\Vie{p}{}{}) \, t_{21}(\Vie{p}{}{}| \Vie{p}{0}{}) \exp(2 i \varphi_s) \, \sum_{n=0}^\infty \left[r_{12}^\mathrm{(F)}(\Vie{p}{}{}) \,  r_{32}^\mathrm{(F)} (\Vie{p}{}{}) \, \exp(2 i \varphi_s) \right]^n \nonumber\\
&= r_{21} (\Vie{p}{}{}|\Vie{p}{0}{}) + \frac{t_{12}^\mathrm{(F)} (\Vie{p}{}{}) \, r_{32}^\mathrm{(F)}(\Vie{p}{}{}) \, t_{21}(\Vie{p}{}{}| \Vie{p}{0}{}) \, \exp(2 i \varphi_s)}{1 - r_{12}^\mathrm{(F)}(\Vie{p}{}{})  \, r_{32}^\mathrm{(F)} (\Vie{p}{}{}) \exp(2 i \varphi_s)} \: ,
\label{eq:toy:rf:0}
\end{align}
where $\varphi_s = 2 \pi \sqrt{\epsilon_2} d \cos \theta_s^{(2)} / \lambda$. The positions of the maxima in the resulting angular intensity distribution $|r (\Vie{p}{}{} | \Vie{p}{0}{})|^2$ are consistent with the predictions given by Lu \etal\cite{Lu1998}. The difference in optical path length between path (0) and (1), and between (1) and (2), and more generally between two such consecutive paths, can be expressed as
\begin{align}
  \Delta = 2\sqrt{\ve_2}d \cos{\theta_{s}^{(2)}},
  \label{eq:delta_s}
\end{align}
where $\theta_s$ in the vacuum is related to $\theta_{s}^{(2)}$ in the film by $\sqrt{\ve_2}\sin{\theta_{s}^{(2)}} = \sqrt{\ve_1} \sin{\theta_s}$ according to Snell's law.
The polar angles of scattering for which the diffusely scattered intensity has local maxima are given by

\begin{subequations}
\begin{align}
  \frac{2 \pi \sqrt{\epsilon_2} d }{\lambda} \: \cos \theta_{s}^{(2)}
  &= \frac{2 \pi d }{\lambda} \: \left( \epsilon_2 - \epsilon_1 \sin^2 \theta_s  \right)^{1/2} =
    (\nu + 1/2) \pi ,
    \label{maxima}
\end{align}
while the positions of the minima are determined from the relation
\begin{align}
  \frac{2 \pi \sqrt{\epsilon_2} d }{\lambda} \: \cos \theta_{s}^{(2)}
  &= \frac{2 \pi d }{\lambda} \: \left( \epsilon_2 - \epsilon_1 \sin^2 \theta_s  \right)^{1/2} =
    \nu \pi,
    \label{minima}
\end{align}
\label{extrema}
\end{subequations}
where $\nu \in \mathbb{Z}$. 
The angular positions of the maxima and minima predicted by Eq.~\eqref{extrema} are indicated by vertical dash-dotted and dotted vertical lines, respectively, in Fig.~\ref{fig:mdrc_cut_vps_bothfigs}, and these predictions agree well with the maxima and minima that can be observed in the in-plane co-polarized mean DRC.
Equation~\eqref{extrema} does not depend on the polar angle of incidence $\theta_0$, which supports the observation that the positions of the maxima and minima of the incoherent components of the mean DRC do not move with angle of incidence for weakly rough films. However, the modulation of the fringes with the angle of incidence cannot be explained if we consider solely the paths depicted in Fig.~\ref{fig:toy}(a). Indeed, additional paths involving a single scattering event may be drawn as illustrated in Fig.~\ref{fig:toy}(b).
It is possible for the incident path not to experience a scattering event when it encounters the top interface for the first time, and it may also bounce within the film an arbitrary number of times before it experiences a scattering event while finally being refracted into the vacuum. Such paths are denoted $(1')$ and $(2')$ in Fig.~\ref{fig:toy}(b). The resulting (partial) reflection amplitude corresponding to the ``single-primed'' paths in Fig.~\ref{fig:toy}(b) reads
\begin{align}
r' (\Vie{p}{}{} | \Vie{p}{0}{}) &= t_{12} (\Vie{p}{}{} | \Vie{p}{0}{}) \, r_{32}^\mathrm{(F)}(\Vie{p}{0}{}) \, t_{21}^\mathrm{(F)}(\Vie{p}{0}{}) \exp(2 i \varphi_0) \, \sum_{n=0}^\infty \left[r_{12}^\mathrm{(F)}(\Vie{p}{0}{}) \,  r_{32}^\mathrm{(F)} (\Vie{p}{0}{}) \, \exp(2 i \varphi_0) \right]^n \nonumber\\
&= \frac{t_{12} (\Vie{p}{}{} | \Vie{p}{0}{}) \, r_{32}^\mathrm{(F)}(\Vie{p}{0}{}) \, t_{21}^\mathrm{(F)}(\Vie{p}{0}{}) \exp(2 i \varphi_0)}{1 - r_{12}^\mathrm{(F)}(\Vie{p}{0}{}) \,  r_{32}^\mathrm{(F)} (\Vie{p}{0}{}) \, \exp(2 i \varphi_0)},
\label{eq:toy:rf:1}
\end{align}
where $\varphi_0 = 2 \pi \sqrt{\epsilon_2} d \cos \theta_0^{(2)} / \lambda$. The difference in optical path length between path ($1'$) and ($2'$) is given by
\begin{align}
  \Delta = 2\sqrt{\ve_2}d \cos{\theta_{0}^{(2)}},
  \label{eq:delta_0}
\end{align}
where $\sqrt{\ve_2}\sin{\theta_{0}^{(2)}} = \sqrt{\ve_1} \sin{\theta_0}$ according to Snell's law.
Hence, we again obtain a series of maxima and minima in the mean DRC if we replace $\theta_{s}^{(2)}$ by $\theta_{0}^{(2)}$ in Eq.~\eqref{extrema}, but this time the positions of the maxima and minima are indeed a function of the polar angle of incidence $\theta_0$.
This interference phenomenon serves to modulate the intensity of the Sel\'{e}nyi interference patterns.
The static fringe pattern and the modulation introduced by the angle of incidence is clearly observed in the in-plane scattered intensities displayed in Figs.~\ref{fig:decomposition}(a) and ~\ref{fig:decomposition}(f). 
However, we still have more optical paths to take into account. Indeed, paths yielding outgoing paths of type $(1')$ and $(2')$ may experience a scattering event while being reflected on the top surface instead of being refracted into the vacuum.
Such a scattering event is indicated by the star in Fig.~\ref{fig:toy}(b), and thereon the path may be reflected within the film an arbitrary number of times before being refracted into the vacuum as depicted by the paths denoted $(1'')$ and $(2'')$ in Fig.~\ref{fig:toy}(b).
In order to obtain the reflection amplitudes corresponding to all such paths, it suffices to multiply the overall reflection amplitude for \emph{all} paths bouncing \emph{any} arbitrary number of times with an angle $\theta_0^{(2)}$ within the film before the scattering event, with the overall reflection amplitude of \emph{all} paths starting from the scattering event and bouncing \emph{any} number of times within the film before being refracted into the vacuum. In this way we obtain the reflection amplitude
\begin{align}
r'' (\Vie{p}{}{} | \Vie{p}{0}{}) &= t_{21}^\mathrm{(F)} (\Vie{p}{0}{}) \, r_{32}^\mathrm{(F)} (\Vie{p}{0}{}) \, \exp(2 i \varphi_0) \, \sum_{n=0}^\infty \left[r_{12}^\mathrm{(F)}(\Vie{p}{0}{}) \,  r_{32}^\mathrm{(F)} (\Vie{p}{0}{}) \, \exp(2 i \varphi_0) \right]^n \nonumber\\
&\quad\times
t_{12}^\mathrm{(F)} (\Vie{p}{}{}) \, r_{32}^\mathrm{(F)} (\Vie{p}{}{}) \, r_{12}(\Vie{p}{}{} | \Vie{p}{0}{}) \, \exp(2 i \varphi_s) \, \sum_{n'=0}^\infty \left[r_{12}^\mathrm{(F)}(\Vie{p}{}{}) \,  r_{32}^\mathrm{(F)} (\Vie{p}{}{}) \, \exp(2 i \varphi_s) \right]^{n'}\nonumber\\
&= \frac{t_{12}^\mathrm{(F)} (\Vie{p}{}{}) \, r_{32}^\mathrm{(F)} (\Vie{p}{}{}) \, r_{12}(\Vie{p}{}{} | \Vie{p}{0}{}) \, r_{32}^\mathrm{(F)} (\Vie{p}{0}{}) \, t_{21}^\mathrm{(F)} (\Vie{p}{0}{}) \, \exp(2 i (\varphi_0 + \varphi_s))}{ \left[ 1 - r_{12}^\mathrm{(F)}(\Vie{p}{}{}) \,  r_{32}^\mathrm{(F)} (\Vie{p}{}{}) \, \exp(2 i \varphi_s) \right] \left[ 1 - r_{12}^\mathrm{(F)}(\Vie{p}{0}{}) \,  r_{32}^\mathrm{(F)} (\Vie{p}{0}{}) \, \exp(2 i \varphi_0) \right]} .
\label{eq:toy:rf:2}
\end{align}
Note that the paths ($1''$) and ($2''$) are somewhat ill-defined in Fig.~\ref{fig:toy}(b). Indeed, each path represents a family of paths with different history prior to the scattering event. For a given path, the path prior to the scattering event consists of a number of specular reflections within the film for which amplitudes dependent on the angle of incidence $\theta_0$, as seen previously for the paths represented by $r'$, while the path that follows after the scattering event consists of a number of specular reflections within the film which are dependent on the angle of scattering $\theta_s$. Therefore, the phase difference between any two such paths will, in general, contain an integer combination of $\varphi_0$ and $\varphi_s$ depending on the number of bounces prior to and after the scattering event. Equation~\eqref{eq:toy:rf:2} hence contains both $\varphi_0$ and $\varphi_s$.
The total reflection amplitude for all possible paths involving a single scattering event for the rough-planar~(RP) film [Figs.~\ref{fig:toy}(a) and ~\ref{fig:toy}(b)] is obtained by summing the amplitudes obtained from all the  previously analyzed diagrams, namely
\begin{equation}
r_\mathrm{RP} (\Vie{p}{}{} | \Vie{p}{0}{}) = r(\Vie{p}{}{} | \Vie{p}{0}{}) + r'(\Vie{p}{}{} | \Vie{p}{0}{}) + r''(\Vie{p}{}{} | \Vie{p}{0}{}) .
\label{eq:toy:rf}
\end{equation}
The intensity distribution corresponding to Eq.~(\ref{eq:toy:rf}) is shown in Fig.~\ref{fig:toy}(d) for normal incidence and $p$-polarized light, and is compared to results based on small amplitude perturbation theory to leading order, Eq.~(\ref{eq:incoMDRC:final}), in the case where only the top interface is rough.
The two results are literally indistinguishable. Similar results were also found in the case of $s$-polarized light, but the results are not shown (in order to keep the figure simple). These findings strongly suggest that the two methods are equivalent.
In particular, this means that the perturbative solution to leading order derived in \ref{AppendixA} can indeed be interpreted as a sum of all paths involving a single scattering event, although this was not obvious from the derivation itself. The model presented here thus justifies this physical picture.
Figure~\ref{fig:toy}(e) shows the incoherent contribution to the in-plane co-polarized mean DRC one would obtain if \emph{only} paths of type ($1''$), ($2''$), and so on were present, in other words the intensity distribution resulting from Eq.~(\ref{eq:toy:rf:2}).
The relative contribution from $r''$ [Fig.~\ref{fig:toy}(e)] to $r_\mathrm{RP}$ [Fig.~\ref{fig:toy}(d)] is so small that it to some approximation may be ignored, as it was in Ref.~\cite{Lu1998}, but we will soon see that this path type is crucial in the case of a system with the rough interface shifted to the bottom of the film.

\smallskip
Let us now analyze the case where only the bottom interface is rough, as illustrated in Fig.~\ref{fig:toy}(c). If we follow paths~(1) and (2) in Fig.~\ref{fig:toy}(c), it becomes evident that a path must first undergo a Snell refraction from vacuum into the film before it may interact with the rough interface.
Following this refraction into the film a given path may undergo an arbitrary number of Snell reflections within the film, now at a polar angle $\theta_0^{(2)}$ with the normal to the mean film interfaces, before it is scattered by the rough interface as indicated by the star in Fig.~\ref{fig:toy}(c). The path then performs an arbitrary number of Snell reflections within the film, now at a polar angle of scattering $\theta_s^{(2)}$ with the normal to the mean film interfaces, before it exits into the vacuum.
All possible paths involving a single scattering event are for the present configuration depicted in Fig.~\ref{fig:toy}(c), and it is now immediately evident that these paths bear close resemblance to those shown in Fig.~\ref{fig:toy}(b) which correspond to the amplitude $r''$.
Consequently the resulting intensity pattern associated with the paths in Fig.~\ref{fig:toy}(c) will exhibit, by construction of the paths, dependencies on \textit{both} the polar angles of incidence and scattering as given by Eqs.~(\ref{eq:delta_0}) and (\ref{eq:delta_s}).
This is supported both by the resulting reflection amplitude [Eq.~\eqref{eq:toy:fr}] and the angular positions of the maxima and minima of the in-plane co-polarized mean DRC displayed in Figs.~\ref{fig:mdrc_cut_vps_bothfigs}(c) and (d), indicated as vertical dashed-dotted and dotted lines, respectively.
Similar to what was done for the paths of type $(1'')$ and $(2'')$ in the configuration depicted in Fig.~\ref{fig:toy}(b), the resulting reflection amplitude for the paths shown in Fig.~\ref{fig:toy}(c) can be expressed as the product of the partial reflection amplitude resulting from all possible paths \emph{prior} to the scattering event and the partial reflection amplitude resulting from all possible paths that may follow \emph{after} the scattering event.
The resulting reflection amplitude for the planar-rough~(PR) film [Fig.~\ref{fig:toy}(c)] obtained in this way reads
\begin{align}
r_\mathrm{PR} (\Vie{p}{}{} | \Vie{p}{0}{}) &= t_{21}^\mathrm{(F)} (\Vie{p}{0}{}) \, \exp(i \varphi_0) \, \sum_{n=0}^\infty \left[r_{12}^\mathrm{(F)}(\Vie{p}{0}{}) \,  r_{32}^\mathrm{(F)} (\Vie{p}{0}{}) \, \exp(2 i \varphi_0) \right]^n \nonumber\\
&\quad \times
t_{12}^\mathrm{(F)} (\Vie{p}{}{}) \, r_{32}(\Vie{p}{}{} | \Vie{p}{0}{}) \, \exp(i \varphi_s) \, \sum_{n'=0}^\infty \left[r_{12}^\mathrm{(F)}(\Vie{p}{}{}) \,  r_{32}^\mathrm{(F)} (\Vie{p}{}{}) \, \exp(2 i \varphi_s) \right]^{n'}\nonumber\\
&= \frac{t_{12}^\mathrm{(F)} (\Vie{p}{}{}) \, r_{32}(\Vie{p}{}{} | \Vie{p}{0}{}) \, t_{21}^\mathrm{(F)} (\Vie{p}{0}{}) \, \exp(i (\varphi_0 + \varphi_s))}{ \left[ 1 - r_{12}^\mathrm{(F)}(\Vie{p}{}{}) \,  r_{32}^\mathrm{(F)} (\Vie{p}{}{}) \, \exp(2 i \varphi_s) \right] \left[ 1 - r_{12}^\mathrm{(F)}(\Vie{p}{0}{}) \,  r_{32}^\mathrm{(F)} (\Vie{p}{0}{}) \, \exp(2 i \varphi_0) \right]} .
\label{eq:toy:fr}
\end{align}
The intensity pattern predicted by Eq.~(\ref{eq:toy:fr}) is presented as a solid line
in Fig.~\ref{fig:toy}(f) for normal incident $p$-polarized light; in the same figure, the filled circles represent the prediction from  Eq.~(\ref{eq:incoMDRC:final}).
As was the case when only the top interface was rough, we find an excellent agreement between the two approaches also when only the bottom interface is rough. A similar agreement was also found when the incident light was $s$-polarized~(results not shown). These findings support our single scattering interpretation of the perturbative solution to leading order.
We have now explained the angular positions of the Sel\'{e}nyi rings and their amplitude modulation with the angle of incidence based on optical path analysis.

\begin{figure}[t]
  \vspace{-1cm} 
  \captionsetup[subfigure]{justification=centering}
  \centering
  \begin{subfigure}{0.5\textwidth} 
    \includegraphics[width=1.0\linewidth,trim= 0.9cm 0.cm 1.4cm .5cm,clip]{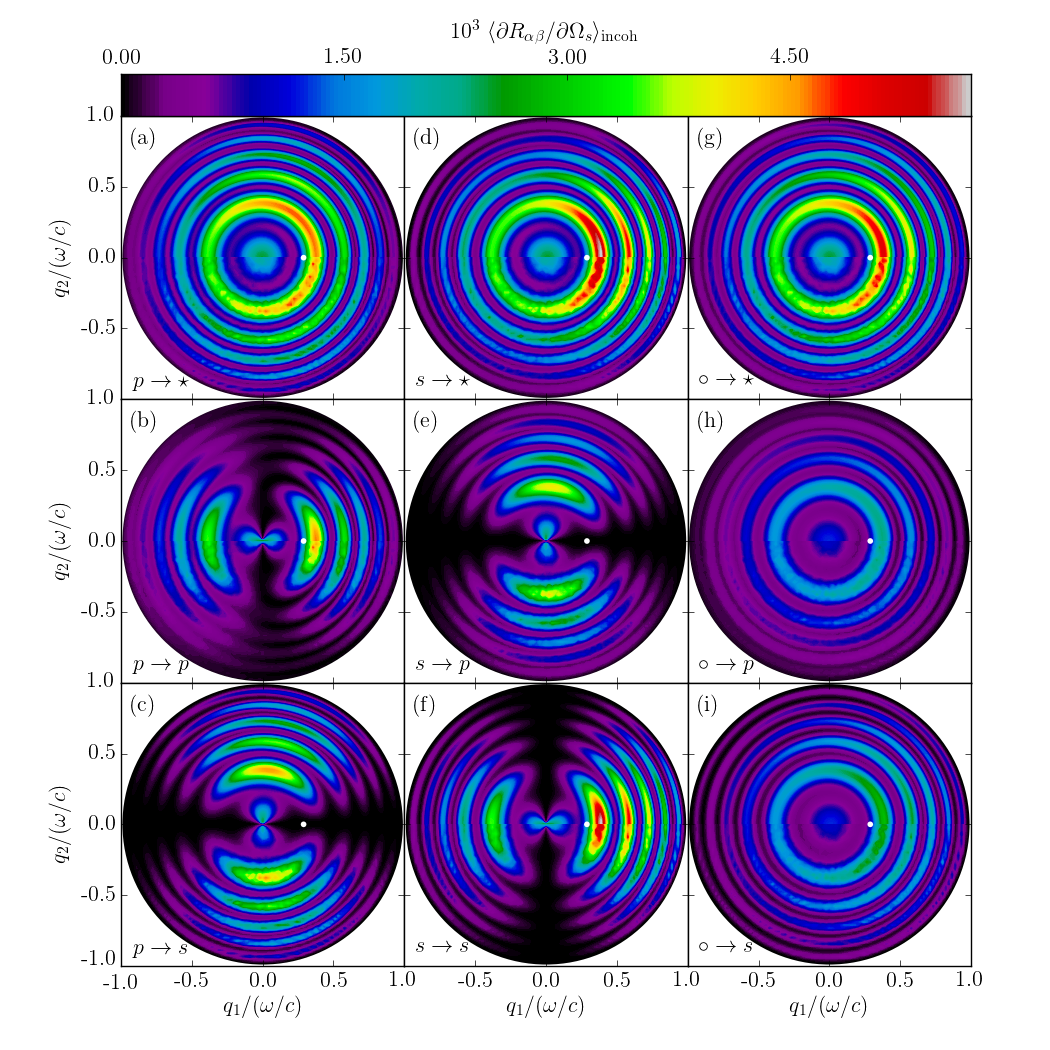}
    \phantomcaption
    \label{fig:2Dmdrc_rough-on-top}
  \end{subfigure}%
  \begin{subfigure}{0.5\textwidth}
    \includegraphics[width=1.0\linewidth,trim= 0.9cm 0.cm 1.4cm .5cm,clip]{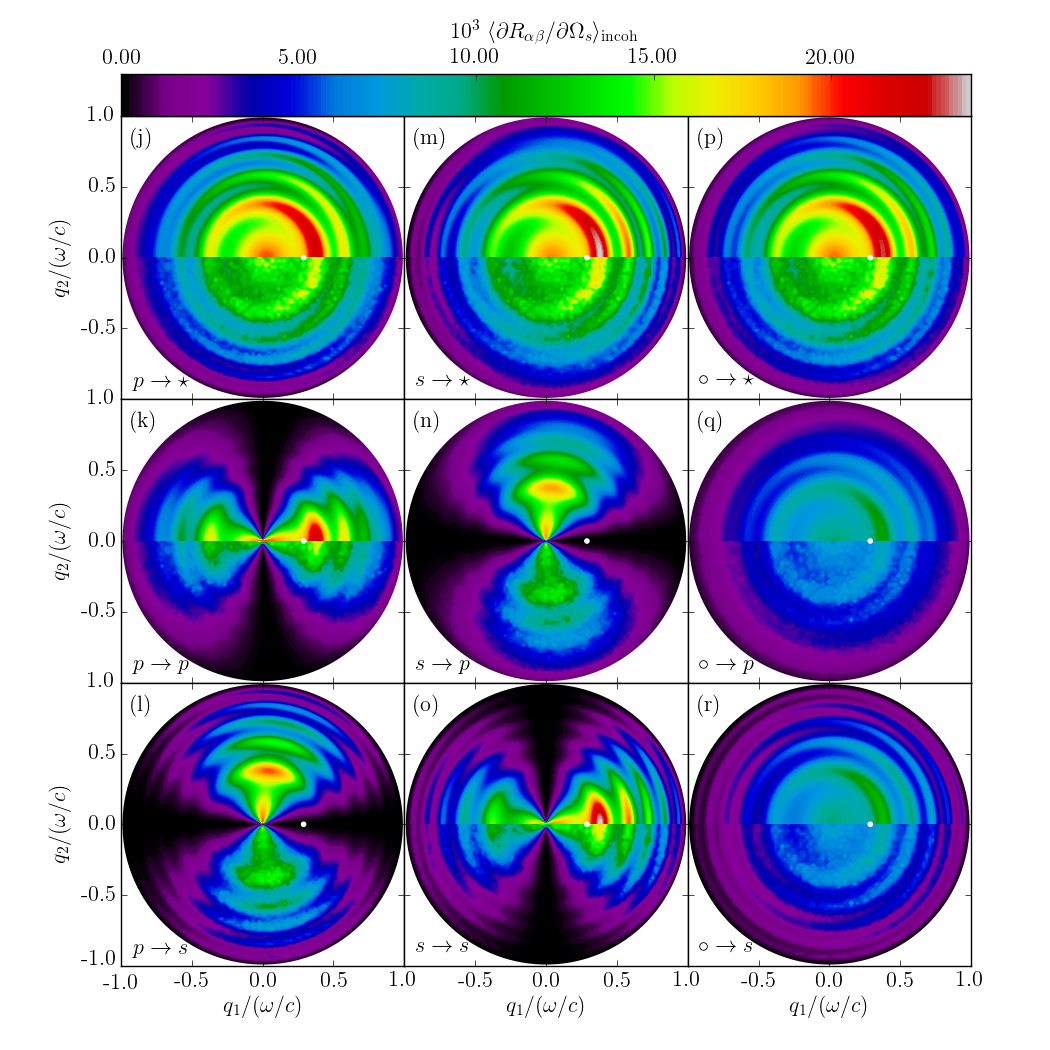}
    \phantomcaption
    \label{fig:2Dmdrc_rough-on-bottom}
\end{subfigure}
\caption{The full angular distribution of the incoherent component of the mean DRC,
  $\left\langle \partial R_{\alpha\beta} / \partial \Omega_s \right\rangle_\mathrm{incoh}$, as function of the lateral wave vector $\mathbf{q}$ of the light that is scattered from a rough  film where either the top interface is rough~[Figs.~\ref{fig:2Dmdrc_vps_bothfigs}(a)--(i)] or the bottom interface is rough~[Figs.~\ref{fig:2Dmdrc_vps_bothfigs}(j)--(r)] and the other interface of the film is planar. The light of wavelength $\lambda = 632.8~\si{nm}$ was incident from vacuum on the rough photoresist film supported by a silicon substrate~[$\ve_1=1.0$, $\ve_2=2.69$, $\ve_3=15.08 + 0.15\im$].
  The rms-roughness of the rough film interface was $\sigma_1=\lambda/30$, $\sigma_2=0$~[Figs.~\ref{fig:2Dmdrc_vps_bothfigs}(a)--(i)] 
  and $\sigma_1=0$, $\sigma_2=\lambda/30$~[Figs.~\ref{fig:2Dmdrc_vps_bothfigs}(j)--(r)].
  The surface-height correlation length was $a=211\si{\nm}=\lambda/3$, the film thickness was $d=5062.4\si{\nm}=8\lambda$ and the angles of incidence were $(\theta_0,\phi_0)=(\ang{16.8},\ang{0})$ for all panels. The positions of the specular directions in reflection are indicated by white dots.
  The remaining parameters assumed for the scattering geometry and used in performing the numerical simulations had values that are identical to those assumed in obtaining the results of Fig.~\protect\ref{fig:mdrc_cut_vps_bothfigs}.
  The upper halves of all panels are results from the small amplitude perturbation method to leading order, while the lower halves show results obtained through the non-perturbative solutions of the RRE.
  The sub-figures in Figs.~\ref{fig:2Dmdrc_vps_bothfigs}(a)--(i) and  \ref{fig:2Dmdrc_vps_bothfigs}(j)--(r) are both organized in the same manner and show how incident  $\beta$-polarized light is scattered by the one-rough-interface film geometry into $\alpha$-polarized light [with $\alpha=p,s$ and $\beta=p,s$] and denoted $\beta\rightarrow\alpha$.
  Moreover, the notation $\circ\rightarrow\star$ is taken to mean that the incident light was unpolarized while the  polarization of the scattered light was not recorded. For instance, this means that the data shown in Fig.~\ref{fig:2Dmdrc_vps_bothfigs}(a) are obtained by adding the data sets presented in Figs.~\ref{fig:2Dmdrc_vps_bothfigs}(b)--(c); similarly, the data shown in Fig.~\ref{fig:2Dmdrc_vps_bothfigs}(g) result from the addition and division by a factor two of the the data sets presented in Figs.~\ref{fig:2Dmdrc_vps_bothfigs}(a) and \ref{fig:2Dmdrc_vps_bothfigs}(d); \emph{etc}.
  Finally, the in-plane intensity variations from Figs.~\protect\ref{fig:2Dmdrc_vps_bothfigs}(b,\,f) and \ref{fig:2Dmdrc_vps_bothfigs}(k,\,o) are the curves depicted in Figs.~\protect\ref{fig:mdrc_cut_vps_bothfigs}(a)--(b) and Figs.~\protect\ref{fig:mdrc_cut_vps_bothfigs}(c)--(d), respectively.
}
\label{fig:2Dmdrc_vps_bothfigs}
\end{figure}

It remains to explain the difference in contrast observed in the interference patterns corresponding to the geometries where the rough surface is either located on the top of the film or at the bottom of the film (with the other film interface planar).
In providing such an explanation, the expressions given by Eqs.~(\ref{eq:toy:rf}) and (\ref{eq:toy:fr}) will prove to be useful alternative representations of the perturbative solutions of the RRE to leading order.
Indeed, we can now investigate the relative contribution from each type of path by artificially removing terms. In our analysis of the type of paths in the two configurations, we have identified that paths of type $(1'')$ and  $(2'')$, in the configuration where the top interface is rough, are similar to paths $(1)$ and $(2)$ for the configuration where the bottom interface is rough.
As was mentioned previously, Fig.~\ref{fig:toy}(e) shows the (diffuse) in-plane mean DRC we would obtain if \emph{only} paths of type $(1'')$, $(2'')$, \emph{etc.} were present; in other words the scattering intensities originating in Eq.~(\ref{eq:toy:rf:2}).
We observe that the curve in Fig.~\ref{fig:toy}(e) exhibits poor contrast, and is very similar to the scattering intensities observed in the case where the bottom interface is rough [Fig.~\ref{fig:toy}(f)]. This clearly hints towards the idea that the poor contrast observed when the bottom film interface is rough is intrinsically linked to the nature of the paths.
Moreover, we have seen that ignoring the contribution from $r''$ in Eq.~(\ref{eq:toy:rf}) gives a result similar to when all terms of the same equation are included. This indicates that the contribution from $r''$ can be neglected relative to the other two terms in Eq.~(\ref{eq:toy:rf}).
However, since paths similar to $(1'')$, $(2'')$, \emph{etc.} are the \emph{only} paths allowed for the configuration where the bottom interface is rough, the contrast is poor by default. In both cases, and as we have seen, a typical path must undergo a number of non-scattering reflections within the film both before \textit{and} after the scattering event occurs.
Consequently, the phase difference between any two such paths will in general involve integer combinations of $\varphi_0$ and $\varphi_s$, as can be seen from Eqs.~\eqref{eq:toy:rf:2} and \eqref{eq:toy:fr}. This phase mixing is the fundamental reason for the difference in contrast found in the contributions to the total intensity $r_\mathrm{RP}$ from the three components of Eq.~\eqref{eq:toy:rf}.
The difference in contrast can also be investigated mathematically by estimating the contrast directly, as explained in \ref{AppendixC}.

We now turn to the full angular distributions for the mean DRC.
Figures~\ref{fig:2Dmdrc_vps_bothfigs}(a)--(i) and \ref{fig:2Dmdrc_vps_bothfigs}(j)--(r) show the full angular distributions of the incoherent contribution to the mean DRC, for simulation parameters corresponding to those assumed in obtaining the results of Figures~\ref{fig:mdrc_cut_vps_bothfigs}(a)--(b) and \ref{fig:mdrc_cut_vps_bothfigs}(c)--(d), respectively.
In fact, the non-perturbative results presented in Figs.~\ref{fig:mdrc_cut_vps_bothfigs}(a)--(b) and \ref{fig:mdrc_cut_vps_bothfigs}(c)--(d) correspond to in-plane cuts along the $q_1$ axis from Figs.~\ref{fig:2Dmdrc_vps_bothfigs}(b,\,f,\,k,\,o).
The results of Fig.~\ref{fig:2Dmdrc_vps_bothfigs} show that, in addition to the interference phenomena already mentioned, the distributions of the incoherent contributions to the mean DRC are also weighted by the shifted power spectrum of the rough interface.
In the current work this is a Gaussian envelope centered at the angle of specular reflection, where the width of the envelope is directly influenced by the surface-height correlation length $a$.
This is shown explicitly in the case of small amplitude perturbation theory to leading order as the term $g(\Vie{p}{}{} -\Vie{p}{0}{})$ in Eq.~\eqref{eq:incoMDRC:final}, and its impact on the scattering distributions should not be confused with the interference phenomena.

The reader may verify that the maxima and minima are located at the same positions as predicted for Fig.~\ref{fig:mdrc_cut_vps_bothfigs}, as is predicted by Eq.~\eqref{extrema}.
However, for Figs.~\ref{fig:2Dmdrc_vps_bothfigs}(j)--(r) the contrast in the oscillations of the incoherent contribution to the mean DRC is now less pronounced, as explained for in-plane scattering.

The lower left $2 \times 2$ panels in each of the panel collections in Fig.~\ref{fig:2Dmdrc_vps_bothfigs} display overall dipole-like patterns oriented along the plane of incidence for co-polarized scattering and perpendicular to it for cross-polarized scattering. These features are consequences of the definition used for the polarization vectors of our system.
They are similar to the scattered intensity patterns obtained in recent studies of light scattering from single two-dimensional randomly rough surfaces~\cite{Nordam2013a,Hetland2016a,Hetland2017,Simonsen2010a,Simonsen2011,Simonsen2009-9}.

\subsection{Two rough interfaces}\label{sec:two:rough}

We will now turn to the discussion of the geometry where \emph{both}  the top and bottom interfaces of the film are rough. In the following it will be assumed that these rough interfaces are characterized by Eq.~\eqref{eq:sumup:cross-covariance}, and for simplicity it will be assumed that their rms-roughness are the same and equal to $\sigma_1=\sigma_2=\lambda/30$. The cross-correlation between these two interfaces is characterized by the parameter $\gamma$ which is allowed to take values in the interval from $-1$ to  $1$. All the remaining experimental parameters are identical to those assumed in the preceding sections of this paper.

For the case where only one of the two interfaces of the film was rough, we demonstrated that good agreement exists between the results obtained by a purely numerical solution of the RRE and those obtained on the basis of a perturbative solution of the same equation~[SAPT]. A purely numerical solution of the RRE associated with a film geometry where more than one of the interfaces are rough is a challenging task that requires extensive computational resources to obtain, and to the best of our knowledge such a purely numerical solution has not yet been reported.
Therefore, for film geometries where both interfaces are rough we will only solve the corresponding RRE through SAPT to obtain the incoherent component of the mean DRC to second order in products of the surface profile functions, for which the relevant expression is given by Eq.~\eqref{eq:incoMDRC:final}. In the following it will be assumed that for the level of surface roughness that we consider here, which provided accurate results for the corresponding one-rough-interface film geometry considered in the preceding subsection, such a perturbative solution method is sufficiently accurate to adequately describe the physics of the problem under investigation.

\begin{figure}[t]
  \vspace{-1cm}
  \centering
  \includegraphics[width=0.49\linewidth,trim= 0.cm 0.cm 0.cm 0.cm,clip]{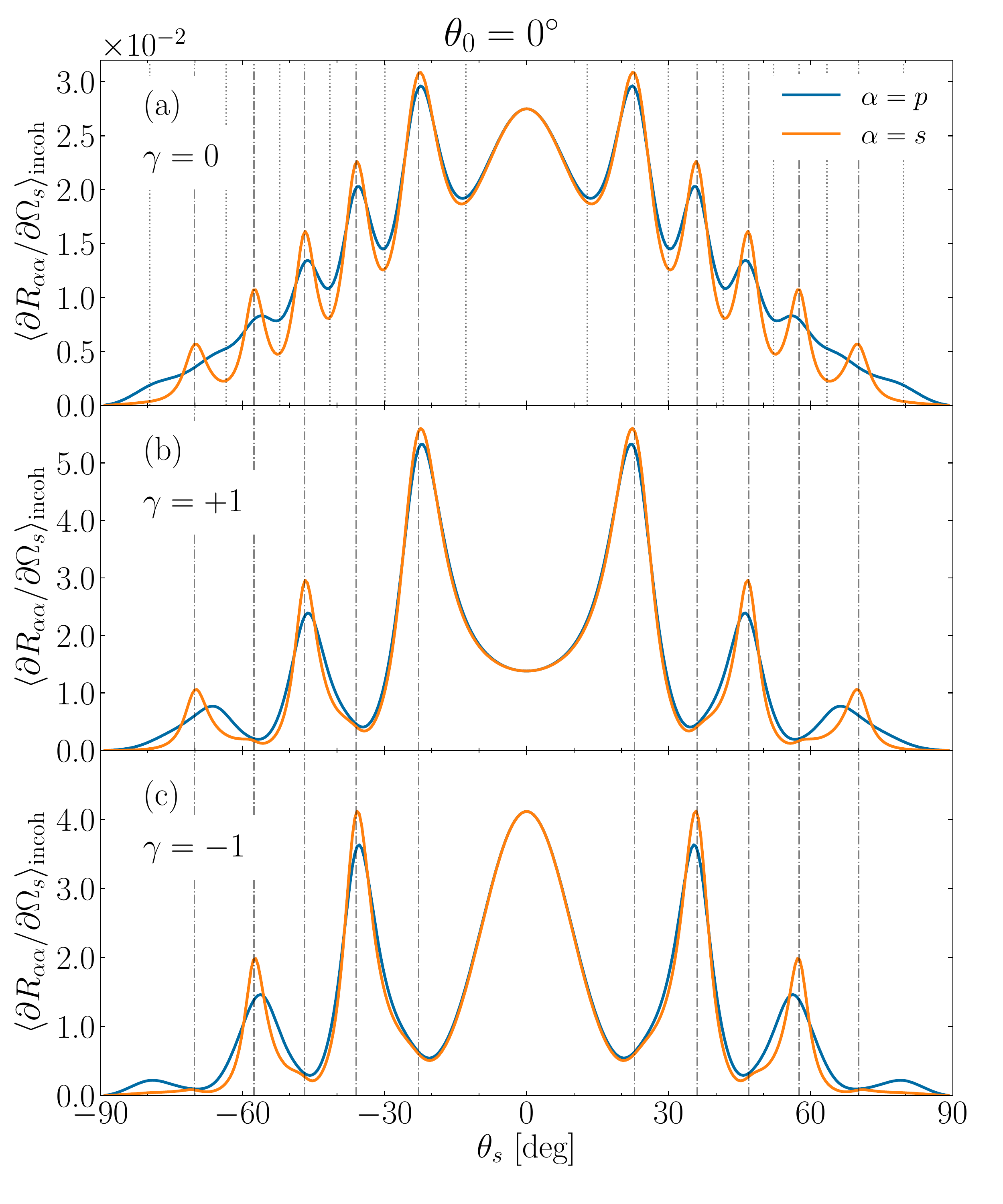}
  \includegraphics[width=0.49\linewidth,trim= 0.cm 0.cm 0.cm 0.cm,clip]{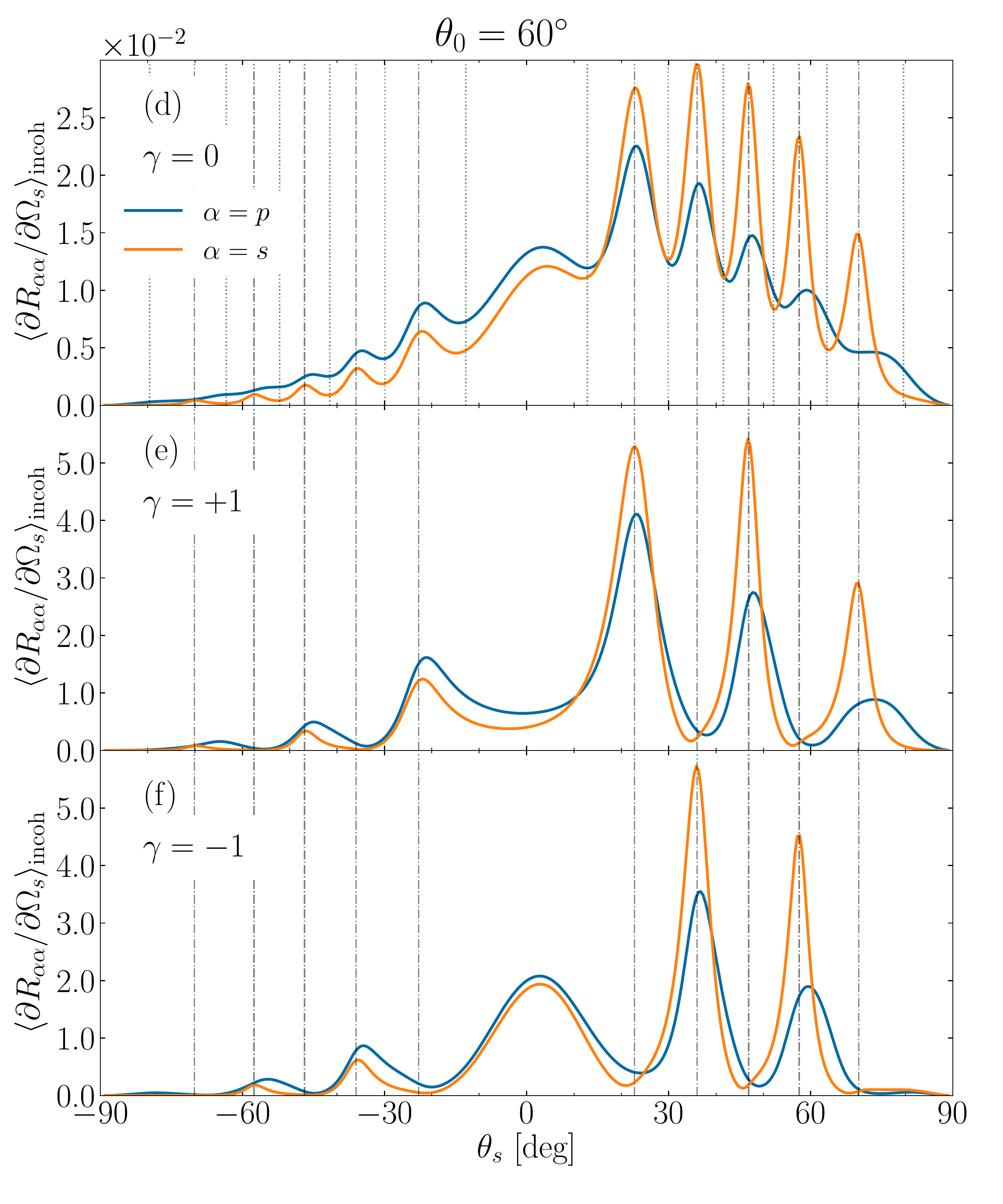}
\caption{
  Incoherent components of the mean differential reflection coefficients  $\left\langle \partial R_{\alpha \alpha} / \partial \Omega_s \right\rangle_\mathrm{incoh}$ for in-plane co-polarized scattering from a two-rough-interface film geometry for the polar angle of incidence $\theta_0=\ang{0}$~[Figs.~\protect\ref{MDRC:cut:normal_inc}(a)--(c)] and
  $\theta_0=\ang{60}$~[Figs.~\protect\ref{MDRC:cut:normal_inc}(d)--(e)].
  The wavelength of the incident light was  $\lambda = \SI{632}{\nm}$, the mean thickness of the film $d = 8 \lambda$, and the dielectric constants of the media were $\ve_1=1.0$, $\ve_2=2.69$, $\ve_3=15.08 + 0.15\im$. The rms-roughness of the interfaces were  $\sigma_1=\sigma_2=\lambda/30$, and the Gaussian correlation functions of each of the surfaces were characterized by the correlation length  $a=\lambda/3$.
  The cross-correlation function between the rough top and rough bottom interface of the film had the form~\eqref{eq:cross_covariance} and was characterized by the parameter $\gamma$ with values as indicated in each of the panels. The vertical dash-dotted and dotted lines indicate the expected angular positions of the maxima and minima of the scattered intensity as predicted by Eq.~\eqref{extrema}, respectively. For reasons of clarity only the expected positions of the minima of the in-plane mean DRCs are indicated in Figs.~\ref{MDRC:cut:normal_inc}(a) and \protect\ref{MDRC:cut:normal_inc}(d).}
\label{MDRC:cut:normal_inc}
\end{figure}

The first set of scattering results for a film bounded by two rough interfaces is presented in Fig.~\ref{MDRC:cut:normal_inc}.  In particular, Figs.~\ref{MDRC:cut:normal_inc}(a)--(c) present the incoherent component of the mean DRC for in-plane co-polarized scattering (\emph{i.e.} $|\hat{\mathbf{p}} \cdot \hat{\mathbf{p}}_0| = 1$ and $\alpha = \beta$) as a function of the polar angle of scattering $\theta_s$, for given polar angle of incidence equal to $\theta_0 = \ang{0}$, and for three extreme values of the cross-correlation parameter $\gamma \in \{0,1,-1\}$.
These three values of $\gamma$ physically correspond to the situations of uncorrelated film interfaces; perfectly positively correlated interfaces so that the film thickness measured along any vertical line segment will be constant and equal to $d$; and perfectly negatively correlated or anti-correlated interfaces, respectively.
From Fig.~\ref{MDRC:cut:normal_inc}(a) one observes that for uncorrelated  interfaces of the film [$\gamma = 0$], the number of interference fringes and their angular positions remain unchanged as compared to what was found when  only one of the two interfaces of the film was rough. This is found to be the case for both $p$- and $s$-polarized incident light. Such behavior can easily be understood in terms of the expression in Eq.~\eqref{eq:incoMDRC:final}; when $\gamma = 0$ only the first two terms in the square brackets on the right-hand-side of this equation contribute. These two terms are the only non-zero contributions to the incoherent component of the mean DRC (to second order) for a film system bounded by two \emph{uncorrelated} rough surfaces.
Moreover, these two contributions are, respectively, identical to the incoherent component of the mean DRC obtained for film geometries where either the top or the bottom interface of the film is rough and the other planar. Summing these two contributions will hence result in summing two similar interference intensity patterns. Consequently, the resulting interference pattern  maintains the same number of fringes at the same positions as the pattern obtained from scattering from the corresponding one-rough interface film  geometry. However, by gradually introducing more cross-correlation between the two rough interfaces of the film [$\gamma\neq0$], one observes that half of the fringes observed for the system for which $\gamma = 0$ are significantly attenuated whereas the other half are enhanced [Figs.~\ref{MDRC:cut:normal_inc}(b) and \ref{MDRC:cut:normal_inc}(c)].
Furthermore, it is observed from the results in Figs.~\ref{MDRC:cut:normal_inc}(a)--(c) that the fringes that are enhanced (attenuated) for the case when $\gamma = 1$ are the fringes being attenuated (enhanced) for the case when $\gamma = -1$.
This phenomenon can be attributed to the last term in the square brackets in Eq.~\eqref{eq:incoMDRC:final} which is linear in $\gamma$ and can take both positive and negative values and hence increase or decrease the value of the intensity pattern resulting from the superposition of the scattering  amplitudes obtained for the two independent aforementioned one-rough-interface film geometries.

The last term in the square brackets of Eq.~\eqref{eq:incoMDRC:final} is an interference term. Physically it can be interpreted as the interference between
a path formed by a single scattering event occurring on the \emph{top} interface of the film such as one depicted in Figs.~\ref{fig:toy}(a-b), and a path consisting of a single scattering event taking place on the \emph{bottom} interface as depiected in Fig.~\ref{fig:toy}(c).
When the two interfaces are uncorrelated, the phase difference between these two optical paths will form an uncorrelated random variable so that the ensemble average of the term where it appears in Eq.~\eqref{eq:incoMDRC:final} will be zero and the mean DRC will equal the sum of the intensities of the two corresponding one-rough-interface geometries, i.e. it will be given by the two first terms of Eq.~\eqref{eq:incoMDRC:final}.
However, when the two interfaces of the film are completely or partially correlated, $|\gamma|>0$, the phase difference of these two paths becomes a correlated random variable so that the interference term --- the last term in \eqref{eq:incoMDRC:final} ---  does not average to zero; this results in an optical interference effect. Consequently, the observed interference pattern for $|\gamma|>0$ will obtain a non-zero contribution from the last term in the square brackets of Eq.~\eqref{eq:incoMDRC:final}, which thus will make it different from the pattern obtained for an uncorrelated film geometry that corresponds to $\gamma=0$.

%
Figures~\ref{MDRC:cut:normal_inc}(d)--(f) present for polar angle of incidence $\theta_0=\ang{60}$  similar results to those presented in Figs.~\ref{MDRC:cut:normal_inc}(a)--(c) for normal incidence. Except for the increased intensity of the light scattered into the forward direction defined by $\theta_s>\ang{0}$ relative to what is scattered into angles $\theta_s<\ang{0}$, and the increased contrast of the fringes observed for $s$-polarized light in the forward direction, the behavior of the mean DRC curves is rather similar for the two angles of incidence. In particular, for the same value of $\gamma$, fringes are observed at the same angular positions for the two angles of incidence.
Moreover, which of the fringes that are enhanced or attenuated by the  introduction of (positive or negative) cross-correlation between the two rough interfaces of the film are also the same for the two angles of incidence. Such behavior is as expected for Sel\'{e}nyi fringes.

%
%

A close inspection of the perturbative results presented in  Fig.~\ref{MDRC:cut:normal_inc} reveals that for both $\theta_0=\ang{0}$ and $\theta_0=\ang{60}$ the angular positions of the maxima of the in-plane, co-polarized mean DRC curves are more accurately predicted by Eq.~\eqref{extrema} for $s$-polarized light than for $p$-polarized light; this seems in particular to be the case for the larger values of $|\theta_s|$. We speculate that such behavior is related to a phase change associated with the Brewster scattering phenomenon \cite{Hetland2016a,Hetland2017,Kawanishi1997} that exists in the case of $p$-polarized light, reminiscent of the well known phase change associated with the Brewster angle found for planar interfaces.

%
So far in our analysis of the two-rough-interface film geometry, we have observed that the enhancement or attenuation of the diffusely scattered co-polarized intensity are localized to regions around the polar angles determined by Eq.~\eqref{maxima}.
In order to make this observation more apparent, Figs.~\ref{fig:decomposition}(a)--(e) present various terms, or combinations of terms, from Eq.~\eqref{eq:incoMDRC:final} when the incident and scattered light is $p$-polarized; Figs.~\ref{fig:decomposition}(f)--(j)  depict similar results for $s$-polarized incident and scattered light. The three first columns of sub-figures that are present in Fig.~\ref{fig:decomposition} --- labeled ``Interface~1'', ``Interface~2'', and ``Cross-correlation'' --- represent the terms in Eq.~\eqref{eq:incoMDRC:final} that contain the factors $\sigma_1^2$, $\sigma_2^2$, and $\sigma_1\sigma_2$, respectively.
The cross-correlation terms, Figs.~\ref{fig:decomposition}(c) and \ref{fig:decomposition}(h), where obtained from the last term of Eq.~\eqref{eq:incoMDRC:final} with $\gamma=1$. Furthermore, a contour plot that appears in the 4th column of Fig.~\ref{fig:decomposition} [labeled ``Total~($\gamma=1$)''], displays the sum of the data used to produce the three first mean DRC contour plots appearing in the same row.
In other words, the results depicted in Figs.~\ref{fig:decomposition}(d) and \ref{fig:decomposition}(i) are the contour plots of the incoherent component of the mean DRC for a film geometry bounded by two perfectly correlated rough interfaces and therefore given by the expression in Eq.~\eqref{eq:incoMDRC:final}  with $\gamma=1$. Similarly, the incoherent component of the mean DRCs for a geometry where the two rough interfaces of the film are perfectly anti-correlated are displayed in the last column of Fig.~\ref{fig:decomposition} [Figs.~\ref{fig:decomposition}(e) and \ref{fig:decomposition}(j)] and labeled ``Total~($\gamma=-1$)''.
These latter results correspond to Eq.~\eqref{eq:incoMDRC:final} with $\gamma=-1$, and can be obtained by summing the results of the two first columns and subtracting the result of the third column of Fig.~\ref{fig:decomposition}.

The contour plots of the cross-correlation terms presented in Figs.~\ref{fig:decomposition}(c) and \ref{fig:decomposition}(h), which are obtained under the assumption that $\gamma=1$, display extrema localized on the same grid of points in the
$(\theta_0,\theta_s)$-plane as the extrema of the incoherent component of the mean DRC obtained when only one of the film interfaces is rough [Figs.~\ref{fig:decomposition}(a)--(b) and \ref{fig:decomposition}(f)--(g)]. An important observation should be made from these results. The minima of the former (the cross-correlation terms) are negative while the latter are always non-negative.
Hence, the incoherent component of the mean DRC for $\gamma=1$, which according to Eq.~\eqref{eq:incoMDRC:final} corresponds to the addition of the results used to produce the three first columns of each row of Fig.~\ref{fig:decomposition}, will cause fringes localized at the minima of the cross-correlation terms to be \emph{attenuated} (or disappear) and those localized at the maxima of the cross-correlation terms to be \emph{enhanced} [see Figs.~\ref{fig:decomposition}(d) and \ref{fig:decomposition}(i)].


%
%

%
\begin{figure*}[t]
  \vspace{-1cm}
  \centering 
  \includegraphics[width=0.95\linewidth,trim= 0.cm 0.cm 0.cm 0.cm,clip]{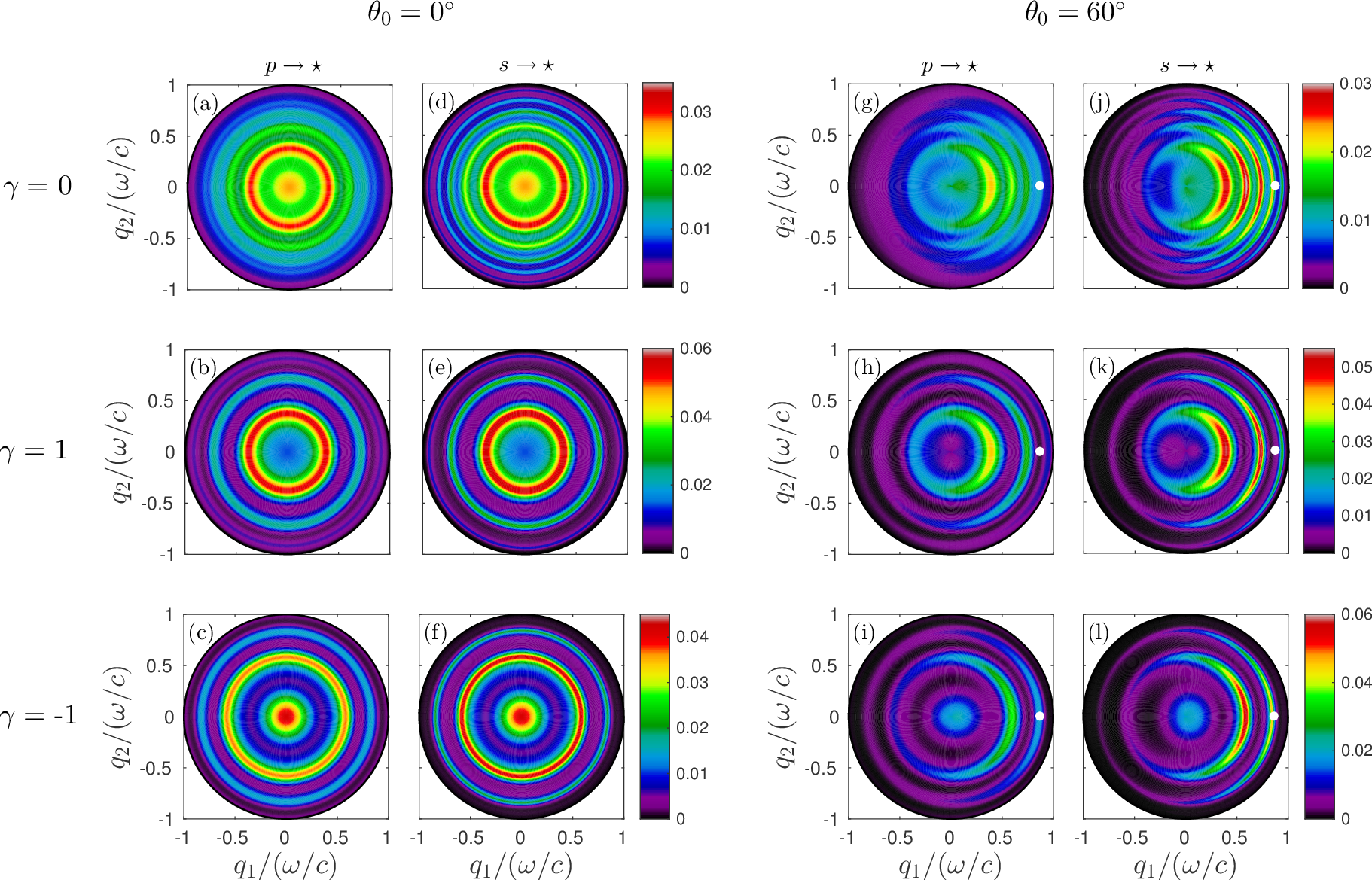}
  \caption{The full angular distribution of the incoherent component of the mean DRC, $\left\langle \partial R_{\alpha \beta} / \partial \Omega_s \right\rangle_\mathrm{incoh}$, for incident  $\beta$-polarized light that is scattered by a two-rough-interface film geometry into $\alpha$-polarized light [with $\alpha=p,s$ and $\beta=p,s$]. When the polarization of the scattered light is not observed, the relevant mean DRC quantity is $\sum_{\alpha=p,s} \left\langle \partial R_{\alpha \beta} / \partial \Omega_s \right\rangle_\mathrm{incoh}$ and this situation is labeled as $\beta\rightarrow\star$.
  The reported results were obtained on the basis of SAPT, Eq.~\eqref{eq:incoMDRC:final}, and the polar angles of incidence were $\theta_0=\ang{0}$~[Fig.~\ref{Fig:FullAngualDistribution-two-rough-film-interfaces}(a)--(f)] and $\theta_0=\ang{60}$~[Fig.~\ref{Fig:FullAngualDistribution-two-rough-film-interfaces}(g)--(l)].
  The incident in-plane wave vector is indicated by the white dot for non-normal incidence [Fig.~\ref{Fig:FullAngualDistribution-two-rough-film-interfaces}(g)--(l)]. The cross-correlation function between the rough top and rough bottom interface of the film had the form~\eqref{eq:cross_covariance} and was characterized by the parameter $\gamma$ as indicated in the figure (and constant for each row of sub-figure). The remaining roughness parameters are identical to those assumed in producing the results presented in Fig.~\ref{MDRC:cut:normal_inc}.
}
\label{Fig:FullAngualDistribution-two-rough-film-interfaces}
\end{figure*}
%
\smallskip
The preceding discussion stays valid when considering the full angular distribution of the incoherent component of the mean DRC. Figure~\ref{Fig:FullAngualDistribution-two-rough-film-interfaces} presents the full angular distribution of the incoherent component of the mean DRC,  obtained on the basis of Eq.~\eqref{eq:incoMDRC:final}, for the two polar angles of incidence $\theta_0=\ang{0}$~[Figs.~\ref{Fig:FullAngualDistribution-two-rough-film-interfaces}(a)--(f)] and  $\theta_0=\ang{60}$~[Figs.~\ref{Fig:FullAngualDistribution-two-rough-film-interfaces}(g)--(l)]. In this figure, each column formed by the sub-plots corresponds to either $p$- or $s$-polarized incident light, and in all cases the polarization of the scattered light was not recorded.
Moreover, each of the three rows of sub-figures that are present in Fig.~\ref{Fig:FullAngualDistribution-two-rough-film-interfaces} corresponds to different values for the cross-correlation parameter $\gamma \in \{0, 1,-1\}$ as indicated in the figure.
From the results presented in Fig.~\ref{Fig:FullAngualDistribution-two-rough-film-interfaces} it should be apparent that what appear as fringes in the in-plane angular dependence of the mean DRCs indeed are expressed as interference rings in the full-angular distribution of the same quantity; this is particularly apparent for normal incidence where the intensity of the (Sel\'{e}nyi) interference rings is independent of the azimuthal angle of scattering $\phi_s$ (due to the rotational invariance of the system and the source).
The angular distributions in Figs.~\ref{Fig:FullAngualDistribution-two-rough-film-interfaces}(a)--(f) also demonstrate very clearly how the possible interference rings present for uncorrelated interfaces of the film~[$\gamma=0$] are enhanced or attenuated when $|\gamma|\neq 0$, \textit{i.e.} when cross-correlation exists between the two rough interfaces of the film.

Figures~\ref{Fig:FullAngualDistribution-two-rough-film-interfaces}(g)--(l) show that interference rings are also present for non-normal incidence and that they are present for the same polar scattering angles $\theta_s$ as was found for normal incidence. However, for non-normal incidence the intensity of the rings does depend on the azimuthal angle of scattering. It is found that the intensity of the interference rings are concentrated to the forward scattering plane [$|\phi_s-\phi_0|<\ang{90}$].

%
\begin{figure*}[t]
\centering 
\includegraphics[width=0.95\linewidth,trim= 0.cm 0.cm 0.cm 0.cm,clip]{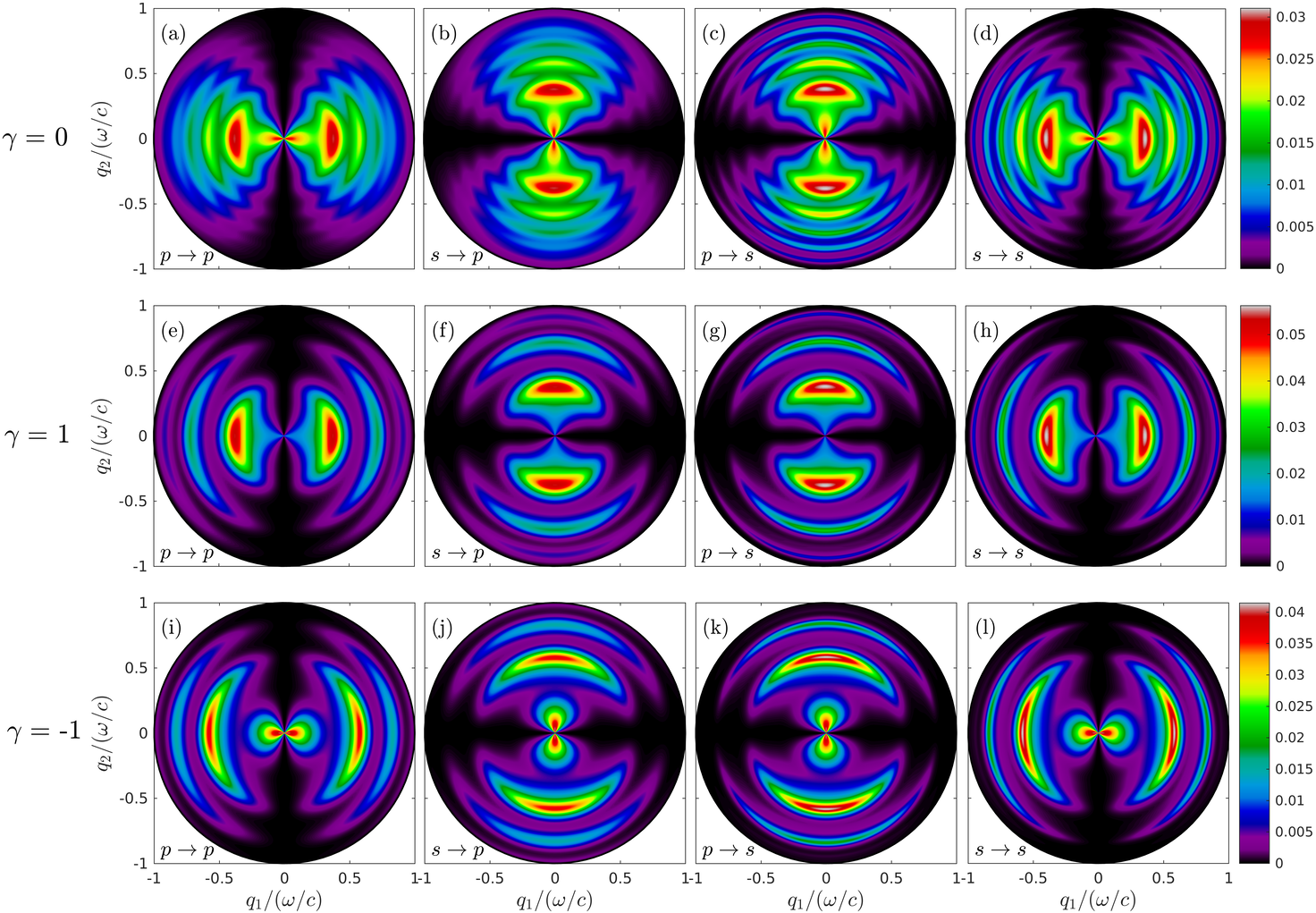}
\caption{
  The full angular distribution of the incoherent component of the mean DRC, $\left\langle \partial R_{\alpha \beta} / \partial \Omega_s \right\rangle_\mathrm{incoh}$, for incident
  $\beta$-polarized light of polar angle $\theta_0=\ang{0}$ that is scattered by a two-rough-interface film geometry into $\alpha$-polarized light [with $\alpha=p,s$ and $\beta=p,s$] and labeled $\beta\rightarrow\alpha$ in the sub-figures. The cross-correlation function between the rough top and rough bottom interface of the film had the form~\eqref{eq:cross_covariance} and was characterized by the parameter $\gamma$ as marked in the figure. The reported results were obtained on the basis of SAPT, Eq.~\eqref{eq:incoMDRC:final}. The remaining experimental and roughness parameters are identical to those assumed in producing the results presented in Figs.~\ref{MDRC:cut:normal_inc} and \ref{Fig:FullAngualDistribution-two-rough-film-interfaces}.
  }
\label{fig:mdrc:q:th0_equals_0}
\end{figure*}
%

For normal incidence Fig.~\ref{fig:mdrc:q:th0_equals_0} presents, for completeness, the full angular distribution of $\left<\partial R_{\alpha\beta}/\partial\Omega_s\right>_\textrm{incoh}$  for all possible linear polarization couplings, \textit{i.e.} from incident $\beta$-polarized light to scattered $\alpha$-polarized light. The values assumed for the cross-correlation parameter in obtaining these results were $\gamma\in\{0,1,-1\}$.
It should be observed from the results of Fig.~\ref{fig:mdrc:q:th0_equals_0} that interference structures are observed but they are \emph{not} ring structures of a constant amplitude as was seen in Fig.~\ref{Fig:FullAngualDistribution-two-rough-film-interfaces}(a)--(f).
The reason for this difference is that in the results presented in  Fig.~\ref{fig:mdrc:q:th0_equals_0} only scattered light of a given linear polarization was observed; this contrasts with the situation assumed in producing  Fig.~\ref{Fig:FullAngualDistribution-two-rough-film-interfaces} where all scattered light was observed and not only scattered light of a given linear polarization.

\smallskip
We have here only shown the extreme cases of cross-correlation, but one may also consider intermediate values for the cross-correlation parameter $\gamma$. The effect found for $\gamma=\pm 1$ remains also for $0 < |\gamma| < 1$ but with less pronounced enhancement and attenuation of the rings. The reader is invited to take a look at the animations in the Supplementary Materials, where the contour plots of the incoherent component of the mean DRCs are featured for smoothly varying  cross-correlation parameter $\gamma$ over the interval from  $-1$ to $1$, for both normal incidence and for $\theta_0=\ang{60}$ incidence.

\subsection{Transmitted light}\label{sec:transmission}

Finally, we would like to briefly comment on what would be observed in transmission if a non-absorbing medium was chosen, such as silica. No results will be presented here, but we have observed that interference rings are also observed in the diffusely transmitted light and that the effect of enhancement and attenuation of the rings induced by the interface cross-correlation still holds.
Furthermore, additional features attributed to the so-called Brewster scattering angles and Yoneda effects in the diffusely transmitted light would then be present. As presented in Ref.~\cite{Hetland2017} for scattering systems of comparable surface roughness and materials, the diffusely transmitted intensity as a function of angle of transmission will be modulated by a typical Yoneda intensity pattern.
At normal incidence this pattern exhibits a peak at some critical angle of scattering for $s$-polarized light and a vanishing intensity for $p$-polarized light (see Ref.~\cite{Hetland2017} for details).
However, we observed that not only did the overall intensity distribution undergo such modulation: the angular positions of the fringes were also affected compared to the predictions provided by naive optical path arguments, analogous to what was presented in this paper for reflection.
The angular positions of the fringes predicted by optical path arguments leading to equations similar to Eq.~(\ref{extrema}) still hold for scattering angles below the Yoneda critical angle, but must be corrected for scattering angles larger than the Yoneda critical angle. We speculate that this is due to a gradual phase shift that occurs above the critical angle, and that it is associated with the Yoneda phenomenon.
Note that this phenomenon is also observed in the diffusely reflected light if the medium of incidence has a higher refractive index than that of the substrate (i.e. $\epsilon_1 > \epsilon_3$) \cite{Hetland2016a,Gonzalez-Alcalde2016}.
Moreover, we have also observed that when scattered to larger polar angles than the Brewster scattering angle the $p$-polarized transmitted light exhibits an additional phase shift, as compared to $s$-polarized transmitted light, resulting in a switch in the positions for the maxima and minima. These and other features of the interference rings in the diffusely transmitted light will be discussed in more detail in a dedicated paper.

\section{Conclusion}
\label{Sec:Conculusions}
Based on both non-perturbative and perturbative solutions of the reduced Rayleigh equation, we have in this paper demonstrated that for systems composed of two-dimensional weakly rough dielectric films, Sel\'{e}nyi rings can be observed in the diffusely scattered light.
These rings make up a static interference pattern that is modulated by the polar angle of incidence.
We have illustrated that the interference mechanism at play can be explained by simple optical path arguments, leading to a simple model capable of predicting both the angular positions of the rings and the expected difference in contrast of the rings for film geometries where either the top or the bottom  interface of the film is rough (but not both interfaces).

Furthermore, by investigating the influence of the cross-correlation between the film interfaces when both interfaces are rough, we have shown that a selective enhancement or attenuation of the interference rings in the diffusely scattered light can be observed.
This suggests that the positions and the amplitudes of Sel\'{e}nyi rings can, when combined with reflectivity and/or ellipsometry measurements, in principle enable the determination of the mean film thickness, the  dielectric constant of the film material and the statistical properties of the interfaces.
In particular, numerical experiments show that the cross-correlation between interfaces can be assessed. Alternatively, film geometries consisting of cross-correlated interfaces can be designed to control the intensity pattern of the diffusely scattered light that they give rise to.
Sensors can also be designed in such a way that the interference rings observed for a clean system with known cross-correlated interfaces will be modified by the adsorption of a substance or nano-particles onto the first interface, hence partially destroying the effective cross-correlation between the interfaces.
These possibilities are, however, likely to be limited by the ordering of length scales $d > \lambda > \sigma$, which expresses the fact that the film thickness must be on the order of a few wavelengths to observe interference rings in the diffusely scattered light and that the rms-roughness of the interfaces should be small compared to the wavelength. Such a length scale ordering combined with controlled interface cross-correlation may be challenging to achieve experimentally.

While the main results presented in this paper considered the diffusely scattered light, the theoretical framework that it presents also allows for the investigation into the light transmitted diffusely through transparent film structures with one or several rough interfaces. The developed theoretical framework is readily generalized to the case of an arbitrary number of correlated layers and allows, for example, for the study of the effect of gradually changing cross-correlations over many interfaces.

We hope that the results presented in this paper can motivate experimental investigations into the scattering of light from rough film systems so that the predictions that are reported here based on theoretical grounds can be confirmed experimentally.


\section*{Acknowledgment}
This research was supported in part by The Research Council of Norway Contract No. 216699. The research of I.S. was supported in part by the French National Research Agency (ANR-15-CHIN-0003).
This research was supported in part by NTNU and the Norwegian metacenter for High Performance Computing (NOTUR) by the allocation of computer time.

%
\section*{References}
\bibliography{scatterbib_hetland}

\clearpage
\appendix
\section{Perturbative solution}\label{AppendixA}

We present here a method known as small amplitude perturbation theory that we apply to find an approximate solution of the reduced Rayleigh equations. We will illustrate the method considering a system made of a stack of three media separated by two randomly rough interfaces, like the one depicted in Fig.~\ref{fig:system}. Using the notation introduced in Sec.~\ref{sec:theory}, we know that the reduced Rayleigh equations for the reflection amplitude is given by Eqs.~\eqref{RREint} and \eqref{matrices}
\begin{equation}
\int \Vie{\Theta}{3,1}{+,+} \ofuipi{p}{}{q}{} \: \Vie{R}{}{} \ofuipi{q}{}{p}{0} \dtwopi{q}{}= - \Vie{\Theta}{3,1}{+,-} \ofuipi{p}{}{p}{0} \: \mathrm{,}
\label{multiRRErefl}
\end{equation}
where we recall that the \emph{forward two-interface transfer kernel} is defined as
%
%
\begin{equation}
\Vie{\Theta}{3,1}{a_{3},a_{1}} \ofuipi{p}{3}{p}{1} = \sum_{a_{2}=\pm} a_{2} \int \Vie{\Theta}{3,2}{a_{3},a_{2}} \ofuipi{p}{3}{p}{2} \,  \Vie{\Theta}{2,1}{a_{2},a_1} \ofuipi{p}{2}{p}{1} \: \dtwopi{p}{2} \:,
\end{equation}
with the \emph{single-interface kernels} $\Vie{\Theta}{l,m}{b,a}$ defined for successive media, i.e. $l, m \in \{ 1, 3 \}$ such that $|l-m| = 1$, $a, b \in \{\pm\}$, as
\begin{equation}
\Vie{\Theta}{l,m}{b,a} \ofuipi{p}{}{q}{} = \alpha^{-1}_l(\mathbf{p}) \Cie{J}{l,m}{b,a} \ofuipi{p}{}{q}{} \: \Vie{M}{l,m}{b,a}  \ofuipi{p}{}{q}{}.
\end{equation}
The perturbative method consists in expanding each single-interface kernel in a series of Fourier moments. In order to avoid unnecessary lengthy expansion, we first introduce some notations that will allow us to keep a compact derivation and proved to be useful for generalizing to an arbitrary number of layers and for numerical implementation. We define
\begin{align}
  \Vie{\tilde{\Theta}}{3,1}{a_{3},a_{1},(\mathbf{m})} \ofuipiMulti{p}{3}{p}{2}{p}{1}
  =& \sum_{a_{2}=\pm} a_{2}  \,
     \alpha_{3}^{-1}(\Vie{p}{3}{}) \left[a_{3} \alpha_{3} (\Vie{p}{3}{}) - a_{2} \alpha_{2} (\Vie{p}{2}{}) \right]^{m_2-1} \,
     \exp\left[-i \left\{a_{3} \alpha_{3} (\Vie{p}{3}{}) -  a_{2} \alpha_{2} (\Vie{p}{2}{})\right\}  d_2\right]
     \nonumber \\
   & \qquad \times
     \alpha_{2}^{-1}(\Vie{p}{2}{}) \left[a_{2} \alpha_{2} (\Vie{p}{2}{}) - a_{1} \alpha_{1} (\Vie{p}{1}{}) \right]^{m_1-1} \,
     \exp\left[-i \left\{a_{2} \alpha_{2} (\Vie{p}{2}{}) - a_{1} \alpha_{1} (\Vie{p}{1}{}) \right\} d_1 \right]
     \nonumber \\
  & \qquad \times
    \Vie{M}{3,2}{a_{3},a_{2}}\!\!\ofuipi{p}{3}{p}{2} \,
    \Vie{M}{2,1}{a_{2},a_{1}}\!\ofuipi{p}{2}{p}{1} \: ,
\label{def:thetaOrder}
\end{align}
where $\mathbf{m} = (m_1, m_2) \in \mathbb{N}^2$ is a pair-index (i.e. a two component multi-index). Here, we have made the choice of factorizing the phase factor $e^{-i(a_{j+1} \alpha_{j+1} (\Vie{p}{j+1}{}) - a_{j} \alpha_{j} (\Vie{p}{j}{}) ) d_j}$, with $d_j = \langle \zeta_j \rangle$ being the offset height of the $j^\mathrm{th}$ interface, from each factor $\Cie{J}{j+1,j}{a_{j+1},a_{j}} \ofuipi{p}{j+1}{p}{j}$ for later convenience. Given this definition, an expansion of the two-interface kernel in Fourier moments is given by
%
%
\begin{align}
\Vie{\Theta}{3,1}{a_{3},a_{1}} \ofuipi{p}{3}{p}{1} &= \sum_{\mathbf{m} = 0}^{\infty} \frac{(-i)^{|\mathbf{m}|}}{\mathbf{m}!} \, \int \hat{h}_2^{(m_2)}(\Vie{p}{3}{} -\Vie{p}{2}{}) \: \hat{h}_1^{(m_1)}(\Vie{p}{2}{} -\Vie{p}{1}{}) \Vie{\tilde{\Theta}}{3,1}{a_{3},a_{1},(\mathbf{m})} \ofuipiMulti{p}{3}{p}{2}{p}{1} \dtwopi{p}{2}\nonumber \\
& = \sum_{\mathbf{m} = 0}^{\infty} \frac{(-i)^{|\mathbf{m}|}}{\mathbf{m}!} \: \Vie{Z}{3,1}{a_{3},a_{1},(\mathbf{m})} \ofuipi{p}{3}{p}{1} \: ,
\label{kernelExpansionMulti}
\end{align}
%
where $\sum_{\mathbf{m} = 0}^{\infty} \equiv \sum_{m_1 = 0}^{\infty}  \sum_{m_2 = 0}^{\infty}$, $|\mathbf{m}| = m_1 + m_2$ is the \emph{length} of the pair-index, and $\mathbf{m}! = m_1 ! \, m_2 !$, and for all $j \in \{1,2\}$,
\begin{equation}
  \hat{h}_j^{(m_j)}(\Vie{q}{}{})
  = \int  \exp \left[-i \Vie{q}{}{}\cdot\mathbf{x} \right]\: \left[\zeta_j \of{x} - d_j\right]^{m_j} \mathrm{d}^2 x \: ,
\end{equation}
is the Fourier moment of $h_j = \zeta_j - d_j$ of order $m_j$. It is then clear that $\Vie{Z}{3,1}{a_{3},a_{1},(\mathbf{m})} \ofuipi{p}{3}{p}{1}$ is a term of order $|\mathbf{m}|$ in product of surface profiles. The reflection amplitude can be expanded as
\begin{equation}
\Vie{R}{}{} \ofuipi{q}{}{p}{0} = \sum_{j=0}^\infty \frac{(-i)^j}{j!} \Vie{R}{}{(j)} \ofuipi{q}{}{p}{0} \: ,
\label{reflExpansion}
\end{equation}
where the term $\Vie{R}{}{(j)} \ofuipi{q}{}{p}{0}$ is of order $j$ in product of surface profiles. We are now ready to start the derivation of the perturbative expansion. By plugging Eqs.~(\ref{kernelExpansionMulti} and \ref{reflExpansion}) into Eq.~(\ref{multiRRErefl}) we obtain
%
%
\begin{equation}
\sum_{\substack{\mathbf{m}' = 0 \\ j=0}}^{\infty} \frac{(-i)^{|\mathbf{m}'|+j}}{\mathbf{m}'! \, j!} \int   \, \Vie{Z}{3,1}{+,+,(\mathbf{m}')} \ofuipi{p}{}{q}{} \: \Vie{R}{}{(j)} \ofuipi{q}{}{p}{0} \dtwopi{q}{} = - \sum_{\mathbf{m} = 0}^{\infty} \frac{(-i)^{|\mathbf{m}|}}{\mathbf{m}!} \Vie{Z}{3,1}{+,-,(\mathbf{m})} \ofuipi{p}{}{p}{0}.
\end{equation}
Summing over all multi-index $\mathbf{m}$ is equivalent to summing over subsets $\Cie{S}{m}{} = \{ \mathbf{m} \in \mathbb{N}^2 \st |\mathbf{m}| = m \}$ of multi-index of constant length $m$, i.e. that we have $\sum_{\mathbf{m} = 0}^{\infty} \equiv \sum_{m = 0}^{\infty} \sum_{\mathbf{m} \in \Cie{S}{m}{}}$, therefore the previous equation can be re-written as
%
%
\begin{equation}
\sum_{\substack{m' = 0 \\ j=0}}^{\infty} \sum_{\mathbf{m}' \in \Cie{S}{m'}{}} \frac{(-i)^{m'+j}}{\mathbf{m}'! \, j!} \int \, \Vie{Z}{3,1}{+,+,(\mathbf{m}')} \ofuipi{p}{}{q}{} \: \Vie{R}{}{(j)} \ofuipi{q}{}{p}{0} \dtwopi{q}{} = - \sum_{m = 0}^{\infty} \sum_{\mathbf{m} \in \Cie{S}{m}{}} \frac{(-i)^{m}}{\mathbf{m}!} \Vie{Z}{3,1}{+,-,(\mathbf{m})} \ofuipi{p}{}{p}{0} .
\end{equation}
We then use the definition of the multinomial coefficient in multi-index form as $|\mathbf{m}|!/\mathbf{m}! =  \binom{|\mathbf{m}|}{\mathbf{m}}$ to obtain
%
%
\begin{align}
\sum_{\substack{m' = 0 \\ j=0}}^{\infty}  \frac{(-i)^{m'+j}}{m'! \, j!} & \sum_{\mathbf{m}' \in \Cie{S}{m'}{}} \binom{m'}{\mathbf{m}'} \int   \, \Vie{Z}{3,1}{+,+,(\mathbf{m}')} \ofuipi{p}{}{q}{} \: \Vie{R}{}{(j)} \ofuipi{q}{}{p}{0} \dtwopi{q}{}  \nonumber \\
&=- \sum_{m = 0}^{\infty}  \frac{(-i)^{m}}{m!} \sum_{\mathbf{m} \in \Cie{S}{m}{}} \binom{m}{\mathbf{m}} \Vie{Z}{3,1}{+,-,(\mathbf{m})} \ofuipi{p}{}{p}{0}.
\end{align}
%
We now make a change of summation index $j \leftrightarrow m - m'$ on the left hand side of the above equation. This makes clearly appear terms of order $m$ in product of surface profiles. We obtain
%
%
%
\begin{align}
\sum_{m=0}^{\infty} \sum_{m'=0}^{m} & \frac{(-i)^{m}}{m'! \, (m-m')!} \sum_{\mathbf{m}' \in \Cie{S}{m'}{}} \binom{m'}{\mathbf{m}'} \int   \, \Vie{Z}{3,1}{+,+,(\mathbf{m}')} \ofuipi{p}{}{q}{} \: \Vie{R}{}{(m-m')} \ofuipi{q}{}{p}{0} \dtwopi{q}{}  \nonumber \\
&= - \sum_{m = 0}^{\infty}  \frac{(-i)^{m}}{m!} \sum_{\mathbf{m} \in \Cie{S}{m}{}} \binom{m}{\mathbf{m}} \Vie{Z}{3,1}{+,-,(\mathbf{m})} \ofuipi{p}{}{p}{0} \: ,
\end{align}
%
which can be re-written with the use of the definition of the binomial coefficient $\binom{m}{m'}$ as
%
%
\begin{align*}
\sum_{m=0}^{\infty} \frac{(-i)^{m}}{m!} & \sum_{m'=0}^{m} \binom{m}{m'} \sum_{\mathbf{m}' \in \Cie{S}{m'}{}} \binom{m'}{\mathbf{m}'} \int  \, \Vie{Z}{3,1}{+,+,(\mathbf{m}')} \ofuipi{p}{}{q}{} \: \Vie{R}{}{(m-m')} \ofuipi{q}{}{p}{0} \dtwopi{q}{} \\ &= - \sum_{m = 0}^{\infty}  \frac{(-i)^{m}}{m!} \sum_{\mathbf{m} \in \Cie{S}{m}{}} \binom{m}{\mathbf{m}} \Vie{Z}{3,1}{+,-,(\mathbf{m})} \ofuipi{p}{}{p}{0} . \nonumber
\end{align*}
%
It is now time to identify terms of same orders in the left and right hand sides. For $m=0$, only the term for $\mathbf{m}' = (0,0)$ remains in the left hand side, only the term $\mathbf{m} = (0,0)$ remains in the right hand side and we have
\begin{equation}
\int  \, \Vie{Z}{3,1}{+,+,(\mathbf{0})} \ofuipi{p}{}{q}{} \: \Vie{R}{}{(0)} \ofuipi{q}{}{p}{0} \dtwopi{q}{}= - \Vie{Z}{3,1}{+,-,(\mathbf{0})} \ofuipi{p}{}{p}{0} .
\end{equation}
which, when expanded, reads
%
%
\begin{align}
\iint  \hat{h}_2^{(0)}&(\Vie{p}{}{} -\Vie{p}{2}{})  \, \hat{h}_1^{(0)}(\Vie{p}{2}{} -\Vie{q}{}{}) \Vie{\tilde{\Theta}}{3,1}{+,+,(\mathbf{0})} \ofuipiMulti{p}{}{p}{2}{q}{} \dtwopi{p}{2} \: \Vie{R}{}{(0)} \ofuipi{q}{}{p}{0} \dtwopi{q}{} \nonumber \\   & = - \int \hat{h}_2^{(0)}(\Vie{p}{}{} -\Vie{p}{2}{})  \, \hat{h}_1^{(0)}(\Vie{p}{2}{} -\Vie{p}{0}{}) \Vie{\tilde{\Theta}}{3,1}{+,-,(\mathbf{0})} \ofuipiMulti{p}{}{p}{2}{p}{0} \dtwopi{p}{2} .
\end{align}
%
From the definition of the zero order Fourier moment, we have $\hat{h}_j^{(0)} (\Vie{q}{}{}) = (2 \pi)^2 \, \delta(\Vie{q}{}{})$, which  yields
%
\begin{equation}
\Vie{\tilde{\Theta}}{3,1}{+,+,(\mathbf{0})} \ofuipiMulti{p}{}{p}{}{p}{} \: \Vie{R}{}{(0)} \ofuipi{p}{}{p}{0} =  - (2 \pi)^2 \Vie{\tilde{\Theta}}{3,1}{+,-,(\mathbf{0})} \ofuipiMulti{p}{0}{p}{0}{p}{0} \, \delta(\Vie{p}{}{} -\Vie{p}{0}{}) .
\end{equation}
%
Here, the reader may recognize the solution of the reflection problem for a stack of layers with flat interfaces, i.e. Fresnel amplitudes
\begin{equation}
\Vie{R}{}{(0)} (\Vie{p}{}{}|\Vie{p}{0}{}) = - \left[ \Vie{\bar{\Theta}}{3,1}{+,+} (\Vie{p}{0}{}) \right]^{-1} \: \Vie{\bar{\Theta}}{3,1}{+,-} (\Vie{p}{0}{}) \: (2 \pi)^2 \delta (\Vie{p}{}{} - \Vie{p}{0}{}) = - \Vie{\boldsymbol \rho}{0}{} (\Vie{p}{0}{}) \: (2 \pi)^2 \delta (\Vie{p}{}{} - \Vie{p}{0}{})\: ,
\label{fresnel:r}
\end{equation}
where $\Vie{\bar{\Theta}}{3,1}{+,+} (\Vie{p}{0}{}) \equiv \Vie{\tilde{\Theta}}{3,1}{+,+,(\mathbf{0})} \ofuipiMulti{p}{0}{p}{0}{p}{0}$ and
$ \Vie{\bar{\Theta}}{3,1}{+,-} (\Vie{p}{0}{}) \equiv \Vie{\tilde{\Theta}}{3,1}{+,-,(\mathbf{0})} \ofuipiMulti{p}{0}{p}{0}{p}{0}$.
Thus, the order zero of the perturbative expansion corresponds to the Fresnel coefficients for the corresponding system with flat interfaces. For orders $m \geq 1$, we have
%
%
\begin{equation*}
\sum_{m'=0}^{m} \binom{m}{m'} \sum_{\mathbf{m}' \in \Cie{S}{m'}{}} \binom{m'}{\mathbf{m}'} \int   \, \Vie{Z}{3,1}{+,+,(\mathbf{m}')} \ofuipi{p}{}{q}{} \: \Vie{R}{}{(m-m')} \ofuipi{q}{}{p}{0} \dtwopi{q}{} = -\sum_{\mathbf{m} \in \Cie{S}{m}{}} \binom{m}{\mathbf{m}} \Vie{Z}{3,1}{+,-,(\mathbf{m})} \ofuipi{p}{}{p}{0} .
\end{equation*}
%
By isolating the term corresponding to $m'=0$, hence $\mathbf{m}' = (0,0)$ and using that for all $j \in \{1,2\}$ we have $\hat{h}_j^{(0)} (\Vie{q}{}{}) = (2 \pi)^2 \, \delta(\Vie{q}{}{})$, we obtain
%
%
%
%
%
%
%
\begin{align}
\Vie{R}{}{(m)} \ofuipi{p}{}{p}{0} = &- \left[ \Vie{\bar{\Theta}}{3,1}{+,+} (\Vie{p}{}{}) \right]^{-1} \: \left[ \sum_{\mathbf{m} \in \Cie{S}{m}{}} \binom{m}{\mathbf{m}}  \Vie{Z}{3,1}{+,-,(\mathbf{m})} \ofuipi{p}{}{p}{0} \right. \nonumber \\
&\left. + \sum_{m'=1}^{m} \binom{m}{m'}  \int \sum_{\mathbf{m}' \in \Cie{S}{m'}{}} \binom{m'}{\mathbf{m}'}\, \Vie{Z}{3,1}{+,+,(\mathbf{m}')} \ofuipi{p}{}{q}{} \Vie{R}{}{(m-m')} \ofuipi{q}{}{p}{0}  \dtwopi{q}{}  \right] .
\label{recursive:layerPerturbation}
\end{align}
%
We have finally obtained a recursive expression giving the $m^\mathrm{th}$ order term in the reflection amplitude expansion as a function of lower order terms. For weakly rough surfaces, an approximation based on a truncation of the expansion of the reflection amplitude Eq.~(\ref{reflExpansion}) to the first non-trivial order often yields accurate physical insights. For $m = 1$, we obtain that
%
%
\begin{align}
  \Vie{R}{}{(1)} \ofuipi{p}{}{p}{0} =
  &- \left[ \Vie{\bar{\Theta}}{3,1}{+,+} (\Vie{p}{}{}) \right]^{-1} \: \left[ \Vie{Z}{3,1}{+,-,(1,0)} \ofuipi{p}{}{p}{0} + \Vie{Z}{3,1}{+,-,(0,1)} \ofuipi{p}{}{p}{0} \right.
    \nonumber \\
  &\left. + \int \left( \Vie{Z}{3,1}{+,+,(1,0)} \ofuipi{p}{}{q}{} + \Vie{Z}{3,1}{+,+,(0,1)} \ofuipi{p}{}{q}{} \right) \Vie{R}{}{(0)} \ofuipi{q}{}{p}{0}  \dtwopi{q}{}  \right]
    \nonumber\\
 =& - \left[ \Vie{\bar{\Theta}}{3,1}{+,+} (\Vie{p}{}{}) \right]^{-1} \: \left[ \Vie{Z}{3,1}{+,-,(1,0)} \ofuipi{p}{}{p}{0} + \Vie{Z}{3,1}{+,-,(0,1)} \ofuipi{p}{}{p}{0} \right.
    \nonumber \\
  &\left. - \left( \Vie{Z}{3,1}{+,+,(1,0)} \ofuipi{p}{}{p}{0} + \Vie{Z}{3,1}{+,+,(0,1)} \ofuipi{p}{}{p}{0} \right) \left[ \Vie{\bar{\Theta}}{3,1}{+,+} (\Vie{p}{0}{}) \right]^{-1} \: \Vie{\bar{\Theta}}{3,1}{+,-} (\Vie{p}{0}{}) \right] .
\label{R1}
\end{align}
%
Where we have used the previously obtained expression for $\Vie{R}{}{(0)} \ofuipi{q}{}{p}{0}$ in Eq.~(\ref{fresnel:r}), and in particular the fundamental property of the Dirac delta. From the definition of $\Vie{Z}{3,1}{a_3,a_1,(\mathbf{m})}$ [Eq.~(\ref{kernelExpansionMulti})] it is clear that for $\mathbf{m} = (1,0)$ or $(0,1)$  the integration reduces to
%
%
\begin{subequations}
\begin{align}
\Vie{Z}{3,1}{a_3,a_1,(1,0)} \ofuipi{p}{}{p}{0} &= \hat{h}_1^{(1)}(\Vie{p}{}{} -\Vie{p}{0}{}) \Vie{\tilde{\Theta}}{3,1}{a_{3},a_{1},(1,0)} \ofuipiMulti{p}{}{p}{}{p}{0} \\
\Vie{Z}{3,1}{a_3,a_1,(0,1)} \ofuipi{p}{}{p}{0} &= \hat{h}_2^{(1)}(\Vie{p}{}{} -\Vie{p}{0}{}) \Vie{\tilde{\Theta}}{3,1}{a_{3},a_{1},(0,1)} \ofuipiMulti{p}{}{p}{0}{p}{0} .
\end{align}
\end{subequations}
It is convenient to group terms with common factor $\hat{h}_j \equiv \hat{h}_j^{(1)}$ in Eq.~(\ref{R1}), which leads to
\begin{equation}
\Vie{R}{}{(1)} \ofuipi{p}{}{p}{0} = \hat{h}_1(\Vie{p}{}{} -\Vie{p}{0}{}) \Vie{\boldsymbol \rho}{1}{} \ofuipi{p}{}{p}{0} + \, \hat{h}_2 (\Vie{p}{}{} -\Vie{p}{0}{}) \Vie{\boldsymbol \rho}{2}{} \ofuipi{p}{}{p}{0} \: ,
\label{r1}
\end{equation}
with
\begin{subequations}
\begin{align}
\Vie{\boldsymbol \rho}{1}{} \ofuipi{p}{}{p}{0}  &= \left[ \Vie{\bar{\Theta}}{3,1}{+,+} (\Vie{p}{}{}) \right]^{-1} \: \left[ \Vie{\tilde{\Theta}}{3,1}{+,+,(1,0)} \ofuipiMulti{p}{}{p}{}{p}{0} \Vie{\boldsymbol \rho}{0}{} (\Vie{p}{0}{}) -  \Vie{\tilde{\Theta}}{3,1}{+,-,(1,0)} \ofuipiMulti{p}{}{p}{}{p}{0} \right] \\
\Vie{\boldsymbol \rho}{2}{} \ofuipi{p}{}{p}{0}  &= \left[ \Vie{\bar{\Theta}}{3,1}{+,+} (\Vie{p}{}{}) \right]^{-1} \: \left[ \Vie{\tilde{\Theta}}{3,1}{+,+,(0,1)} \ofuipiMulti{p}{}{p}{0}{p}{0} \Vie{\boldsymbol \rho}{0}{} (\Vie{p}{0}{}) -  \Vie{\tilde{\Theta}}{3,1}{+,-,(0,1)} \ofuipiMulti{p}{}{p}{0}{p}{0} \right] .
\end{align}
\label{def:rho}%
\end{subequations}
%
%
%
We have treated the case of reflection as a representative example, but the same method applies for transmission.

\section{Differential reflection coefficient}\label{AppendixB}

Assuming we have obtained the reflection amplitudes $R_{\alpha \beta} (\Vie{p}{}{} \st \Vie{p}{0}{} )$ either by using the perturbative approach or by the purely numerical simulation, we can now proceed to express the differential reflection coefficient~(DRC) defined as the time-averaged flux radiated around a given scattering direction $(\theta_s,\phi_s)$ per unit solid angle and per unit incident flux and denoted $\partial R / \partial \Omega_s (\Vie{p}{}{} \st \Vie{p}{0}{} )$. Let a virtual hemisphere of radius $r \gg c/\omega$ lie on the plane $x_3 = 0$ on top of the scattering system. The support of this hemisphere is a disk of area $S = \pi r^2$. We consider the scattering from a \emph{truncated} version of the scattering system in which the surface profiles are set to be flat outside the disk support. Consequently, the field amplitudes we will manipulate are not strictly speaking those of the full system of interest but will converge to them as $r \to \infty$. We will nevertheless keep the same notations as that from the full system introduced in Section~\ref{sec:theory} for simplicity. The time-averaged flux incident on this disk is given by
\begin{align}
P_{\mathrm{inc}/S} &= - \mathrm{Re} \frac{c}{8 \pi} \int_S  \left[ \Vie{E}{0}{*} (\Vie{p}{0}{}) \times \left( \frac{c}{\omega}  \, \Vie{k}{1}{-} (\Vie{p}{0}{}) \times \Vie{E}{0}{} (\Vie{p}{0}{}) \right) \right] \cdot \hat{\mathbf{e}}_3 \: \exp \left[ -i (\Vie{k}{1}{-*} (\Vie{p}{0}{}) - \Vie{k}{1}{-} (\Vie{p}{0}{})) \cdot \Vie{x}{}{} \right] \dtwox \nonumber \\
&= - \frac{c^2}{8 \pi \omega} \, \mathrm{Re}  \int_S  \left[ | \Vie{E}{0}{} (\Vie{p}{0}{}) |^2 \, \Vie{k}{1}{-} (\Vie{p}{0}{})   -
\left( \Vie{E}{0}{*} (\Vie{p}{0}{})  \cdot \Vie{k}{1}{-} (\Vie{p}{0}{}) \right) \cdot \Vie{E}{0}{} (\Vie{p}{0}{})  \right] \cdot \hat{\mathbf{e}}_3 \dtwox \nonumber \\
                   &= S \, \frac{c^2}{8 \pi \omega}  \, \alpha_1 (\Vie{p}{0}{}) \, | \Vie{E}{0}{} (\Vie{p}{0}{}) |^2
\nonumber \\
&= S \, \frac{c^2}{8 \pi \omega}  \, \alpha_1 (\Vie{p}{0}{}) \, \left[ \left|\Cie{E}{0,p}{}\right|^2  + \left|\Cie{E}{0,s}{}\right|^2 \right].
\end{align}
Here, the $^*$ denotes the complex conjugate, and incident field amplitude $\Vie{E}{0}{} (\Vie{p}{0}{}) = \Cie{E}{0,p}{} \, \hat{\mathbf{e}}_{p,1}^{\mathnormal{-}} (\Vie{p}{0}{})  + \Cie{E}{0,s}{} \, \hat{\mathbf{e}}_s  (\Vie{p}{0}{})$ as defined in Eq.~(\ref{incField}), the vector identity
$\mathbf{a}\times(\mathbf{b}\times\mathbf{c}) = (\mathbf{a}\cdot\mathbf{c})\mathbf{b} - (\mathbf{a}\cdot\mathbf{b})\mathbf{c}$ and the orthogonality between the field and the wave vector $\Vie{E}{0}{*} (\Vie{p}{0}{})  \cdot \Vie{k}{1}{-} (\Vie{p}{0}{}) = 0$ have been used.
Note that the flux incident on the disk is proportional to the disk area. Let us now consider the outgoing flux crossing an elementary surface $\mathrm{d}\sigma = r^2 \sin \theta_s \mathrm{d}\theta_s \mathrm{d}\phi_s = r^2 \mathrm{d}\Omega_s$ around a point $\Vie{r}{}{} = r \, (\sin \theta_s \cos \phi_s \, \hat{\mathbf{e}}_1 + \sin \theta_s \sin \phi_s \, \hat{\mathbf{e}}_2 + \cos \theta_s \, \hat{\mathbf{e}}_3 ) = r \, \hat{\mathbf{n}}$. The flux crossing this elementary surface is given by
\begin{equation}
P_{\mathrm{d}\sigma} = \frac{c}{8 \pi} \, \mathrm{Re}  \left[ \Vie{E}{1}{+*} (\Vie{r}{}{}) \times \Vie{H}{1}{+} (\Vie{r}{}{}) \right] \cdot \hat{\mathbf{n}} \, \mathrm{d}\sigma .
\label{outpower}
\end{equation}
We then use the well-known asymptotic expansion of the field in the far-field given by (see Refs.~\cite{Agarwal1977,Miyamoto1962})
\begin{subequations}
\begin{align}
\Vie{E}{1}{+} (\Vie{r}{}{}) &\sim - i \, \epsilon_1^{1/2} \frac{\omega}{2 \pi \, c} \cos \theta_s \, \frac{\exp (i \epsilon_1^{1/2} \frac{\omega}{c} r)}{r} \, \Vie{E}{1}{+} (\Vie{p}{}{})
\\
\Vie{H}{1}{+} (\Vie{r}{}{}) &\sim - i \, \epsilon_1 \frac{\omega}{2 \pi \, c} \cos \theta_s \, \frac{\exp (i \epsilon_1^{1/2} \frac{\omega}{c} r)}{r} \, \hat{\mathbf{n}} \times \Vie{E}{1}{+} (\Vie{p}{}{})
\end{align}
\label{asymptotic:field}%
\end{subequations}
where $\Vie{p}{}{} = \sqrt{\epsilon_1} \frac{\omega}{c} (\sin \theta_s \cos \phi_s \, \hat{\mathbf{e}}_1 + \sin \theta_s \sin \phi_s \, \hat{\mathbf{e}}_2 )$. This asymptotic approximation will become more and more accurate as we let $r \to \infty$. Plugging Eq.~(\ref{asymptotic:field}) into Eq.~(\ref{outpower}) we obtain
\begin{equation}
P_{\mathrm{d}\sigma} = \epsilon_1^{3/2}  \left(\frac{\omega}{2 \pi \, c}\right)^2 \, \cos^2 \theta_s \, \frac{c}{8 \pi} \, | \Vie{E}{1}{+} (\Vie{p}{}{})|^2 \, \mathrm{d}\Omega_s = \epsilon_1^{3/2}  \left(\frac{\omega}{2 \pi \, c}\right)^2 \, \cos^2 \theta_s \, \frac{c}{8 \pi} \, \left( | \Cie{E}{1,p}{+} (\Vie{p}{}{})|^2  + | \Cie{E}{1,s}{+} (\Vie{p}{}{})|^2\right) \, \mathrm{d}\Omega_s .
\end{equation}
The total differential reflection coefficient is then given by
\begin{equation}
  \frac{\partial R}{\partial \Omega_s} (\Vie{p}{}{} \st \Vie{p}{0}{} )
  =
  \lim_{r\to\infty} \, \frac{P_{\mathrm{d}\sigma}}{P_{\mathrm{inc}/S} \, \mathrm{d}\Omega_s}
  =
  \lim_{r\to\infty} \,
  \frac{\epsilon_1}{S}
  \left(\frac{\omega}{2 \pi \, c}\right)^2
  \frac{ \cos^2 \theta_s }{ \cos \theta_0 }
  %
  \, \frac{| \Cie{E}{1,p}{+} (\Vie{p}{}{})|^2  + | \Cie{E}{1,s}{+} (\Vie{p}{}{})|^2}{| \Cie{E}{0,p}{} |^2  + | \Cie{E}{0,s}{} |^2} .
\label{total:DRC}
\end{equation}
From the total differential reflection coefficient given by Eq.~(\ref{total:DRC}), we deduce the differential reflection coefficient when an incident plane wave of polarization $\beta$, with in-plane wave vector $\Vie{p}{0}{}$ is reflected into a plane wave of polarization $\alpha$ with in-plane wave vector $\Vie{p}{}{}$ given as
\begin{equation}
  \frac{\partial R_{\alpha \beta}}{\partial \Omega_s} (\Vie{p}{}{} \st \Vie{p}{0}{} )
  =
  \lim_{r\to\infty} \,
  \frac{\epsilon_1}{S}
  \left(\frac{\omega}{2 \pi \, c}\right)^2
  \frac{ \cos^2 \theta_s }{ \cos \theta_0 }
  %
  \, \left|R_{\alpha \beta} (\Vie{p}{}{} \st \Vie{p}{0}{})\right|^2
  =
  \lim_{r\to\infty} \, \frac{\partial R_{\alpha \beta}^{(S)}}{\partial \Omega_s} (\Vie{p}{}{} \st \Vie{p}{0}{} ) .
\label{ab:DRC}
\end{equation}
As we are interested in averaging the optical response over realizations of the surface profiles, we consider the following ensemble average
\begin{equation}
  \bigg\langle \frac{\partial R_{\alpha \beta}^{(S)}}{\partial \Omega_s} (\Vie{p}{}{} \st \Vie{p}{0}{} ) \bigg\rangle
  =
  \frac{\epsilon_1}{S}
  \left(\frac{\omega}{2 \pi \, c}\right)^2
  \frac{ \cos^2 \theta_s }{ \cos \theta_0 }
  %
  \, \left\langle |R_{\alpha \beta} (\Vie{p}{}{} \st \Vie{p}{0}{})|^2 \right\rangle  .
\end{equation}

By decomposing the reflection amplitudes as the sum of the mean and fluctuation (deviation from the mean)
\begin{equation}
R_{\alpha \beta} (\Vie{p}{}{} \st \Vie{p}{0}{}) = \left\langle R_{\alpha \beta} (\Vie{p}{}{} \st \Vie{p}{0}{}) \right\rangle  + \left[ R_{\alpha \beta} (\Vie{p}{}{} \st \Vie{p}{0}{}) - \left\langle R_{\alpha \beta} (\Vie{p}{}{} \st \Vie{p}{0}{}) \right\rangle \right] \: ,
\end{equation}
we can decompose the mean differential reflection coefficient as the sum of a coherent term and an incoherent term
\begin{equation}
  \bigg\langle \frac{\partial R_{\alpha \beta}^{(S)}}{\partial \Omega_s} (\Vie{p}{}{} \st \Vie{p}{0}{} ) \bigg\rangle
  = \bigg\langle \frac{\partial R_{\alpha \beta}^{(S)}}{\partial \Omega_s} (\Vie{p}{}{} \st \Vie{p}{0}{} ) \bigg\rangle_\mathrm{coh}
  +
  \bigg\langle \frac{\partial R_{\alpha \beta}^{(S)}}{\partial \Omega_s} (\Vie{p}{}{} \st \Vie{p}{0}{} ) \bigg\rangle _\mathrm{incoh}\: ,
\end{equation}
where
\begin{subequations}
\begin{align}
  \bigg\langle \frac{\partial R_{\alpha \beta}^{(S)}}{\partial \Omega_s} (\Vie{p}{}{} \st \Vie{p}{0}{} ) \bigg\rangle_\mathrm{coh}
  &=
  \frac{\epsilon_1}{S}
  \left(\frac{\omega}{2 \pi \, c}\right)^2
  \frac{ \cos^2 \theta_s }{ \cos \theta_0 }
  %
    \, \left| \left\langle R_{\alpha \beta} (\Vie{p}{}{} \st \Vie{p}{0}{}) \right\rangle \right|^2
  \\
  \bigg\langle \frac{\partial R_{\alpha \beta}^{(S)}}{\partial \Omega_s} (\Vie{p}{}{} \st \Vie{p}{0}{} ) \bigg\rangle_\mathrm{incoh}
  &=
  \frac{\epsilon_1}{S}
  \left(\frac{\omega}{2 \pi \, c}\right)^2
  \frac{ \cos^2 \theta_s }{ \cos \theta_0 }
  %
    \, \left[ \left\langle |R_{\alpha \beta} (\Vie{p}{}{} \st \Vie{p}{0}{})|^2 \right\rangle - \left| \left\langle R_{\alpha \beta} (\Vie{p}{}{} \st \Vie{p}{0}{}) \right\rangle \right|^2 \right].
\label{mdrc:inco}%
\end{align}
\end{subequations}
If we now use the expression found in \ref{AppendixA} for the reflection amplitudes to first order in the product of surface profiles,
\begin{equation}
\mathbf{R} \ofuipi{p}{}{p}{0} \approx  \Vie{R}{}{(0)} \ofuipi{p}{}{p}{0} - i \Vie{R}{}{(1)} \ofuipi{p}{}{p}{0} \: ,
\end{equation}
where $\Vie{R}{}{(0)} \ofuipi{p}{}{p}{0}$ is the response from the corresponding system with flat interfaces (i.e. that of a Fabry-Perot interferometer), Eq.~(\ref{fresnel:r}), and $\Vie{R}{}{(1)} \ofuipi{p}{}{p}{0}$ is given in Eq.~(\ref{r1}), we obtain that the factor in the square bracket in Eq.~(\ref{mdrc:inco}) reads
\begin{align}
  \left\langle |R_{\alpha \beta} (\Vie{p}{}{} \st \Vie{p}{0}{})|^2 \right\rangle - \left| \left\langle R_{\alpha \beta} (\Vie{p}{}{} \st \Vie{p}{0}{}) \right\rangle \right|^2
  &= \left\langle \left|R_{\alpha \beta}^{(1)} (\Vie{p}{}{} \st \Vie{p}{0}{})\right|^2 \right\rangle
    \nonumber \\
&= \left\langle | \hat{h}_{1,S} (\Vie{p}{}{} - \Vie{p}{0}{}) |^2 \right\rangle \: |\rho_{1,\alpha \beta} \ofuipi{p}{}{p}{0} |^2  + \left\langle | \hat{h}_{2,S} (\Vie{p}{}{} - \Vie{p}{0}{}) |^2 \right\rangle \: |\rho_{2,\alpha \beta} \ofuipi{p}{}{p}{0} |^2  \nonumber\\
&\quad + 2 \: \mathrm{Re}  \left\langle \hat{h}_{1,S} (\Vie{p}{}{} - \Vie{p}{0}{}) \hat{h}_{2,S}^{*} (\Vie{p}{}{} - \Vie{p}{0}{}) \right\rangle  \left( \rho_{1,\alpha \beta} \ofuipi{p}{}{p}{0} \rho_{2,\alpha \beta}^* \ofuipi{p}{}{p}{0}  \right)  .
\end{align}
Note here that we are still dealing with a scattering system whose surface profiles are flat outside the disk of radius $r$, hence the subscript $S$.  For the statistical properties attributed to the surface profiles in Sec.~\ref{sec:scatt:sys}, we have
\begin{align}
\left\langle \hat{h}_{i,S} (\Vie{q}{}{}) \hat{h}_{j,S}^* (\Vie{q}{}{}) \right\rangle &= \left\langle \int_S \int_S h_{i} (\Vie{x}{}{}) h_{j} (\mathbf{x}') \exp \left[i \Vie{q}{}{} \cdot (\Vie{x}{}{} - \mathbf{x}') \right] \: \mathrm{d}^2x \:  \mathrm{d}^2x' \right\rangle \nonumber \\
&=  \int_S \int_S \left\langle h_{i} (\Vie{x}{}{}) h_{j} (\mathbf{x}') \right\rangle \: \exp \left[i \Vie{q}{}{} \cdot (\Vie{x}{}{} - \mathbf{x}') \right]  \: \mathrm{d}^2x \:  \mathrm{d}^2x'  \nonumber \\
&=  \int_S \int_S \gamma_{ij} \: W(\Vie{x}{}{} - \mathbf{x}') \: \exp \left[i \Vie{q}{}{} \cdot (\Vie{x}{}{} - \mathbf{x}') \right] \: \mathrm{d}^2x \:  \mathrm{d}^2x'  .
\end{align}
Here we have used the definition of the Fourier transform, and the fact that ensemble average commutes with the integration of the surfaces and the definition of the correlation function. We have also introduced the shorthand $\gamma_{ij} = \left[\delta_{ij} + \gamma (1-\delta_{ij}) \right] \, \sigma_i \, \sigma_j$. Via the change of variable $\Vie{u}{}{} = \Vie{x}{}{} - \Vie{x}{}{}'$ we obtain
\begin{equation}
\left\langle \hat{h}_{i,S} (\Vie{q}{}{}) \hat{h}_{j,S}^* (\Vie{q}{}{}) \right\rangle = S \: \gamma_{ij} \: \int_S W(\Vie{u}{}{}) \: \exp (i \Vie{q}{}{} \cdot \Vie{u}{}{} ) \: \mathrm{d}^2u  = S \: \gamma_{ij} \: g_S (\Vie{q}{}{})  .
\label{pow:spec:trunc}
\end{equation}
Thus
\begin{align}
  \left\langle |R_{\alpha \beta} (\Vie{p}{}{} \st \Vie{p}{0}{})|^2 \right\rangle - \left| \left\langle R_{\alpha \beta} (\Vie{p}{}{} \st \Vie{p}{0}{}) \right\rangle \right|^2
  = S \: g_S (\Vie{p}{}{} -\Vie{p}{0}{}) \, \Big[ &\sigma_1^2 \: \left|\rho_{1,\alpha \beta} \ofuipi{p}{}{p}{0} \right|^2  + \sigma_2^2 \: \left|\rho_{2,\alpha \beta} \ofuipi{p}{}{p}{0} \right|^2 \nonumber \\
& + 2 \gamma \sigma_1 \sigma_2 \: \mathrm{Re} \left\{ \rho_{1,\alpha \beta} \ofuipi{p}{}{p}{0} \rho_{2,\alpha \beta}^* \ofuipi{p}{}{p}{0}  \right\}  \Big]
\end{align}
Finally, by plugging the above equation into Eq.~(\ref{mdrc:inco}), the surface area $S$ cancels and letting $r \to \infty$, $g_S \to g$ (where we remind the reader that $g$ is the power spectrum of the surface profiles) and we finally obtain the expression for the incoherent component of the mean differential reflection coefficient for the entire (infinite) system under the first order approximation of the reflected amplitudes in product of the surface profiles
\begin{align}
\left\langle \frac{\partial R_{\alpha \beta} (\Vie{p}{}{} | \Vie{p}{0}{})}{\partial \Omega_s} \right\rangle_{\mathrm{incoh}} =& \: \epsilon_1 \left(\frac{\omega}{2 \pi c} \right)^2 \: \frac{\cos^2 \theta_s}{\cos \theta_0} \: g (\Vie{p}{}{} -\Vie{p}{0}{}) \:
\Big[ \sigma_1^2 \: \left|\rho_{1,\alpha \beta} \ofuipi{p}{}{p}{0} \right|^2  + \sigma_2^2 \: \left|\rho_{2,\alpha \beta} \ofuipi{p}{}{p}{0} \right|^2  \nonumber \\ &+ 2 \gamma \sigma_1 \sigma_2 \: \mathrm{Re} \left\{ \rho_{1,\alpha \beta} \ofuipi{p}{}{p}{0} \rho_{2,\alpha \beta}^* \ofuipi{p}{}{p}{0}  \right\}  \Big] .
\label{eq:incoMDRC:final:app}
\end{align}

\section{Contrast estimates}\label{AppendixC}

\noindent We propose here to motivate mathematically that the phase mixing in paths of type (1''), (2'') \emph{etc.}, from Fig.~\ref{fig:toy}(b) and those from Fig.~\ref{fig:toy}(c) intrinsically leads to poorer contrast in the interference pattern found in the incoherent contribution to the mean DRC than, for example, paths of type (1), (2) in Fig.~\ref{fig:toy}(a), where no phase mixing is allowed.
As a prototypical reflection amplitude for a sum of paths that involves phase mixing and a sum of paths that does not (and will serve as reference), let us have respectively
\begin{subequations}
  \begin{align}
    r_{\mathrm{mix} \varphi}
    =&
       \frac{\tilde{r}}{\left[1- r_0 \exp (2 i \varphi_0) \right] \, \left[1 - r_s \exp(2 i \varphi_s) \right]}
    \\
    r_{\mathrm{ref}}
    =&
       \frac{\tilde{r}}{1 - r_s \exp(2 i \varphi_s)} .
  \end{align}
  \label{eq:r:mimic}%
\end{subequations}
These reflection amplitudes mimic the structure from Eqs.~(\ref{eq:toy:fr}) and Eq.~(\ref{eq:toy:rf:0}) respectively, but we will see that the precise expressions for the numerators do not matter for the contrast, and are hence denoted by the same symbol $\tilde{r}$. Note that all the reflection amplitudes in Eq.~\eqref{eq:r:mimic} depend on angles of incidence and scattering, but for clarity we drop these arguments.
Our first step consists in taking the square modulus of Eq.~(\ref{eq:r:mimic})
\begin{subequations}
\begin{align}
  I_{\mathrm{mix} \varphi}
  =&
     \frac{|\tilde{r}|^2}{ \left|1- r_0 \exp (2 i \varphi_0)\right|^2
     \, \left| 1 - r_s \exp(2 i \varphi_s) \right|^2}
  \\
 I_{\mathrm{ref}} =& \frac{|\tilde{r}|^2}{| 1 - r_s \exp(2 i \varphi_s )|^2} \: ,
\end{align}
\label{eq:I:mimic}%
\end{subequations}
and in bounding the intensity by using the triangular inequality
\begin{subequations}
\begin{align}
\frac{|\tilde{r}|^2}{( 1 + |r_0| )^2 \, ( 1 + | r_s | )^2} \quad \leq & \quad I_{\mathrm{mix} \varphi} \quad \leq \quad \frac{|\tilde{r}|^2}{( 1 - |r_0| )^2 \, ( 1 - | r_s | )^2} \\
\frac{|\tilde{r}|^2}{( 1 + | r_s | )^2}\quad \leq & \quad I_{\mathrm{ref}} \quad \leq \quad \frac{|\tilde{r}|^2}{( 1 - | r_s | )^2} .
\end{align}
\label{eq:I:mimic:bound}%
\end{subequations}
It is clear from Eq.~(\ref{eq:I:mimic:bound}) that the intensity lies between two bounding curves.
A fair estimate for the trend, i.e. the intensity without the oscillations would be given by $|\tilde{r}|^2$, and we thus estimate, or rather define, the \emph{inverse contrast} as
\begin{subequations}
\begin{align}
\eta_{\mathrm{mix} \varphi}^{-1} &= ( 1 + |r_0| )^2 \, ( 1 + | r_s | )^2  - ( 1 - |r_0| )^2 \, ( 1 - | r_s | )^2 \\
\eta_{\mathrm{ref}}^{-1} &= ( 1 + | r_s | )^2 - ( 1 - | r_s | )^2 .
\end{align}
\label{eq:eta}%
\end{subequations}
This may not be the most \emph{natural} definition for the contrast, but we choose this one since it is easier to work with and will not change the conclusion. By re-writing Eq.~\ref{eq:eta} by using straightforward algebra, we obtain
\begin{subequations}
  \begin{align}
    \eta_{\mathrm{mix} \varphi}^{-1}
    &=
      4 | r_s | + 4 |r_0| + 4 |r_0| |r_s| + 4 |r_0|^2 |r_s|
      \label{eta:mix}
    \\
    \eta_{\mathrm{ref}}^{-1}
    &=
      4 | r_s | .
\end{align}
\label{eq:eta:end}%
\end{subequations}
This shows that the inverse contrast for phase mixing is larger than that of the reference, i.e. that the \emph{contrast} in the case of phase mixing is smaller than that of the reference. Indeed, the two last terms in Eq.~(\ref{eta:mix}) are cross-terms resulting directly from the phase mixing nature of the initial reflection amplitude.
Note that the choice for the reference was arbitrary and one could choose to study paths of type (1'), (2'), \emph{etc.}, in Fig.~\ref{fig:toy}(b), and hence replace $r_s \exp(2 i \varphi_s)$ in Eq.~(\ref{eq:r:mimic}) by $r_0 \exp(2 i \varphi_0)$, and the conclusion would still hold.

\end{document}